\documentclass[11pt]{article}
\usepackage[T1]{fontenc}
\usepackage[margin=1in]{geometry}
\usepackage{lmodern,amsmath,amssymb,amsthm,booktabs,graphicx,array,longtable}
\usepackage[authoryear,round]{natbib}
\usepackage{xcolor}
\usepackage[colorlinks=true,allcolors=blue]
{hyperref}
\usepackage{microtype}
\usepackage{needspace}
\usepackage{placeins}
\newtheorem{theorem}{Theorem}
\newtheorem{proposition}{Proposition}
\newtheorem{lemma}{Lemma}
\newtheorem{corollary}{Corollary}
\newtheorem{example}{Example}
\theoremstyle{remark}\newtheorem{remark}{Remark}
\theoremstyle{definition}\newtheorem{assumption}{Assumption}
\newcommand{\E}{\mathbb E}
\newcommand{\Var}{\operatorname{Var}}
\newcommand{\Cov}{\operatorname{Cov}}
\newcommand{\ind}{\mathbf 1}
\hypersetup{pdftitle={Double Machine Learning for Panel Data: Identification, Specification, and a Bank Regulation Effect},pdfauthor={Ali Habibnia; Esfandiar Maasoumi; Gaurav Shah}}
\title{Double Machine Learning for Panel Data:\\ Identification, Specification, and a Bank Regulation Effect}
\author{Ali Habibnia$^{a,d}$ \and Esfandiar Maasoumi$^{b}$ \and Gaurav Shah$^{c,d}$}
\date{\small $^a$Department of Economics, Virginia Tech\\
$^b$Department of Economics, Emory University\\
$^c$Department of Computer Science, Virginia Tech\\
$^d$Dataism Laboratory for Quantitative Finance
}
\begin{document}
\maketitle
\begin{abstract}
This paper examines robustness and causal interpretation of panel estimates using double/debiased machine learning. The prototype treatment setup is the 2018 US banking regulation. Under common treatment timing, we characterize identification and nuisance restrictions and derive exact bank-level representations of specified "within" and "first-difference" scores. Simulations assess panel model transformation-induced specification bias and inference under weaker overlap. Empirically, estimates vary with asset-size classification and prior growth. Prior-growth adjustment reduces overlap-weighted estimates across four Machine Learners (ML); doubly robust estimates depend on the fitted propensity and concentrated weights. The banking application compares changes in risk-weighted-asset growth across two groups exposed to different regulatory relief. The findings highlight specification issues and the relation between functional-form robustness and causal identification.

\end{abstract}
\par\vfill\noindent\rule{2in}{0.3pt}\par\noindent{\footnotesize Email addresses (in author order): \href{mailto:habibnia@vt.edu}{habibnia@vt.edu}; \href{mailto:esfandiar.maasoumi@emory.edu}{esfandiar.maasoumi@emory.edu}; \href{mailto:gauravs19@vt.edu}{gauravs19@vt.edu}.}\par

\clearpage
\section{Introduction}\label{sec:introduction}
The 2018 US banking reform (of Dodd-Frank) relaxed regulatory requirements for bank holding companies, with different implications across asset-size groups. We reassess differences in quarterly risk-weighted-asset growth around the reform. The panel model of \citet{chronopoulos2023} serves as the benchmark for the application of double/debiased machine learning (DML) to examine robustness to nonlinear transformations. In panel models, Fixed Effects (FE), First Differences (FD) and Correlated Random Effects (CRE) can produce different "nuisance" functions for DML, require different restrictions and even target different treatment effects. We study these distinctions to determine what robustness across estimators establishes and which conditions support a causal interpretation.

The application compares 91 US bank holding companies over twenty quarters, 2015Q2 through 2020Q1. Banks with 2018Q1 assets of \$50--250 billion are compared with banks of \$10--50 billion. Both groups received relief, so the comparison concerns differential regulatory exposure. Its causal interpretation requires the identifying assumptions stated below. Based on several measures of risk behaviour, we provide a quarterly growth analysis using matched specifications, take explicit account of treatment-effect targets, and investigate the impact of bank classification, prior growth and outcome construction.

DML embodies the Neyman Orthogonality (NO) idea for flexible estimation of nuisance functions for (first order) robust estimation of target objects in estimating equations \citep{chernozhukov2018}. In partially linear regression models, nuisance functions are usually conditional means of the outcome and treatment variables. The first step auxiliary (ML) fitted residuals determine the coefficient of interest in a Neyman orthogonal score that removes first-order nuisance estimation errors. Sample splitting and cross-fitting separate nuisance training from score evaluation and address overfitting. In our panel context, this means block splitting bank histories for training and test samples. Subject to appropriate nuisance rates and sampling conditions, these procedures limit regularisation and nuisance-estimation bias. This DML robustification assumes identification objects and conditions are invariant and given, an issue worthy of care with panel data transformations, as we make clear.

DML also accommodates ML estimation with expanded, potentially high-dimensional covariate representations. Our application includes polynomial LASSO with 33 baseline terms (42 when prior growth feature is added), linear/logistic fitting, forests and boosted trees. The 91 bank histories (rather than the 1,820 bank-quarter observations), are the basis for training samples. 

Panel transformations may imply consequential changes in the nuisance functions. In the classic linear panel models, the Within estimator and an appropriately specified Mundlak regression yield the same slopes (inference targets) \citep{mundlak1978,baltagi2023,wooldridge2025}. Transforming the inputs of a nonlinear function, is generally different from the transformed function itself. FD may require current and lagged covariates, while the Within conditional mean can depend on the whole covariate history. \citet[Sections~5.1.1--5.1.2]{clarke2026} study the bank-mean adjustment and nonlinear transformation problem; \citet{fuhr2024double} examine CRE-based DML. Related DiD analyses show that covariate adjustment and transformation would impose substantive new restrictions \citep{caetano2022difference,caetano2024conditional}.

Our formal treatment setup concerns one treated cohort. The literature on variation in adoption dates identifies additional issues involving cohort effects and aggregation \citep{callaway2021,sun2021,goodmanbacon2021,roth2023whats}. We isolate consequences of panel transformations and nuisance restrictions that arise even with a common treatment date.

The paper makes four related contributions. First, we assess the robustness of conventional panel treatment-effect estimates to flexible and regularised nuisance estimation. In matched cases we hold samples, conditioning information, time adjustment and fitting procedures the same where feasible. We distinguish changes in nuisance approximation from changes in the panel model, identifying restrictions and different treatment-effect targets. We replicate prior findings and assess robustness to functional-forms (agreement across machine learners is not evidence sufficient for causal identification).

Second, with panel transformation and bank-mean adjustments similar to \citet{clarke2026}, we characterise identification and treatment-effect targets under common treatment timing. Unrestricted adjustment on treatment means and calendar information can eliminate residual treatment variation, while restricted classes can identify the coefficient under compatible outcome-model and rank conditions. Exact identities reduce specified Within and FD scores to bank-level residual regressions for pre/post and adoption-quarter changes. At the true conditional means, and under the respective dynamic outcome restrictions, the treatment targets are weighted averages of post-treatment and adoption-quarter effects, weighted by conditional variances. Such weighting is noted in \citep{angrist1998,li2018balancing,lal2025representative}; our contribution is its explicit connection to these panel specifications, prediction restrictions and effect horizons. We distinguish these weighted targets from the unweighted average treatment effect on the treated (ATT), using established doubly robust DiD methods as a separate ATT comparator \citep{chang2020,santanna2020}. Our inference results specialise DML arguments to the bank-level ratios, and give the additional nuisance-rate and sampling conditions. We clarify inference with treatment-stratified folds and repeated-partition aggregation.

Third, the banking analysis evidences both stability and specification sensitivity. In the replicated linear benchmark treatment effect is 0.0193, close to the Random Forest (RF) estimate of 0.0181 on the same two-way demeaned inputs. Excluding banks whose endpoint asset bands had changed, the coefficient in the linear case is reduced to 0.0122. In bank-level matched comparisons, including the "prior growth" feature reduces partialling-out estimates from OLS/logit, forests, polynomial LASSO and XGBoost. The DR-DiD estimates are more sensitive to the fit of the propensity score: for loss-minimising polynomial LASSO, the median largest normalised control weight is approximately 87\%, and the one-standard-error penalty rule reduces this concentration. Definition of the "Outcome" variable, influential asset changes, and regulatory timing, further impact the estimates. These findings identify consequential choices in this bank regulation setting.

Our final contribution from the theoretical and empirical findings is to provide practical guidance for implementing and interpreting panel DML. The intended inference target and identifying restrictions guide the choice of panel transformation, nuisance information sets and inference procedure. Matched nonlinear-basis comparisons and simulations show the consequences of misspecifying the transformed regression and of targeting different effect horizons. The coverage experiments indicate under coverage for DR-DiD under weaker overlap in the designs considered. The empirical results explain why cross-fitting and predictive tuning must be accompanied by checks of residual treatment variation, propensity support, control-weight concentration and split sensitivity. These diagnostics inform interpretation; they do not verify the identifying assumptions or guarantee valid confidence intervals.

Section~\ref{sec:setup} introduces the bank regulation treatment setup, the variables and basic identifying assumptions. Section~\ref{sec:panel-specifications} compares the panel transformations, and Section~\ref{sec:method} develops the restrictions, score representations and treatment-effect targets. Section~\ref{sec:estimation} states the estimation and inference conditions. Section~\ref{sec:simulation} presents simulation evidence, and Section~\ref{sec:emp} examines the banking application. Section~\ref{sec:conclusion} concludes. The Supplement contains proofs, additional analyses and reproducibility details.
\section{Bank regulation, data and the panel model}\label{sec:setup}\label{sec:emp:reform}
\subsection{Prior analysis of the regulation}
\citet{chronopoulos2023} examine changes in bank risk decisions around the Economic Growth, Regulatory Relief, and Consumer Protection Act (EGRRCPA), enacted on 24 May 2018. \footnote{We use data supplied by Dimitris K. Chronopoulos and Muhammed H. Yilmaz to revisit their comparison of bank holding companies} Bank holding companies (Banks) with assets of \$50--250 billion and \$10--50 billion are differentially treated groups/states. We examine how the estimated difference in "risk asset growth" depends on panel model specification, nonlinear covariate adjustment and the construction of bank comparison.

Section 401 of the Act provided immediate relief from enhanced prudential standards by size, below \$100 billion, with delayed provisions and Federal Reserve discretion above that threshold \citep{egrrcpa2018}. The treatment effect setting is one of differential regulatory exposure, rather than the overall effect of a reform relative to no reform. The Federal Reserve's July 2018 statement provided company-run stress-test relief for institutions with \$10--100 billion in assets and risk-committee relief for publicly traded holding companies below \$50 billion \citep[pp.~4, 6]{fed2018statement}. Earlier changes also matter: the Federal Reserve proposed removing qualifying large, noncomplex firms from qualitative capital-plan assessment in September 2016 and adopted the rule in January 2017 for that year's assessment \citep{fed2016ccar,fed2017ccar}. The 2019 tailoring framework subsequently incorporated additional risk indicators \citep{fed2019}. The \$10--50 billion comparison banks were not directly affected by the specific increase in the enhanced-prudential-standards asset threshold examined by \citet{chronopoulos2023}, which motivates their use as a comparison group for that provision. However, banks in this group also received relief under other EGRRCPA provisions.

\citet{chronopoulos2023} conduct identification diagnostics and robustness checks beyond their baseline two-way fixed-effects regression. They examine state-by-time fixed effects, propensity-score matching, entropy balancing, alternative samples and event windows, placebo tests, event-study dynamics and an imputation-based DiD estimator \citep[Section~3.2, Figure~1 and Table~4]{chronopoulos2022wp}. They also consider several measures of bank risk; our reassessment concerns quarterly risk-weighted-asset growth. We further assess robustness to nonlinear nuisance estimation and characterise how panel transformations, identifying restrictions and treatment-effect weighting affect estimation and interpretation. Our empirical analysis uses the supplied data and reconstructed benchmark specifications; Supplementary Section~\ref{app:status} documents the data provenance and the unavailable parts of the original data-construction procedure.

\subsection{Sample, variables and treatment coding}\label{sec:emp:data}
The data contain 1,911 bank-quarter records for 91 banks over 21 quarters. Forming quarterly growth removes the first observation for each bank. The estimation panel is balanced, with $N=91$ banks and $T=20$ quarters from 2015Q2 through 2020Q1, giving 1,820 observations. There are 20 banks in the larger group and 71 in the smaller comparison group. Twelve quarters precede the reform and eight follow it. \footnote{One missing liquidity observation is replaced by the estimation-sample median in the retained benchmark; deleting it is a separate sensitivity calculation.}

Let $i=1,\ldots,N$ index banks and $t=1,\ldots,T$ index quarters. Write $G_i=1$ for banks with 2018Q1 assets in the \$50--250 billion band and $G_i=0$ for banks in the \$10--50 billion band. Quarter $t^*=13$ corresponds to 2018Q2. Define
\begin{equation}
P_t=\ind\{t\geq t^*\},\qquad D_{it}=G_iP_t.
\label{eq:common-timing}
\end{equation}
Here $P_t$ is the post-enactment indicator and $D_{it}$ is the treatment variable used in the regression. The coding assigns the same start quarter to every bank in the larger group and keeps $D_{it}=1$ thereafter. This is an enactment-based design, rather than a bank-specific effective-date design. Section 401(d) generally delayed the threshold change for eighteen months, made it immediate for holding companies below \$100 billion, and allowed earlier exemptions at Federal Reserve discretion \citep{egrrcpa2018}. Thus the common start date does not mean simultaneous regulatory relief. Section~\ref{sec:emp:outcome-timing} reports the existing statutory-timing sensitivity analysis.

Let $RWA_{it}$ denote risk-weighted assets, $A_{it}$ total assets, and $\Delta R_{it}=R_{it}-R_{i,t-1}$ the first difference of any variable $R$. The outcome $Y_{it}$ is the supplied winsorised quarterly log change in $RWA_{it}$. It is reproduced exactly by clipping raw growth at $-0.048185002$ and $0.27956116$, the empirical first and ninety-ninth percentiles of the 1,820 observed growth values. This equivalent rule is consistent with pooled full-period winsorisation; the original program and cutoff sample remain unverified. It alters 36 observations, including twelve after enactment and the BB\&T merger quarter in 2019Q4. We retain the supplied outcome in all sample sensitivities. A coefficient of $0.01$ corresponds approximately to one percentage point of quarterly growth. Before winsorisation, the accounting identity
\begin{equation}
\Delta\log RWA_{it}=\Delta\log A_{it}
 +\Delta\log(RWA_{it}/A_{it})
\label{eq:decomp}
\end{equation}
separates balance-sheet growth from changes in regulatory risk density. Risk-weighted-asset growth is therefore not a pure measure of changes in portfolio risk.

The vector $X_{it}$ contains the six contemporaneous controls used in the benchmark literature: deposit funding, provisions, operating efficiency, dividends, derivatives and liquidity ratios. The original variable definitions \citep[Table~1]{chronopoulos2022wp} measure interest-bearing deposits, dividends, trading derivatives and liquid assets relative to total assets; provisions are measured relative to total loans; and operating efficiency is noninterest expense divided by noninterest income plus net interest income. \footnote{We retain the available winsorised variables and their scaling rather than reconstructing these ratios from regulatory filings. The original source columns and the available construction information are listed in Supplementary Section~\ref{app:status}. }

Because these controls may respond to the reform, their inclusion requires a substantive outcome-model justification and does not by itself identify the total effect of regulatory relief.

\Needspace{10\baselineskip}
\subsection{The linear benchmark}
The reference model is
\begin{equation}
Y_{it}=\alpha_i+\lambda_t+\theta_0D_{it}+X_{it}'\beta+U_{it}.
\label{eq:linear-panel-framework}
\end{equation}
The bank effect $\alpha_i$ captures additive time-invariant differences, $\lambda_t$ is a quarter effect common to all banks, $\beta$ is the vector of control coefficients, and $U_{it}$ is the remaining error. The coefficient $\theta_0$ measures the change in the conditional difference between the two bank groups after enactment within this model. Interpreting it as a causal effect requires additional restrictions. The fixed effects account for bank and quarter-specific components within this specified outcome model.

\enlargethispage{2pt}
The benchmark model gives an estimated effect of $0.0193$, with a bank-clustered standard error of $0.0045$.\footnote{CR1 multiplies the uncorrected clustered covariance by $N/(N-1)\times(n-1)/(n-k)$, where $n$ counts bank-quarter observations and $k$ is the full design rank, including fixed effects. Here $N=91$, $n=1{,}820$ and $k=117$. Matched comparisons use a common rank convention. The original benchmark, checked against Table~2, column~(2), of the accessible working-paper version \citep{chronopoulos2022wp}, reports $0.0195$ with standard error $0.0044$ and 1,819 observations. Deleting the single imputed liquidity observation gives $0.01952$ on 1,819 observations, recovering the published coefficient and sample size at the reported precision. Our bank-clustered standard error with the CR1 finite-sample degrees-of-freedom correction is $0.00455$; the corresponding uncorrected cluster standard error is $0.00438$. The latter rounds to the published value, but the historical covariance convention has not been verified. We retain the supplied-data benchmark and report the deletion separately. Section~\ref{sec:emp} compares the estimates on explicitly stated samples and specifications.}

Equation~\eqref{eq:linear-panel-framework} is useful for organising the analysis. Fixed effects (FE) and first differences (FD) remove $\alpha_i$. A correlated random effects (CRE) specification instead models the association between the bank effect and the regressors, commonly by including bank (time) averages of the regressors \citep{mundlak1978}. Section~\ref{sec:panel-specifications} first compares these procedures in the linear model and then replaces $X_{it}'\beta$ by a nonlinear function. This order separates the consequences of a panel transformation from those of fitting the unknown nonlinear functions by machine learning.

\subsection{The exposure contrast and the role of identifying assumptions}
Let $\mathcal T_0$ and $\mathcal T_1$ denote the sets of pre- and post-enactment quarters, and let $q=|\mathcal T_1|/T$ be the post-period share. For bank $i$, define its average pre-period outcome, average post-period outcome and their difference as
\begin{equation}
H_i^{\mathrm{pre}}=\frac1{|\mathcal T_0|}\sum_{t\in\mathcal T_0}Y_{it},\quad
H_i^{\mathrm{post}}=\frac1{|\mathcal T_1|}\sum_{t\in\mathcal T_1}Y_{it},\quad
H_i=H_i^{\mathrm{post}}-H_i^{\mathrm{pre}}.
\label{eq:collapsed}
\end{equation}
Since $Y_{it}$ is quarterly growth, $H_i$ is a change in average growth, not cumulative growth over the sample. Without additional controls, the two-way fixed-effects coefficient on $D_{it}$ equals the difference in mean $H_i$ between the groups. A causal interpretation of this difference requires a restriction on their unobserved outcomes.

To describe the application without assuming that the smaller banks were unaffected, let $H_i^R$ and $H_i^0$ be the change in average growth under the actual reform and under no reform, respectively. Define $\tau_g^R=\E[H_i^R-H_i^0\mid G_i=g]$ for group $g\in\{0,1\}$ and $\kappa=\E[H_i^0\mid G_i=1]-\E[H_i^0\mid G_i=0]$. Then
\begin{equation}
\E[H_i\mid G_i=1]-\E[H_i\mid G_i=0]
=\tau_1^R-\tau_0^R+\kappa.
\label{eq:reform-contrast}
\end{equation}
Parallel trends relative to no reform sets $\kappa=0$; it does not set the comparison group's reform response $\tau_0^R$ to zero. Interpreting an exposure contrast as a causal effect requires well-defined alternative regulatory environments and a restriction justifying the use of comparison banks for the larger banks' reference outcomes. Shared relief does not establish that restriction, and the two regulatory packages need not be nested.

We therefore report the banking estimates as comparisons between regulatory exposures. The causal results below state assumptions under which the procedures identify an effect for a well-defined pair of exposure paths. Covariate adjustment introduces a further restriction. Group membership is deterministic given the 2018Q1 asset-size classification, because the two bands are disjoint. Unrestricted conditioning on that exact classification variable makes the propensity zero or one and eliminates group overlap. Other pre-treatment covariates can potentially account for relevant size-related outcome trends under substantive restrictions. Their inclusion, or the omission of the classification variable, cannot itself establish conditional parallel trends. Residual size-related differences in counterfactual growth remain an identifying concern even when fitted propensities are nondegenerate.

\subsection{Difference-in-differences assumptions for the pre/post comparison}\label{sec:setup:target}
The following assumptions state when a pre/post comparison has a causal interpretation. Let $Z_i$ be a vector of pre-treatment covariates. In the application it first contains the six control ratios averaged over the twelve pre-enactment quarters; a second specification adds $H_i^{\mathrm{pre}}$. Define $p(Z_i)=\Pr(G_i=1\mid Z_i)$, the conditional probability of membership in the larger group, and $\pi=\Pr(G_i=1)$, the population share in the larger group. This is the bank-level propensity score relative to $Z_i$, even though membership is deterministic given the richer information in baseline assets. Its causal use requires conditional parallel trends given the chosen $Z_i$. Later panel transformations have their own treatment conditional means, which depend on the transformation and conditioning variables.

For the causal argument, let $\mathbf d^1=(P_1,\ldots,P_T)$ and $\mathbf d^0=\mathbf 0_T$ denote two well-defined exposure paths. In the banking interpretation, $\mathbf d^1$ indexes the larger-bank regulatory exposure from enactment, while $\mathbf d^0$ indexes the reference regulatory environment represented by the comparison group, including relief applicable to those banks. Zero denotes the reference exposure, not no reform. The binary coding alone does not establish well-defined counterfactual outcomes under both environments or make comparison banks valid counterfactuals for larger banks. Let $Y_{it}(\mathbf d^g)$ denote bank $i$'s potential outcome in quarter $t$ under path $g\in\{0,1\}$. Write $Y_{it}(g)$ as an abbreviation, $\bar Y_{i,\mathcal T_j}(\mathbf d^g)=|\mathcal T_j|^{-1}\sum_{t\in\mathcal T_j}Y_{it}(\mathbf d^g)$ for its period average, and $H_i(g)$ for its post-minus-pre average as in equation~\eqref{eq:collapsed}. The period-specific and average post-period effects on treated banks are
\begin{equation}
\tau_t=\E[Y_{it}(1)-Y_{it}(0)\mid G_i=1],\qquad
\bar\tau=\frac1{|\mathcal T_1|}\sum_{t\in\mathcal T_1}\tau_t.
\label{eq:att}
\end{equation}
Thus $\bar\tau$ is the average treatment effect on the treated (ATT) for the specified paths. In particular, $H_i(0)$ concerns the comparison-group reference environment, whereas $H_i^0$ in equation~\eqref{eq:reform-contrast} concerns no reform. Their equality is not assumed. Using comparison-bank changes to identify the larger banks' reference outcomes requires the conditional parallel-trends restriction below, together with the other stated assumptions.

\begin{assumption}[Consistency and no interference]\label{ass:consistency}
$Y_{it}=Y_{it}(\mathbf d^{G_i})$, and one bank's exposure does not affect another bank's potential outcomes.
\end{assumption}
\begin{assumption}[No anticipation]\label{ass:noanticipation}
$Y_{it}(1)=Y_{it}(0)$ for every pre-treatment quarter $t\in\mathcal T_0$.
\end{assumption}
\begin{assumption}[Conditional parallel trends]\label{ass:pt}
For the stipulated pre-treatment covariates $Z_i$,
\begin{equation}
\E[H_i(0)\mid G_i=1,Z_i]=\E[H_i(0)\mid G_i=0,Z_i].
\label{eq:cpt}
\end{equation}
With no conditioning covariates, the restriction is
\begin{equation}
\E[H_i(0)\mid G_i=1]=\E[H_i(0)\mid G_i=0].
\label{eq:unconditional-pt}
\end{equation}
\end{assumption}
\begin{assumption}[Overlap for the ATT]\label{ass:support}
$\pi\in(0,1)$ and $p(Z_i)<1$ almost surely on the covariate support of treated banks.
\end{assumption}
\begin{assumption}[Sampling]\label{ass:sampling}
Complete bank records $W_i=(Y_{i1},\ldots,Y_{iT},G_i,X_{i1},\ldots,X_{iT})$ are independent draws from a common distribution over a fixed calendar. Dependence within a bank is unrestricted. The displayed potential-outcome contrasts are integrable, and variables used in projections are square integrable.
\end{assumption}

Two timing issues bear on these assumptions in the application. The Senate passed S.2155 on 14 March 2018, within the final quarter classified as pre-enactment \citep{senate2018vote}. Possible anticipation is not confined to that quarter: the Senate Banking Committee advanced the bill on 5 December 2017 \citep{senate2017committee}. Responses to expected relief could therefore affect earlier pre-enactment outcomes and covariate averages; these legislative dates alone do not establish that banks responded. The earlier CCAR rule exempted qualifying firms with assets of \$50--250 billion, nonbank assets below \$75 billion and no global systemically important bank designation from qualitative assessment \citep{fed2017ccar}. Its eligibility was not identical to the asset-band indicator. This earlier differential exposure bears on the maintained trend comparison, rather than mechanically constituting anticipation of the 2018 Act. Section~\ref{sec:emp:outcome-timing} reports existing window comparisons without treating them as isolated policy experiments.

Competitive spillovers between banks would violate the no-interference restriction; their direction is not established here. The sampling condition is separate from the causal restrictions. It allows serial dependence within a bank but excludes remaining dependence across banks, including heterogeneous exposure to unmodelled common shocks. In the application, population overlap and conditional parallel trends require substantive justification beyond the fit of a treatment regression.

Under consistency, no anticipation, conditional parallel trends and overlap,
\begin{equation}
\bar\tau=\E\!\left[\E[H_i\mid G_i=1,Z_i]-\E[H_i\mid G_i=0,Z_i]\,\middle|\,G_i=1\right].
\label{eq:att-functional}
\end{equation}
This is the standard conditional DiD identifying formula \citep{abadie2005,santanna2020}. It permits persistent differences in outcome levels. Its restriction on untreated changes differs from cross-sectional selection on observables, which restricts potential outcomes in levels. Parallel trends also concerns the chosen outcome scale: validity for levels need not imply validity after a nonlinear transformation \citep{roth2023trends}. Here the maintained restriction concerns the supplied growth outcome. Adding pre-period outcomes changes the conditional parallel-trends restriction and therefore needs its own justification \citep{ding2019bracketing}. Supplementary Proposition~\ref{prop:did-benchmark} states the no-control fixed-effects equivalence and its causal scope. Table~\ref{tab:notation} collects the notation used in the application and causal comparison. Each transformation-specific function is defined when introduced in the next section.

\begin{table}[htbp]
\centering\small
\caption{Notation for the banking comparison}\label{tab:notation}
\begin{tabular}{@{}p{.22\linewidth}p{.72\linewidth}@{}}
\toprule
Symbol & Definition\\\midrule
$i,t,N,T$ & Bank and quarter indices; numbers of banks and quarters.\\
$G_i,P_t,D_{it}$ & Larger-bank indicator, post-enactment indicator and coded exposure $D_{it}=G_iP_t$.\\
$t^*,q,\pi$ & First post-enactment quarter, post-period share and population larger-bank share.\\
$RWA_{it},A_{it},\Delta$ & Risk-weighted assets, total assets and the adjacent-period difference operator.\\
$Y_{it},X_{it}$ & Quarterly risk-weighted-asset growth and the six contemporaneous control ratios.\\
$\alpha_i,\lambda_t$ & Additive bank and quarter effects in the panel model.\\
$\theta_0,\beta,U_{it}$ & Treatment coefficient, linear control coefficients and error in the benchmark model.\\
$H_i^{\mathrm{pre}},H_i^{\mathrm{post}},H_i$ & Pre- and post-period average quarterly growth and their difference.\\
$Z_i,p(Z_i)$ & Stipulated pre-treatment covariates and $\Pr(G_i=1\mid Z_i)$.\\
$\mathcal T_0,\mathcal T_1$ & Sets of pre- and post-enactment quarters.\\
$\mathbf d^1,\mathbf d^0;H_i(g)$ & Exposure paths and the potential post-minus-pre average under path $g$.\\
$\tau_g^R,\kappa$ & Group-$g$ reform response relative to no reform, and the between-group difference in no-reform changes.\\
$\tau_t,\bar\tau$ & Period-specific effect on treated banks and its post-period average for those paths.\\
\bottomrule
\end{tabular}
\end{table}
\section{Panel specifications and nonlinear covariates}\label{sec:panel-specifications}
The linear model provides a reference for comparing fixed effects, first differences and correlated random effects. Here we examine how nonlinear covariate adjustment impacts their implementation. Identification and parameter invariance are mathematical properties; consistency and causal inference require additional assumptions.

\subsection{Fixed Effects, First Differences and Correlated Random Effects}
For any panel variable $R_{it}$, define its bank-specific time mean, cross-sectional quarter mean, and overall mean by
\[
\bar R_i=T^{-1}\sum_tR_{it},\qquad
\bar R_t=N^{-1}\sum_iR_{it},\qquad
\bar R=(NT)^{-1}\sum_{i,t}R_{it}.
\]
These definitions apply component by component to vectors of controls. The within-bank transformation is $R_{it}^{w}=R_{it}-\bar R_i$. Applying it to equation~\eqref{eq:linear-panel-framework} gives
\begin{equation}
Y_{it}^{w}=\theta_0D_{it}^{w}+(X_{it}^{w})'\beta
 +(\lambda_t-\bar\lambda)+U_{it}^{w},
\label{eq:linear-within}
\end{equation}
where $\bar\lambda=T^{-1}\sum_t\lambda_t$. Quarter effects remain in this equation and must also be accounted for. On the balanced panel, two-way demeaning uses $\ddot R_{it}=R_{it}-\bar R_i-\bar R_t+\bar R$ and removes both additive effects:
\begin{equation}
\ddot Y_{it}=\theta_0\ddot D_{it}+\ddot X_{it}'\beta+\ddot U_{it}.
\label{eq:linear-twoway}
\end{equation}
Ordinary least squares (OLS) on these transformed variables reproduces the slopes from the regression with bank and quarter indicators.

First differencing instead gives
\begin{equation}
\Delta Y_{it}=\theta_0\Delta D_{it}+\Delta X_{it}'\beta
 +(\lambda_t-\lambda_{t-1})+\Delta U_{it},\qquad t=2,\ldots,T.
\label{eq:linear-fd}
\end{equation}
The differenced regression retains time-effect differences, which quarter dummies absorb. Their omission needs a restriction on the time effects or an orthogonality condition. Here $\Delta D_{it}=G_i(P_t-P_{t-1})$ is nonzero only in the enactment quarter, and $\Delta Y_{it}$ is a change in quarterly growth. FE and FD can estimate the same constant coefficient, with different precision. Under dynamic effects, their targets can differ; Corollary~\ref{cor:dynamic-panel} gives the nonlinear targets, and Supplementary Proposition~\ref{prop:fd} gives the covariate-adjusted linear weights.

The original FD implementation includes differenced controls but no intercept or quarter indicators. Its coefficient therefore contains the projection of omitted common time shocks onto residualised treatment. Control adjustment can spread that residual beyond treated banks at enactment; Supplementary equation~\eqref{eq:fd-omitted-time} gives the exact expression. With no controls or intercept, the coefficient is simply the treated banks' mean change at enactment, including the common shock.

A linear CRE specification retains the observations in levels and includes bank-means of the regressors:
\begin{equation}
Y_{it}=\lambda_t+\theta_0D_{it}+X_{it}'\beta
 +\rho\bar D_i+\bar X_i'\gamma+e_{it}.
\label{eq:linear-cre}
\end{equation}
Here $\rho$ and $\gamma$ are coefficients on the means, $e_{it}$ is the regression error, and $\lambda_t$ denotes unrestricted quarter effects. On a balanced panel, pooled OLS with both treatment and control means reproduces the identified two-way fixed-effects (TWFE) slopes. The construction of \citet{mundlak1978} concerns unit fixed effects; quarter indicators here also account for time effects. A probabilistic CRE interpretation additionally restricts the conditional mean of the bank effect.

The two-way Mundlak formulation of \citet{wooldridge2025} instead includes both bank and quarter means of the regressors. Let $\mathsf R$ be the matrix whose columns contain the stacked treatment and controls, and let $Q$ be the matrix that applies two-way demeaning to a stacked panel vector.
\begin{lemma}[Two-way Mundlak equivalence]\label{lem:mundlak}
On a balanced panel, if $\mathsf R'Q\mathsf R$ is nonsingular, the coefficient on $\mathsf R$ in pooled OLS of $Y$ on $\mathsf R$, the bank means of $\mathsf R$, the quarter means of $\mathsf R$ and a constant equals $(\mathsf R'Q\mathsf R)^{-1}\mathsf R'QY$, the coefficient from regressing $QY$ on $Q\mathsf R$.
\end{lemma}
The lemma concerns full-sample OLS slopes, rather than covariance estimates or cross-fitted procedures. Equation~\eqref{eq:linear-cre} yields the same slopes: its treatment and control residuals are two-way demeaned, and unrestricted quarter indicators already span the quarter means in the lemma.

\citet{arkhangelsky2024} develop generalised Mundlak estimators using group-level balancing statistics under group-level unconfoundedness and restrictions on the joint distribution of treatment and covariates. Their assignment-based identification argument differs from the outcome-model and projection restrictions used here. The common connection is the use of group summaries to account for unobserved heterogeneity; our nonlinear comparison specifies when joint adjustment on those summaries removes treatment variation and when an additive restriction can retain it.

For a constant-coefficient regression interpretation, a sufficient error restriction is strict exogeneity,
\begin{equation}
\E[U_{it}\mid D_i,X_i,\alpha_i]=0\quad\text{for every }t,
\label{eq:panel-strict-exogeneity}
\end{equation}
where $D_i=(D_{i1},\ldots,D_{iT})'$ and $X_i=(X_{i1},\ldots,X_{iT})$ are a bank's treatment and covariate histories. With rank and sampling conditions, strict exogeneity justifies the transformed regression moments. It is stronger than contemporaneous orthogonality and needs justification when controls respond to earlier outcomes or treatment. Conditional parallel trends provides a different identifying restriction.

\subsection{Outcome and treatment regressions in DML}\label{sec:method:dml}
Throughout the population calculations, expectations average over a randomly sampled bank and a uniformly sampled quarter of its fixed calendar (over $T-1$ quarters for FD). This convention does not treat bank-quarter rows as independent. DML estimates the chosen specification using fitted conditional means. Write $Y$ and $D$ for the outcome and treatment, either in levels or transformed, and $C$ for the conditioning variables. The partially linear regression (PLR) of \citet{robinson1988} is
\begin{equation}
Y=\theta_0D+g_0(C)+U,\qquad \E[U\mid D,C]=0.
\label{eq:plr}
\end{equation}
When $Y$ and $D$ have been transformed, $g_0(C)$ in this generic notation denotes the corresponding transformed outcome function; it is not assumed to be the levels function evaluated at transformed covariates. Equation~\eqref{eq:plr} must be justified for the selected inputs.

A \emph{nuisance function} is an unknown function needed to estimate $\theta_0$ but not itself the parameter of interest. The two conditional means used for partialling out are
\begin{equation}
m_0(C)=\E[D\mid C],\qquad
\ell_0(C)=\E[Y\mid C]=\theta_0m_0(C)+g_0(C).
\label{eq:dml-conditional-means}
\end{equation}
The function $g_0$ excludes treatment, whereas $\ell_0$ predicts the observed outcome. The treatment regression $m_0$ is a probability when its dependent variable is binary. Within-bank demeaned treatment is generally nonbinary, but here $\Delta D_{it}=G_i\ind\{t=t^*\}$ is binary, with $t^*$ denoting adoption. Conditional on calendar time, its mean is zero outside adoption and equals the conditional group propensity at adoption. The levels-treatment probability can also be positive subsequently. A learner, such as OLS or a random forest, is the algorithm used to fit a conditional mean.

For example, FD fits $\E[\Delta Y_{it}\mid C_{it}]$ and $\E[\Delta D_{it}\mid C_{it}]$, using a declared information set such as $C_{it}=(X_{it},X_{i,t-1},t)$. Both regressions require justification under the transformed model; applying a levels regression to $\Delta X_{it}$ does not supply it.

For candidate functions $\ell$ and $m$, define the estimating function, conventionally called a score in the DML literature, by
\begin{equation}
\psi(W;\theta,\ell,m)
=(D-m(C))\{Y-\ell(C)-\theta(D-m(C))\},\qquad W=(Y,D,C).
\label{eq:score}
\end{equation}
The population estimating equation at the true nuisance functions is
\begin{equation}
\E[\psi(W;\theta,\ell_0,m_0)]=0.
\label{eq:moment}
\end{equation}
For square-integrable variables and functions, equation~\eqref{eq:plr} gives the unique root $\theta_0$ when $J(C)=\E[(D-m_0(C))^2]>0$. The sample equation regresses outcome residuals on treatment residuals through the origin. Neyman orthogonality means that the expected score has zero directional derivative in every admissible nuisance direction at the true coefficient and functions \citep{chernozhukov2018}. This limits sensitivity to nuisance estimation under suitable rates; identification and the outcome restrictions remain separate requirements.

Cross-fitting evaluates each bank using nuisance functions fitted on other banks, with all quarters of a bank held out together. Section~\ref{sec:estimation} specifies the implementation and its sampling requirements. In particular, full-sample quarter means used for two-way demeaning share observations across folds, which requires additional inferential treatment.

Finally, the interactive regression model of \citet{chernozhukov2018}, $Y=g(D,C)+U$, allows the conditional contrast $g(1,C)-g(0,C)$ to vary with $C$ for binary treatment. It is distinct from an interactive fixed-effects model with latent factors. Its usual treatment-effect interpretation uses selection on observables and overlap; flexibility in $g$ does not remove those assumptions. The DiD comparison in Section~\ref{sec:setup:target} instead uses restrictions on untreated outcome changes. Supplementary Section~\ref{app:interactive} explains the counterfactual-support distinction. The two identifying routes cannot be treated as interchangeable solely because both use machine learning.

\subsection{Transforming a nonlinear outcome model}
Replace the linear covariate term in equation~\eqref{eq:linear-panel-framework} by an unknown function $g_0$:
\begin{equation}
Y_{it}=\alpha_i+\lambda_t+\theta_0D_{it}+g_0(X_{it})+U_{it}.
\label{eq:nonlinear-panel}
\end{equation}
Demeaning or differencing removes $\alpha_i$, while transforming the nonlinear term into
\begin{align}
g_{it}^{w}&=g_0(X_{it})-T^{-1}\sum_{s=1}^Tg_0(X_{is}),
\label{eq:nonlinear-within}\\
g_{it}^{\Delta}&=g_0(X_{it})-g_0(X_{i,t-1}).
\label{eq:nonlinear-fd}
\end{align}
Generally, $g_{it}^{w}\ne g_0(X_{it}^{w})$ and $g_{it}^{\Delta}\ne g_0(\Delta X_{it})$: transforming function values differs from transforming their arguments. For example, with $g_0(x)=x^2$,
\begin{equation}
g_0(X_{it})-g_0(X_{i,t-1})
=(X_{it}+X_{i,t-1})\Delta X_{it}.
\label{eq:quadratic-difference}
\end{equation}
Banks with the same covariate change can therefore have different nuisance values. Flexible fitting on $\Delta X_{it}$ alone need not recover this function.

The pair $(X_{it},X_{i,t-1})$ and time give a sufficient representation for the differenced function; the history $X_i$ and time do so for the within function. Lower-dimensional summaries can suffice under additional structure. These transformation arguments follow \citet[Section~5.1.2]{clarke2026}; the next proposition specialises them to common treatment timing.

CRE requires a separate examination because it specifies a regression in levels rather than removing $\alpha_i$ by differencing. One illustrative nonlinear CRE restriction is
\begin{equation}
\E[\alpha_i\mid D_i,X_i]=r_0(\bar D_i,\bar X_i),
\label{eq:nonlinear-cre-effect}
\end{equation}
where the unknown function $r_0$ describes the conditional mean of the bank effect. This restricts its dependence on the complete regressor histories to their bank averages. Under strict exogeneity, it gives
\begin{equation}
\E[Y_{it}\mid D_i,X_i]
=\lambda_t+\theta_0D_{it}+g_0(X_{it})+r_0(\bar D_i,\bar X_i).
\label{eq:nonlinear-cre-mean}
\end{equation}
The separate functions of current covariates, bank means and time impose additivity, a restriction not established for the banking data. Unrestricted joint adjustment permits additional interactions and can therefore have different identifying content.

\begin{proposition}[Conditional means under common treatment timing]\label{prop:panel-transformations}
Suppose~\eqref{eq:nonlinear-panel} and strict exogeneity~\eqref{eq:panel-strict-exogeneity} hold, all model components are square integrable, and $t^*\in\{2,\ldots,T\}$. Expectations average uniformly over the relevant quarters, independently of the bank record. Write $A_i=(X_{it^*},X_{i,t^*-1})$, $p_\Delta(A_i)=\Pr(G_i=1\mid A_i)$ and $p_W(X_i)=\Pr(G_i=1\mid X_i)$.

\emph{(i) First differences.} For $C^\Delta_{it}=(X_{it},X_{i,t-1},t)$, put $g^\Delta_0(C^\Delta_{it})=\lambda_t-\lambda_{t-1}+g_{it}^\Delta$. Then
\begin{align}
m^\Delta_0(C^\Delta_{it})&=\ind\{t=t^*\}p_\Delta(A_i),&
\ell^\Delta_0(C^\Delta_{it})&=\theta_0m^\Delta_0(C^\Delta_{it})+g^\Delta_0(C^\Delta_{it}).
\label{eq:fd-transformed-means}
\end{align}

\emph{(ii) Within-bank demeaning.} For $C^W_{it}=(X_i,t)$, put $g^W_0(C^W_{it})=\lambda_t-\bar\lambda+g_{it}^w$. Then
\begin{align}
m^W_0(C^W_{it})&=(P_t-q)p_W(X_i),&
\ell^W_0(C^W_{it})&=\theta_0m^W_0(C^W_{it})+g^W_0(C^W_{it}).
\label{eq:within-transformed-means}
\end{align}
The corresponding treatment residuals are $\ind\{t=t^*\}(G_i-p_\Delta)$ and $(P_t-q)(G_i-p_W)$, with second moments
\begin{equation}
J_\Delta=\frac{\E[p_\Delta(1-p_\Delta)]}{T-1},\qquad
J_W=q(1-q)\E[p_W(1-p_W)].
\label{eq:panel-residual-variances}
\end{equation}
The transformed equations satisfy the conditional error restriction in~\eqref{eq:plr}. Each score identifies $\theta_0$ when its displayed second moment is positive.

\emph{(iii) Correlated random effects.} Under~\eqref{eq:nonlinear-cre-effect}, let $C^{\mathrm{CRE}}_{it}=(X_{it},\bar X_i,G_i,t)$ and $g^{\mathrm{CRE}}_0=\lambda_t+g_0(X_{it})+r_0(qG_i,\bar X_i)$. The levels equation has this additive nuisance and a conditionally mean-zero error. Unrestricted conditional-mean adjustment on $C^{\mathrm{CRE}}$ gives $m^{\mathrm{CRE}}_0=D_{it}$ and $J_{\mathrm{CRE}}=0$. This degeneracy concerns unrestricted adjustment; the additive model can identify the coefficient under the rank condition in Corollary~\ref{cor:panel-restricted}.
\end{proposition}

Table~\ref{tab:specification-comparison} summarises the different specifications and constructions. The probabilities $p_\Delta(A_i)$, $p_W(X_i)$ and $p(Z_i)$ concern the same group indicator but different information sets; none can generally substitute for another. The FD expectation uses $T-1$ quarters and the within expectation uses $T$; their treatment scales also differ, so the displayed second moments do not rank efficiency. Calendar time gives the FD treatment residual its adoption-quarter support. A fitted learner need not respect that support unless it is imposed. Full histories are sufficient for the within construction, but are not necessary under every nonlinear model. Conditioning on histories that respond to treatment also requires a different causal justification. Strict exogeneity given histories and conditional parallel trends given pre-treatment $Z$ are distinct assumptions.

DiD permits group differences in untreated levels. CRE can accommodate them through $r_0(qG_i,\bar X_i)$ while retaining additive restrictions. Omitting $G_i$, or its equivalent $\bar D_i$, can leave residual treatment variation only if the remaining information does not already determine treatment. Such omission also requires a valid outcome restriction.

\begin{example}[A trending covariate with a quadratic effect]\label{ex:quadratic-trend}
Let $X_{it}=\nu_i+ct+\varepsilon_{it}$, $c\ne0$, and $g_0(x)=x^2$. Suppose the proposition's assumptions hold, banks are independent and identically distributed, $0<\pi<1$, $\nu_i$ has a finite fourth moment, and the $\varepsilon_{it}$ are independent and identically distributed with mean zero, positive variance and a finite fourth moment, independently of $(G_i,\nu_i)$. Write $\bar t_j$ for the average date in $\mathcal T_j$. As $N\to\infty$ with $T$ fixed, the probability limit of linear TWFE with $X_{it}$, and the population partialling-out coefficient for within-transformed observations given only $(X_{it}^w,t)$, both equal
\begin{equation}
\theta_0+2c(\bar t_1-\bar t_0)
\{\E[\nu_i\mid G_i=1]-\E[\nu_i\mid G_i=0]\}.
\label{eq:quadratic-trend-bias}
\end{equation}
The full-history within coefficient equals $\theta_0$ if $J_W>0$. Including $X_{it}^2$ as an additional regressor in TWFE also makes its population coefficient equal $\theta_0$ under its own residual-treatment rank condition.
\end{example}

Demeaning removes the bank-level component of $X_{it}$, but the quadratic outcome term contains the group-dependent trend $2c\nu_it$. Agreement between linear and flexible adjustment on demeaned inputs therefore need not validate either specification. This example does not establish misspecification in the banking data. Supplementary Section~\ref{app:panel-transformations} gives the proofs.

\begin{table}[!htbp]
\centering\small
\caption{Panel specifications under common treatment timing}\label{tab:specification-comparison}
\begin{tabular}{@{}>{\raggedright\arraybackslash}p{.13\linewidth}>{\raggedright\arraybackslash}p{.33\linewidth}>{\raggedright\arraybackslash}p{.47\linewidth}@{}}
\toprule
Specification & Linear reference & Nonlinear construction and treatment variation\\\midrule
FE (within) & Demean by bank and include quarter effects; equivalent to TWFE. & $(X_i,t)$ is sufficient for unrestricted adjustment; residual $(P_t-q)(G_i-p_W)$. Demeaned functions provide a restricted alternative.\\\addlinespace
FD & Regress $\Delta Y$ on $\Delta D$, $\Delta X$ and quarter indicators. & $(X_{it},X_{i,t-1},t)$ is sufficient; residual $\ind\{t=t^*\}(G_i-p_\Delta)$. Differences of covariates alone need further justification.\\\addlinespace
CRE & Levels, quarter indicators and bank means of treatment and controls reproduce TWFE slopes. & Under~\eqref{eq:nonlinear-cre-effect}, unrestricted joint adjustment predicts treatment exactly. Additive adjustment instead requires the outcome and rank restrictions of Corollary~\ref{cor:panel-restricted}.\\
\bottomrule
\end{tabular}
\end{table}
\section{Treatment adjustment and causal interpretation}\label{sec:method}
The choice of conditioning variables and function class determines residual treatment variation. The outcome  model restrictions determine whether the target coefficient has a causal interpretation. We first establish the projection identities and then connect specified panel scores to difference-in-differences (DiD).

\subsection{Treatment means in a nonlinear regression}\label{sec:method:revealed}
For the partially linear model in equation~\eqref{eq:plr}, let $\mathcal C$ denote the information in the conditioning variables $C$, and recall $J(C)=\E[(D-m_0(C))^2]=\E[\Var(D\mid C)]$. At the true conditional means, the derivative of the expected score with respect to the coefficient is $-J(C)$. A unique partially linear coefficient therefore requires $J(C)>0$.

\begin{proposition}[Deterministic treatment and loss of identification]\label{prop:support}
If $\mathcal C_1\subseteq\mathcal C_2$, then
\begin{equation}
J(\mathcal C_1)-J(\mathcal C_2)
=\E\bigl[(\E[D\mid\mathcal C_2]-\E[D\mid\mathcal C_1])^2\bigr]\geq0.\label{eq:nested}
\end{equation}
Moreover, $J(\mathcal C)=0$ if and only if $D$ is $\mathcal C$-measurable. If $D=h(C)$ almost surely, then $\theta_0$ is not identified in the PLR~\eqref{eq:plr} with $g_0$ unrestricted: every $\theta$ gives the same observational model once $g_0(C)$ is replaced by $g_0(C)+(\theta_0-\theta)h(C)$.
\end{proposition}

This conditional-variance decomposition matters here because treatment means are standard regressors in linear CRE. Let $q=|\mathcal T_1|/T$ and $\widehat\pi=N^{-1}\sum_iG_i$ denote the post-period share and sample share of larger banks.

\begin{corollary}[Treatment means and calendar information]\label{cor:absorbing}
In a balanced panel with common treatment timing and $D_{it}=G_iP_t$, realised treated share $\widehat\pi=N^{-1}\sum_iG_i\in(0,1)$ and post share $q\in(0,1)$,
\begin{equation}
\bar D_i=qG_i,\qquad \bar D_t=\widehat\pi P_t,\qquad
D_{it}=\bar D_i\bar D_t/(q\widehat\pi).\label{eq:identity}
\end{equation}
More generally, for absorbing treatment $D_{it}=\ind\{t\geq t_i^*\}$ with $t_i^*\in\{1,\ldots,T+1\}$ and $t_i^*=T+1$ for never-treated units, $t_i^*=T+1-T\bar D_i$, so the unit treatment mean together with calendar time reconstructs the treatment path, including under staggered adoption.
\end{corollary}

In the application, $q=8/20$ and $\widehat\pi=20/91$. Hence a nonlinear function of the bank treatment mean and calendar time can reproduce $D_{it}$ exactly. At the population conditional means the partialling-out score is then zero for every coefficient value. A numerically positive denominator produced by imperfect prediction does not identify a coefficient in that unrestricted model. It does not rule out identification under the different restrictions of DiD.

\subsection{Restriction on the nuisance function}\label{sec:framework}
A linear CRE regression includes treatment means additively and excludes their product with time. More generally, a restricted set of functions can leave nonzero residual treatment variation after projection onto that set. The outcome restriction must be stated along with the chosen nuisance-function class.

Let $\mathcal M$ be a closed linear subspace of square-integrable functions of $C$. Constants may be included but are not required; a within-transformed class can have zero time mean. ``Linear'' means closed under linear combinations: spline basis functions, for example, can be nonlinear in individual covariates. Write $\|f\|_2^2=\E[f(C)^2]$ for a function's squared $L^2$ norm. Let $\Pi_{\mathcal M}$ denote least-squares projection onto this space and define
\begin{equation}
 m_{\mathcal M}=\Pi_{\mathcal M}D,\qquad
 \ell_{\mathcal M}=\Pi_{\mathcal M}Y,\qquad
 J_{\mathcal M}=\E[(D-m_{\mathcal M})^2].
\end{equation}

\begin{proposition}[Restricted adjustment and its identifying content]\label{prop:restricted}
Suppose $D,Y$ are square integrable and the representation $Y=\theta_0D+g_0(C)+U$ satisfies $\E[U\mid D,C]=0$. Then
\begin{equation}
J_{\mathcal M}=J(C)+\|m_0-m_{\mathcal M}\|_2^2.\label{eq:restricted-info}
\end{equation}
If $J_{\mathcal M}>0$, the population residual-ratio coefficient
$\theta_{\mathcal M}=\E[(D-m_{\mathcal M})(Y-\ell_{\mathcal M})]/J_{\mathcal M}$ satisfies
\begin{equation}
\theta_{\mathcal M}-\theta_0
=\frac{\E[(m_0-m_{\mathcal M})(g_0-\Pi_{\mathcal M}g_0)]}
{J_{\mathcal M}},
\qquad
|\theta_{\mathcal M}-\theta_0|\leq\frac{\|m_0-m_{\mathcal M}\|_2\,\|g_0-\Pi_{\mathcal M}g_0\|_2}{J_{\mathcal M}}.\label{eq:restricted-bias}
\end{equation}
Consequently the model with the maintained restriction $g_0\in\mathcal M$ identifies $\theta_0$ if and only if $J_{\mathcal M}>0$, even when $J(C)=0$. If $g_0\notin\mathcal M$, $\theta_{\mathcal M}$ is a projection coefficient with the discrepancy in~\eqref{eq:restricted-bias}, which can still be zero when the two projection errors are orthogonal. At $(\ell_{\mathcal M},m_{\mathcal M})$, the nuisance derivatives of the expected score~\eqref{eq:score} vanish in every direction within $\mathcal M$, for any $\theta$. This projection orthogonality does not establish the causal interpretation of its root.
\end{proposition}

With only group and period indicators, unrestricted adjustment on $(G_i,P_t)$ predicts $D_{it}=G_iP_t$ perfectly. Additive adjustment leaves $(G_i-\pi)(P_t-q)$, where $\pi=\Pr(G_i=1)$, with variance $\pi(1-\pi)q(1-q)>0$. Under the maintained additive untreated-outcome model and mean-zero error restriction, the coefficient recovers a constant effect. An unrestricted group-by-post outcome component is observationally indistinguishable from treatment.

In Random Forest, forest depth or minimum leaf size does not by itself define one of these projection spaces, and a positive denominator does not verify the outcome restriction. Supplementary Corollaries~\ref{cor:twfe} and \ref{cor:heterogeneous} give finite-design and heterogeneous-effect decompositions, drawing on the regression-weighting results of \citet{angrist1998}, \citet{aronow2016} and \citet{sloczynski2022}.

\begin{corollary}[Restricted within and additive CRE classes]\label{cor:panel-restricted}
Maintain the model and exogeneity conditions of Proposition~\ref{prop:panel-transformations}. For $\bar f_i=T^{-1}\sum_sf(X_{is})$, define the closed linear spaces
\begin{align}
\mathcal M_W&=\overline{\{f(X_{it})-\bar f_i+a_t:\ \textstyle\sum_ta_t=0\}}^{\,L^2},
\label{eq:within-class}\\
\mathcal M_{\mathrm{CRE}}&=\overline{\{f(X_{it})+h(G_i,\bar X_i)+a_t\}}^{\,L^2}.
\label{eq:additive-cre-class}
\end{align}
Here $f$ and $h$ range over functions with square-integrable evaluations, $a_t$ over deterministic quarter effects, and the closure includes mean-square limits. Project $(Y^w,D^w)$ onto $\mathcal M_W$, or $(Y,D)$ onto $\mathcal M_{\mathrm{CRE}}$ under~\eqref{eq:nonlinear-cre-effect}. The corresponding population residual-ratio coefficient equals $\theta_0$ whenever its $J_{\mathcal M}$ is positive.
\end{corollary}

The within class evaluates functions before bank demeaning; demeaned spline basis functions and centred quarter indicators provide a finite-dimensional approximation. With a fixed basis and OLS, this is a standard series fixed-effects projection. The order-of-transformation issue concerns the outcome specification itself and arises before choosing a DML implementation. A forest on $X_{it}^w$ alone need not implement this class. CRE permits a group-dependent bank-mean component additively with the current-covariate and quarter components. Both constructions still require positive $J_{\mathcal M}$ and, for fitted estimators, suitable convergence rates. Class membership alone does not imply conditional parallel trends; that implication requires the untreated-outcome model and conditional-mean restrictions discussed below.

Thus a CRE-based DML procedure, as studied by \citet{fuhr2024double}, must be assessed through its actual regressors and nuisance class. Corollary~\ref{cor:absorbing} concerns unrestricted adjustment that determines treatment; Corollary~\ref{cor:panel-restricted} instead gives sufficient restrictions for a particular additive class. The two results do not imply failure of every CRE implementation.

\subsection{Within residuals, bank-mean correction and pre/post changes}\label{sec:method:structured}
\citet[Section~5.1.1]{clarke2026} adjust preliminary treatment predictions so that each bank's treatment residuals have mean zero. This is a particular CRE construction, distinct from allowing a general nonlinear regression to use the treatment mean as a covariate. For a held-out bank, let $\widetilde m_{it}$ and $\widetilde\ell_{it}$ be preliminary treatment and outcome predictions obtained from other banks. Their time averages are denoted by overbars. The adjustment replaces $\widetilde m_{it}$ by
\begin{equation}
\widehat m_{it}^{*}=\widetilde m_{it}+\bar D_i-\overline{\widetilde m}_i,
\qquad v_{it}^{*}=D_{it}-\widehat m_{it}^{*}.
\label{eq:mean-correction}
\end{equation}
Consequently $\sum_t v_{it}^{*}=0$, so an additive bank component drops out of $\sum_t v_{it}^{*}(Y_{it}-\widetilde\ell_{it})$. The corrected prediction need not lie in $[0,1]$. Supplementary Proposition~\ref{prop:structured} gives the projection identities and the limited linear equivalence of this correction.

For the design $D_{it}=G_iP_t$, suppose the preliminary treatment fit predicts group membership once per bank and then multiplies by the post indicator: $\widetilde m_{it}=P_ta_i$. Let $b_i$ be the difference between the post- and pre-period averages of $\widetilde\ell_{it}$. The next result connects this specified unweighted panel score to a regression using one change per bank. It does not cover every nonlinear FE or CRE estimator. Conditional-variance and overlap weighting are established results \citep{angrist1998,li2018balancing,vansteelandt2022}; here they arise from the pre/post representation of the specified panel score under DiD.

\begin{theorem}[Pre/post representation and treatment-effect weighting]\label{thm:design}
\emph{(i) Sample identities.} In a balanced panel with common treatment timing and post share $q\in(0,1)$, let $a_i$ be a unit-level assignment prediction for held-out unit $i$, set $\widetilde m_{it}=P_ta_i$, and apply the bank-mean prediction adjustment of Proposition~\ref{prop:structured}. With $b_i=\bar{\widetilde\ell}_{i,\mathcal T_1}-\bar{\widetilde\ell}_{i,\mathcal T_0}$ the fitted counterpart of $H_i$, for arbitrary fitted predictions,
\begin{align}
v^*_{it}&=(P_t-q)(G_i-a_i),\label{eq:design-resid}\\
\frac{1}{NT}\sum_{i,t}(v^*_{it})^2
&=q(1-q)\frac{1}{N}\sum_i(G_i-a_i)^2,\label{eq:design-sample-risk}\\
\widehat\tau_{\mathrm{ov}}&=
\frac{\sum_{i,t}v_{it}^{*}(Y_{it}-\widetilde\ell_{it})}{\sum_{i,t}(v_{it}^{*})^2}
=\frac{\sum_i(G_i-a_i)(H_i-b_i)}{\sum_i(G_i-a_i)^2},\label{eq:design-collapse}
\end{align}
whenever the denominator is positive. The last expression is a regression through the origin; no intercept is added.

\emph{(ii) Prediction risk with known and estimated propensities.} Let $Z$ contain pre-treatment unit characteristics, condition on an independent training sample so that $a=a(Z)$ is fixed on validation units, and put $J_G=\E[p(Z)\{1-p(Z)\}]$. Under Assumption~\ref{ass:sampling},
\begin{equation}
\E[(v^*)^2\mid\mathrm{training}]
=q(1-q)\{J_G+\|a-p\|_2^2\}.
\label{eq:design-risk}
\end{equation}
For $a=p$ and $\pi\in(0,1)$, the population residual second moment $\E[(v^*)^2]$, divided by $\Var(D)=\pi q(1-\pi q)$, is at most $R_{\mathrm{add}}=(1-\pi)(1-q)/(1-\pi q)$, the additive-regression reference of Supplementary Section~\ref{app:classes}, with equality if and only if $p(Z)$ is constant.

\emph{(iii) Causal target.} Under Assumptions~\ref{ass:consistency} to \ref{ass:sampling}, define $\tau(Z)=\E[\bar Y_{\mathcal T_1}(\mathbf d^1)-\bar Y_{\mathcal T_1}(\mathbf d^0)\mid G=1,Z]$. If $J_G>0$, the population version of~\eqref{eq:design-collapse} with $a=p$ and any square-integrable $b=b(Z)$ equals
\begin{equation}
\tau_{\mathrm{ov}}=
\frac{\E[p(Z)\{1-p(Z)\}\tau(Z)]}{J_G}
=\frac{\E[\{1-p(Z)\}\tau(Z)\mid G=1]}
{\E[1-p(Z)\mid G=1]},
\label{eq:overlap-target}
\end{equation}
which differs from the ATT by
\begin{equation}
\tau_{\mathrm{ov}}-\bar\tau
=-\frac{\Cov(p(Z),\tau(Z)\mid G=1)}
{\E[1-p(Z)\mid G=1]}.
\label{eq:overlap-att-gap}
\end{equation}
\end{theorem}

The identity also holds for a within score with $\widehat m^W_{it}=(P_t-q)a_i$ and $\widehat\ell^W_{it}=\widetilde\ell_{it}-\overline{\widetilde\ell}_i$, since its residual is $v^*_{it}$ and outcome bank means drop out. Proposition~\ref{prop:panel-transformations} estimates $p_W(X_i)$ from full histories. The application estimates $p(Z_i)$ from pre-reform characteristics under conditional parallel trends given $Z_i$. The algebra agrees, but the information sets and identifying restrictions differ.

Part (i) is sample algebra; part (iii) adds causal assumptions and the true propensity. Among treated banks, $1-p(Z)$ gives greater weight to profiles with more comparable controls. The weighted effect equals the ATT under homogeneous effects. Supplementary Example~\ref{ex:overlap-att} illustrates their difference under heterogeneity.

The factorised prediction is a sufficient construction for the identity, rather than a necessary condition for panel estimation or inference. Adding any bank-constant fitted term to $P_ta_i$ gives the same corrected residuals because the correction removes that term. Similarly, a first-difference score admits a bank-level representation when its treatment fit is restricted to the enactment quarter.

\begin{corollary}[First-difference score at the treatment date]\label{cor:fd-collapse}
In the design of Theorem~\ref{thm:design}, define $R_i=\Delta Y_{it^*}$ and suppose the fitted treatment regression is $\widehat m^\Delta_{it}=\ind\{t=t^*\}a_i$. Let $\widehat\ell^\Delta_{it}$ be fitted outcome predictions and put $b_i=\widehat\ell^\Delta_{it^*}$. Without a bank-mean correction of the differenced predictions, the treatment residual is $\ind\{t=t^*\}(G_i-a_i)$ and
\begin{equation}
\widehat\theta_\Delta
=\frac{\sum_{i,t\geq2}(\Delta D_{it}-\widehat m^\Delta_{it})(\Delta Y_{it}-\widehat\ell^\Delta_{it})}
{\sum_{i,t\geq2}(\Delta D_{it}-\widehat m^\Delta_{it})^2}
=\frac{\sum_i(G_i-a_i)(R_i-b_i)}{\sum_i(G_i-a_i)^2},
\label{eq:fd-collapse}
\end{equation}
whenever the denominator is positive.
\end{corollary}

Treatment residuals vanish outside $t^*$, so only that quarter's outcome predictions enter. A feasible implementation cross-fits $\Pr(G=1\mid B)$ and $\E[R\mid B]$ across banks for declared characteristics $B$; Supplementary Corollary~\ref{cor:fd-inference} gives inference conditions. The sample identity itself requires neither cross-fitting nor causal assumptions.

For inference, the outcome prediction in the change regression targets $\ell_H(Z)=\E[H\mid Z]$. The corresponding score is
\begin{equation}
\psi_{\mathrm{ov}}(W;\theta,\ell_H,p)
=(G-p(Z))\{H-\ell_H(Z)-\theta(G-p(Z))\},\qquad W=(H,G,Z).
\label{eq:overlap-score}
\end{equation}
At the true $p$, subtracting any square-integrable function of $Z$ leaves the same population root. Orthogonality with respect to estimation of both functions, however, uses the true $\ell_H$. The difference matters when comparing outcome fits.

More explicitly, hold square-integrable working functions $a(Z)$ and $b(Z)$ fixed and let $\theta_H(b,a)$ be the root of the expected overlap score. Under the assumptions of Theorem~\ref{thm:design}(iii),
\begin{equation}
\theta_H(b,a)=
\frac{J_G\tau_{\mathrm{ov}}+\E[(a-p)(b-\ell_H)]}
{J_G+\|a-p\|_2^2}.
\label{eq:overlap-working-root}
\end{equation}
This follows by expanding both residual products and conditioning on $Z$. With $b=\ell_H$, propensity error attenuates the root by $J_G/(J_G+\|a-p\|_2^2)$. With both functions estimated, the error product can offset or reverse that change. Thus the score is orthogonal at the true functions but is not doubly robust in the pair $(p,\ell_H)$. Assignment prediction loss estimates $J_G+\|a-p\|_2^2$; it does not separately measure the error term or provide an attenuation correction.

\subsection{Dynamic effects and the choice of transformation}\label{sec:method:dynamic}
For dynamic effects, the two transformed conditional-mean scores can target different periods and covariate distributions. This comparison requires an untreated-outcome panel model beyond parallel trends for pre/post averages.

\begin{corollary}[First-difference and within targets under dynamic effects]\label{cor:dynamic-panel}
Write $Y_{it}(d)=Y_{it}(\mathbf d^d)$, $d\in\{0,1\}$, for the two exposure paths defined in Section~\ref{sec:setup:target}. Suppose consistency and no anticipation hold, and reference-path outcomes satisfy
\[
Y_{it}(0)=\alpha_i+\lambda_t+g_0(X_{it})+U_{it},\qquad
\E[U_{it}\mid G_i,X_i,\alpha_i]=0\quad\text{for every }t.
\]
Assume the covariate history is unchanged by the two exposure paths and all outcome components and effects are square integrable. Write $\beta_{it}=Y_{it}(1)-Y_{it}(0)$ for post-period effects and $\bar\beta_i=|\mathcal T_1|^{-1}\sum_{t\in\mathcal T_1}\beta_{it}$. Define $\tau_\Delta(A_i)=\E[\beta_{it^*}\mid G_i=1,A_i]$ and $\tau_W(X_i)=\E[\bar\beta_i\mid G_i=1,X_i]$. Using the information sets in Proposition~\ref{prop:panel-transformations} and their true outcome and treatment conditional means under this dynamic model, the population roots, when $J_\Delta,J_W>0$, are
\begin{equation}
\theta_\Delta=\frac{\E[p_\Delta(1-p_\Delta)\tau_\Delta]}{\E[p_\Delta(1-p_\Delta)]},\qquad
\theta_W=\frac{\E[p_W(1-p_W)\tau_W]}{\E[p_W(1-p_W)]}.
\label{eq:dynamic-panel-targets}
\end{equation}
\end{corollary}

FD averages adoption-quarter effects, while within demeaning averages post-period effects, generally with different weights. Other conditioning sets can yield different targets: covariate-adjusted linear FD, for example, can put nonzero weight on later periods (Supplementary Proposition~\ref{prop:fd}). Supplementary Section~\ref{app:panel-transformations} proves the corollary.

The positive denominators in Corollary~\ref{cor:dynamic-panel} identify the displayed weighted coefficients, not automatically the unweighted impact or average post-period ATT. A stratum containing only treated banks receives zero conditional-variance weight, although its effects enter the ATT. Identifying those effects requires treated-support overlap or a further extrapolation restriction; equality of a weighted coefficient and the ATT additionally requires a restriction on the relation between effects and weights. Supplementary Section~\ref{app:dynamic-att-support} gives a simple observational-equivalence example.

Under this corollary's untreated-outcome model, strict exogeneity implies conditional parallel trends given the full invariant history $X_i$: the conditional mean of $H_i(0)$ is the pre/post difference in $\lambda_t+g_0(X_{it})$ in both groups. Consequently $\theta_W$ is the overlap-weighted DiD effect for that history. This implication does not generally survive replacing $X_i$ by a smaller pre-treatment vector $Z_i$, because the omitted history can have different conditional distributions across groups. The application's parallel-trends restriction given $Z_i$ remains a separate assumption.

\subsection{A doubly robust ATT comparison}\label{sec:method:att}
To retain the treated-population average as the target, we also use the established doubly robust DiD estimator of \citet{santanna2020}, with cross-fitting as in \citet{chang2020}. Define the control-outcome regression $\mu_0(Z)=\E[H\mid G=0,Z]$. It differs from the all-bank outcome regression $\ell_H$. The population score is
\begin{equation}
\phi(W;\tau,\mu_0,p)=\frac{G}{\pi}\{H-\mu_0(Z)-\tau\}
-\frac{1-G}{\pi}\frac{p(Z)}{1-p(Z)}\{H-\mu_0(Z)\}.
\label{eq:drdid}
\end{equation}
Under the causal assumptions above, its expectation at the ATT is zero if either the group propensity or the control-outcome regression is correctly specified. This double-robustness property does not remove the parallel-trends requirement and does not itself guarantee valid inference.

Partition the banks into $K\geq2$ folds. Let $k(i)$ index the fold containing bank $i$ and write $\widehat p_i=\widehat p_{-k(i)}(Z_i)$. Define the held-out outcome residual and control weight by
\[
r_i=H_i-\widehat\mu_{0,-k(i)}(Z_i),\qquad
w_i=(1-G_i)\frac{\widehat p_i}{1-\widehat p_i}.
\]
Pooling the held-out observations gives the normalised estimator
\begin{equation}
\widehat\tau_{\mathrm{DR}}=\frac{\sum_iG_ir_i}{\sum_iG_i}
-\frac{\sum_iw_ir_i}{\sum_iw_i}.
\label{eq:normalized}
\end{equation}
For the banking estimates, the propensity in $w_i$ is clipped to $\min\{0.99,\max\{0.01,\widehat p_i\}\}$. The overlap score uses the unmodified predictions.

The overlap and DR scores use the same bank-level changes but generally target different effects and require different outcome regressions. We use DR-DiD as an established ATT benchmark. The control effective sample size, $\mathrm{ESS}_0=(\sum_iw_i)^2/\sum_iw_i^2$, describes concentration of its odds weights. Supplementary Section~\ref{app:dr} develops normalisation, inference and clipping; empirical weight diagnostics are reported in Supplementary Section~\ref{app:clipping}.

\paragraph{Notation for the main coefficients and treatment variation.}
The following lookup collects objects defined above; each retains the assumptions and conditioning set of its defining result. Tables label the estimator based on~\eqref{eq:overlap-score} as ``Overlap PLR'' and the normalised doubly robust estimator as ``DR''.
\begin{center}\small
\begin{tabular}{@{}>{\raggedright\arraybackslash}p{.21\linewidth}>{\raggedright\arraybackslash}p{.72\linewidth}@{}}
\toprule
Notation & Meaning and defining result\\\midrule
$\theta_0$ & Constant coefficient in the maintained partially linear model~\eqref{eq:plr}.\\
$\theta_{\mathcal M}$ & Residual-ratio projection coefficient under the restricted nuisance class; Proposition~\ref{prop:restricted}.\\
$\theta_\Delta,\theta_W$ & Conditional-variance-weighted adoption-quarter and average post-period effects under Corollary~\ref{cor:dynamic-panel}.\\
$\bar\tau,\tau_{\mathrm{ov}}$ & Average effect on treated units and its overlap-weighted counterpart; equations~\eqref{eq:att} and~\eqref{eq:overlap-target}.\\
$\theta_H(b,a)$ & Population root at working outcome and group predictions $b,a$; equation~\eqref{eq:overlap-working-root}.\\
$J(C),J_{\mathcal M}$ & Population residual treatment second moments under conditional-mean or restricted-projection adjustment; Proposition~\ref{prop:restricted}.\\
$J_G$ & Residual group variance $\E[p(Z)\{1-p(Z)\}]$ in the pre/post construction; Theorem~\ref{thm:design}.\\
$J_\Delta,J_W$ & Transformed residual treatment variances using adoption covariates and full histories, respectively; equation~\eqref{eq:panel-residual-variances}.\\
\bottomrule
\end{tabular}\end{center}
\section{Estimation and inference}\label{sec:estimation}
We cross-fit complete bank records: fit nuisance functions on $K-1$ folds and evaluate them on the remaining fold, keeping each bank's quarters together. Scaling, tuning and fitted preprocessing use training data only. The estimator pools held-out predictions before taking each ratio \citep{clarke2026,chernozhukov2018}.

\subsection{Conditions for the two change-based estimators}
For the overlap estimator, the nuisance functions are the group propensity $p(Z)$ and the all-bank outcome regression $\ell_H(Z)=\E[H\mid Z]$. For DR-DiD, they are $p(Z)$ and the control-outcome regression $\mu_0(Z)=\E[H\mid G=0,Z]$. Let $r_p$, $r_\ell$ and $r_\mu$ denote the largest $L^2$ estimation errors across folds, evaluated at a fresh bank while holding each trained function fixed. Both nuisance functions for the estimator in question must be consistent, with
\begin{equation}
r_p(r_\ell+r_p)=o_p(N^{-1/2})\quad\text{for overlap},\qquad
r_p r_\mu=o_p(N^{-1/2})\quad\text{for DR}.
\label{eq:rates}
\end{equation}
Neyman orthogonality removes first-order nuisance errors; these rates control the remainder. The overlap condition includes $r_p=o_p(N^{-1/4})$, because equation~\eqref{eq:overlap-working-root} contains squared propensity error as well as a product. DR inference here requires consistency of both functions and the product condition. Population double robustness alone is insufficient.

Supplementary Theorem~\ref{thm:collapsed-inference} states the overlap result for i.i.d.\ bank records, fixed $T$ and $K$, $\E[H^2]<\infty$, fitted propensities in $[0,1]$, and fixed positive residual group variation $J_G=\E[p(Z)\{1-p(Z)\}]$. It does not require $p$ to be uniformly bounded away from one: the score uses bounded treatment residuals rather than propensity odds. With $V=G-p(Z)$, one bank's first-order contribution is
\begin{equation}
\varphi_{\mathrm{ov}}(W)=
\frac{V\{H-\ell_H(Z)-\tau_{\mathrm{ov}}V\}}{J_G}.
\label{eq:main-overlap-if}
\end{equation}
The average influence function gives the leading estimation error; its fitted sample variance divided by $N$ estimates sampling variance. The theorem holds $J_G$ fixed as $N$ grows. Without causal assumptions, the same argument applies to the statistical residual-ratio target; conditional parallel trends and the other identifying restrictions give its treatment-effect interpretation.

Supplementary Proposition~\ref{prop:dr-inference} gives the corresponding result for normalised DR. It additionally requires $p(Z)\leq1-\varepsilon$ for a fixed $\varepsilon>0$ and fitted propensities at most $1-c$, with $0<c<\varepsilon$, to bound the control odds weights. No lower propensity bound is needed. Its influence function accounts for estimation of both denominators in equation~\eqref{eq:normalized}.

Both results allow treatment-stratified folds and a fixed number of repeated partitions under the supplementary conditions. With stratified allocation, nuisance-rate assumptions concern the actual training-sample law conditional on group counts; an i.i.d.-training rate theorem requires adaptation. Their mean and median aggregation rules specialise established DML arguments \citep{chernozhukov2018,chang2020,santanna2020}. A partition range describes split sensitivity, whereas influence calculations estimate sampling variance under the stated conditions. For paired specifications, the relevant influence value is the within-bank difference of their influence values. Section~\ref{app:inference-statement} states the joint conditions and aggregation convention used in Table~\ref{tab:covered-uncertainty}.

The sample identity in Theorem~\ref{thm:design} transfers the bank-level inference result only when its induced predictions $a_i$ and $b_i$ estimate the matching $p(Z_i)$ and $\ell_H(Z_i)$ and satisfy that result's information-set, sampling, preprocessing and rate conditions. Exact equality of estimating equations alone does not supply these conditions. The requirements concern the induced bank-level predictions; intermediate panel predictions may contain components that cancel in the reduction. For FD with treatment predictions restricted to enactment, Corollary~\ref{cor:fd-collapse} and Supplementary Corollary~\ref{cor:fd-inference} give the corresponding ratio and inference conditions for $R_i=\Delta Y_{it^*}$. Factorisation is a sufficient route to these results; other panel procedures can admit different inference arguments.

\subsection{Limits of the empirical inference}
The upper propensity bound for DR inference is stronger than Assumption~\ref{ass:support}. Clipping estimated propensities can limit numerical instability, but cannot create comparable controls. A fixed clipping rule preserves propensity consistency only if it leaves the population propensity unchanged; Supplementary equation~\eqref{eq:clipped-dr-root} gives the discrepancy otherwise.

For the fixed-basis comparisons in Tables~\ref{tab:matched-basis} and \ref{tab:model-linked-fits}, Supplementary Proposition~\ref{prop:finite-basis-inference} gives bank-clustered influence-based inference for the corresponding population projection coefficients. Both nuisance regressions use the same fixed basis and OLS, so their estimation errors have parametric rates under the stated moment and rank conditions. The result does not require the basis to represent the true outcome function; interpreting its coefficient causally does require additional restrictions. Full-sample two-way demeaning instead shares estimated quarter means across banks and needs a corresponding expansion. Table~\ref{tab:estimator-scope} distinguishes these procedures.

Conventional fixed-dimensional OLS projection inference does not require cross-fitting. We retain common folds in the basis comparisons to match the other implementations; their fitted coefficients need not equal full-sample OLS coefficients.

The empirical standard errors treat the supplied analysis-ready bank records as independent observations and omit uncertainty from upstream preprocessing. In particular, the outcome can be reconstructed using sample-pooled winsorisation cutoffs. Estimating such cutoffs can contribute first-order sampling variation; neither bank-level folds nor orthogonality to the fitted regressions removes that contribution automatically.

A group-specific post-enactment aggregate shock cannot be separated from the exposure effect in a two-group comparison with one treatment date; bank clustering does not resolve that identification problem. Neither independent-bank sampling nor the forest nuisance rates is established for the banking data. Forest convergence requires additional regularity and fitting conditions \citep{wager2018}. Linear comparisons use the stated bank-clustered covariance convention. Table~\ref{tab:covered-uncertainty} reports nominal influence-based intervals for selected bank-level ratios; the fixed-basis tables report nominal standard errors for their projection targets. Other nonlinear comparisons report split sensitivity. Section~\ref{sec:simulation} assesses finite-sample performance with known targets.

\begin{table}[!htbp]
\centering\small
\caption{Identified objects and inference across panel specifications}\label{tab:estimator-scope}
\begin{tabular}{@{}>{\raggedright\arraybackslash}p{.22\linewidth}>{\raggedright\arraybackslash}p{.39\linewidth}>{\raggedright\arraybackslash}p{.32\linewidth}@{}}
\toprule
Procedure & Population coefficient and ATT status & Inference requirements\\\midrule
Linear FE, FD and CRE & Constant-effect interpretation requires the stated outcome and error restrictions. With heterogeneous effects, projection weights need not give the ATT; omitting FD time effects can retain a common shock. & Bank-clustered linear inference under the working sampling model; Supplementary Proposition~\ref{prop:fd} states FD weights.\\\addlinespace
Unrestricted adjustment revealing treatment & $J(C)=0$: the partially linear coefficient is unidentified. This does not rule out ATT identification under different DiD restrictions. & Imperfect fitted prediction cannot restore population identification.\\\addlinespace
Generic transformed-input or mean-augmented forests & A causal target is not established for these fitted rows merely by applying the transformation. They need not implement the restricted classes or exact reductions. & Bank splitting alone does not supply the required nuisance limits or handle shared quarter-mean estimation.\\\addlinespace
Within regression on demeaned level basis functions & Identified projection when its residual variance is positive. A causal interpretation additionally requires basis sufficiency and the outcome restrictions. & Proposition~\ref{prop:finite-basis-inference} gives inference for the fixed-basis projection.\\\addlinespace
Factorised FD; full-history within & Under Corollary~\ref{cor:dynamic-panel}, conditional-variance-weighted impact and average post-period effects, respectively. Positive residual variance alone does not identify the unweighted ATT. & The bank-level reduction must satisfy the relevant information, sampling and nuisance-rate conditions.\\\addlinespace
Pre/post partialling out on $Z$ & Under Assumptions~\ref{ass:consistency}--\ref{ass:sampling}, the ATT is identified, but this score targets $\tau_{\mathrm{ov}}$, generally different from the ATT. & Theorem~\ref{thm:collapsed-inference}: matching nuisance limits and the product-plus-square rate.\\\addlinespace
Normalised doubly robust DiD & ATT under consistency, no anticipation, conditional parallel trends and treated-support overlap, with a correct propensity or control-outcome regression. & Proposition~\ref{prop:dr-inference}: both nuisance limits, bounded odds and a product rate.\\
\bottomrule
\end{tabular}
\par\smallskip\begin{minipage}{.97\linewidth}\footnotesize
Notes: The dynamic weighted targets can be defined when some treated strata lack controls. Identifying an unweighted ATT requires support or an additional extrapolation restriction; equality with a weighted coefficient requires further restrictions on effects and weights. Reported uncertainty treats analysis-ready bank records as independent and omits upstream preprocessing uncertainty. Formal results state the corresponding moment and rank conditions.
\end{minipage}
\end{table}
\FloatBarrier
\section{Simulation evidence}\label{sec:simulation}
We examine the transformation results in a design with a known nonlinear outcome function and group-correlated covariates. We assess interval coverage when comparison weights are concentrated. The sample sizes and treatment timing resemble the empirical application, but the covariate and outcome distributions are synthetic. Each experiment examines a specified mechanism or procedure; together they do not provide a general ranking of FE, FD and CRE estimators.

\subsection{Nonlinear transformations and effect horizons}\label{sec:simulation:transformations}
Example~\ref{ex:quadratic-trend} separates a correctly transformed outcome function from a function of demeaned inputs. We implement it with twenty quarters, treatment beginning in quarter thirteen, and a constant effect of 0.2. A bank's scalar covariate is the sum of a group-correlated bank component, a common linear trend and an independent disturbance; the outcome function is quadratic. Unit effects can be correlated with the bank component and group membership, while the outcome error is independent of the covariate history and serially correlated. Supplementary Section~\ref{app:transformed-panel-validation} specifies every distribution and nuisance function.

The two fitted within specifications use the same linear and squared covariate terms, nineteen centred quarter indicators, five bank folds and OLS fitting rules. The first evaluates the quadratic basis in levels and then removes its bank means. The second evaluates the same basis on demeaned covariates and then re-centres its columns. Both nuisance regressions use the selected basis. In this design the first specification has population coefficient 0.2, while the second has population coefficient 0.6 because demeaning the inputs removes information needed to represent the quadratic outcome function. With 1,000 replications at each sample size, the corresponding means are 0.203 and 0.602 at $N=91$, and 0.198 and 0.597 at $N=364$ (Table~\ref{tab:transformed-panel-validation}). Increasing the sample size reduces dispersion around these different limits.

A second calculation on the same generated panels lets the treatment effect increase from 0.2 at adoption by 0.04 in each subsequent quarter. Its average post-treatment value is 0.34. Here we use the exact conditional means for the respective FD and full-history within information sets, rather than the finite-basis projections. The oracle FD means are 0.203 and 0.200 at the two sample sizes, and the within means are 0.341 and 0.338. Each is close to its own effect horizon within Monte Carlo uncertainty. This calculation evaluates Corollary~\ref{cor:dynamic-panel}; it does not assign those targets to arbitrary fitted panel regressions.

\begin{table}[!htbp]
\centering\small
\caption{Known-target validation of transformed-panel specifications}\label{tab:transformed-panel-validation}
\begin{tabular}{@{}rlrrrr@{}}
\toprule
$N$ & Specification & Target & Mean (MCSE) & Bias & RMSE\\\midrule
91 & Within: demean level basis & 0.200 & 0.203 (0.002) & 0.003 & 0.077\\
91 & Within: basis of demeaned inputs & 0.200 & 0.602 (0.005) & 0.402 & 0.435\\
91 & Dynamic: oracle first differences & 0.200 & 0.203 (0.005) & 0.003 & 0.151\\
91 & Dynamic: oracle within & 0.340 & 0.341 (0.003) & 0.001 & 0.086\\
\addlinespace
364 & Within: demean level basis & 0.200 & 0.198 (0.001) & -0.002 & 0.039\\
364 & Within: basis of demeaned inputs & 0.200 & 0.597 (0.003) & 0.397 & 0.405\\
364 & Dynamic: oracle first differences & 0.200 & 0.200 (0.002) & 0.000 & 0.073\\
364 & Dynamic: oracle within & 0.340 & 0.338 (0.001) & -0.002 & 0.044\\
\bottomrule\end{tabular}
\par\smallskip\begin{minipage}{.97\linewidth}\footnotesize
Notes: 1,000 independent panels per sample size, with 20 quarters and adoption in quarter 13. The constant-effect target is 0.2. Both fitted within specifications use the same two quadratic-basis terms, 19 transformed quarter indicators, five bank folds and OLS fitting rule. Only the order of basis evaluation and bank demeaning changes; the incorrectly transformed specification has probability limit 0.6. The dynamic rows use the true conditional means for the respective pair and full-history information sets. Their targets are the adoption effect 0.2 and average post-period effect 0.34. Bias and RMSE use the displayed target, not the misspecified probability limit. MCSE denotes the Monte Carlo standard error of the mean. Supplementary Section~\ref{app:transformed-panel-validation} gives the complete design.
\end{minipage}
\end{table}

\subsection{Finite-sample inference for bank-level changes}\label{sec:simulation:collapsed}
A separate experiment evaluates the overlap and doubly robust ATT intervals at $N=91$, using 400 replications for each of two propensity specifications with expected treated share $20/91$. Both satisfy population overlap; their propensity ranges are approximately $[0.081,0.450]$ and $[0.0112,0.883]$. Each bank has six independent uniform covariates, a nonlinear untreated conditional mean and a heterogeneous effect. Forest settings and ten treatment-stratified five-fold partitions match the application. Each score is evaluated against its own known target. Supplementary Section~\ref{app:collapsed-inference-calibration} gives the design, aggregation rules and full results.

For median aggregation, as used in the principal banking tables, nominal 95\% coverage is 92.25\% for the overlap estimator and 95.50\% for the doubly robust estimator under the first design. Under weaker overlap, coverage is 92.75\% and 89.50\%, respectively. The corresponding Monte Carlo standard errors are 1.34, 1.04, 1.30 and 1.53 percentage points. Supplementary Table~\ref{tab:collapsed-inference-calibration} also retains the mean-aggregation results, whose weaker-overlap DR coverage is 87.00\%. Oracle doubly robust coverage also falls to 88.75\%, so nuisance estimation is not the only source of undercoverage. Mean oracle control effective sample size falls from 48.46 to 9.42; the conditional centring calculation in Lemma~\ref{lem:weight-centring} explains part of the downward discrepancy in the estimated control variance component under this homoskedastic design. It supplies a diagnostic, rather than a generally valid variance correction.

Supplementary Sections~\ref{app:additional-simulation} and~\ref{app:heterogeneous-sim} report complementary experiments on within-bank dependence, nuisance-error interactions and heterogeneous-effect targets. They show why bank-level sampling, joint nuisance accuracy and the distinction between overlap effects and the ATT matter. None of these experiments establishes nominal coverage for the banking data.
\FloatBarrier
\section{The Empirical Bank Regulation Effect across panel specifications}\label{sec:emp}
We first reproduce the linear banking benchmark and assess its sensitivity to sample construction. We then compare flexible specifications, holding time adjustment and fitting rules fixed where possible. The final comparisons use pre-enactment information to examine average changes in bank growth and the consequences of conditioning on prior growth.

The outcome in the level model is quarterly log RWA growth, so 0.01 is approximately one percentage point of quarterly growth. First differences compare adjacent-quarter changes in growth, whereas the bank-level outcome compares its pre- and post-enactment averages. Linear standard errors cluster by bank and use the stated CR1 correction. Flexible point estimates are medians across repeated partitions of complete banks, with nuisance functions fitted on training banks and evaluated on held-out banks. Partition ranges describe sample-splitting variation. Table~\ref{tab:estimator-scope} states the relationship between these implementations and the formal results.

\subsection{The replicated benchmark and sample composition}\label{sec:emp:benchmark}
Panel A of Table~\ref{tab:linear-specifications} reproduces the original specifications, using an independent Python implementation;\footnote {the original R program was inspected.}. The two-way fixed-effects benchmark for \citet{chronopoulos2023} is 0.0193, with bank-clustered standard error 0.0045. The original within, correlated random effects (CRE) and first-difference (FD) results also reproduce well. The original CRE omits an intercept. Adding one to the specification with bank and quarter means recovers the fixed-effects coefficient. Panel B shows that within regression and the linear Mundlak specification with unrestricted quarter effects also give 0.0193. This equality is algebraic on the balanced sample; a common parameter-count correction aligns their standard errors.

\begin{table}[htbp]
\centering\small
\caption{Replicated linear regressions and common-specification comparisons}\label{tab:linear-specifications}
\begin{tabular}{@{}lrrr@{}}
\toprule
Specification & Coefficient & Bank-clustered s.e. & Observations\\\midrule
\multicolumn{4}{@{}l}{\textit{A. Original supplied implementations}}\\\addlinespace
Two-way fixed effects & 0.0193 & 0.0045 & 1,820\\
Two-way demeaning, no intercept & 0.0193 & 0.0044 & 1,820\\
CRE with bank means, no intercept & 0.0184 & 0.0037 & 1,820\\
CRE with bank and quarter means, no intercept & 0.0194 & 0.0044 & 1,820\\
First differences, no intercept or quarter effects & 0.0189 & 0.0054 & 1,729\\\addlinespace
\multicolumn{4}{@{}l}{\textit{B. Common two-way specification}}\\\addlinespace
Two-way fixed effects & 0.0193 & 0.0045 & 1,820\\
Within transformation & 0.0193 & 0.0045 & 1,820\\
Linear Mundlak CRE with quarter effects & 0.0193 & 0.0045 & 1,820\\
First differences with quarter effects & 0.0135 & 0.0095 & 1,729\\
\bottomrule\end{tabular}
\par\smallskip\begin{minipage}{.97\linewidth}\footnotesize
Notes: All rows use the same 91 banks and six supplied contemporaneous controls. Panel A uses an independent Python implementation of the inspected original R specifications; the R program was not executed. The mean-based CRE regressions omit an intercept and unrestricted quarter indicators, and the FD regression omits an intercept and quarter indicators. Bank and quarter means include treatment and all six controls. Panel B includes unrestricted quarter effects. Its CRE regression also includes an intercept and bank means of treatment and controls; within estimation removes bank means from outcome, treatment, controls and quarter indicators. FD uses adjacent-quarter differences. Standard errors use bank clustering and the finite-sample correction CR1. Panel A uses each regression's parameter count (117, 7, 14, 21 and 7); Panel B uses the same 117-parameter correction for its three equivalent level regressions and 26 for FD. The equivalent level regressions also have identical uncorrected covariance. Neither coefficient equality nor its absence establishes causal identification.
\end{minipage}\end{table}

The original FD regression omits an intercept and quarter indicators and gives 0.0189 (0.0054). Including quarter effects gives 0.0135 (0.0095). Reconstructed adjacent differences agree with the supplied differenced data, so this difference reflects specification. Because $\Delta D_{it}$ is nonzero only for larger banks in 2018Q2, the quarter-adjusted FD coefficient principally concerns the immediate change. Its design weights in Proposition~\ref{prop:fd} put 0.9988 on the adoption-period effect; the later weights sum to 0.0012 and include both signs.

Size classification materially affects the comparison. Of the 71 comparison banks classified using 2018Q1 assets, 27 were below \$10 billion in 2015Q1. Three of the twenty larger banks were then below \$50 billion. The former banks had average pre-enactment growth of 0.0444, compared with 0.0221 among the other comparison banks. Table~\ref{tab:sample-sensitivity} excludes the thirty banks in different endpoint asset bands and obtains 0.0122 (0.0046). Classifying the retained banks by 2015Q1 assets and restricting those assets to \$10--250 billion gives 0.0107 (0.0045). The latter exercise does not recover banks absent from the supplied sample and therefore does not reconstruct a complete earlier banking population.

\begin{table}[htbp]
\centering\small\setlength{\tabcolsep}{4pt}
\caption{Sample construction and influential observations}\label{tab:sample-sensitivity}
\begin{tabular}{@{}p{.49\linewidth}rrrr@{}}
\toprule
Specification & Coefficient & CR1 SE & Banks & Observations\\\midrule
Original sample & 0.0193 & 0.0045 & 91 & 1,820 \\
Exclude banks whose asset band changed & 0.0122 & 0.0046 & 61 & 1,220 \\
Classify by 2015Q1 assets; retain \$10--250bn & 0.0107 & 0.0045 & 64 & 1,280 \\
Delete BB\&T, 2019Q4 observation & 0.0178 & 0.0044 & 91 & 1,819 \\
Delete BB\&T throughout & 0.0181 & 0.0045 & 90 & 1,800 \\
Exclude quarters with $|\Delta\log A|>0.25$ & 0.0166 & 0.0033 & 91 & 1,791 \\
Exclude quarters with $|\Delta\log A|>0.15$ & 0.0123 & 0.0030 & 91 & 1,762 \\
\bottomrule\end{tabular}
\par\smallskip\begin{minipage}{.97\linewidth}\footnotesize
Notes: Each row refits the two-way fixed-effects benchmark with the same six contemporaneous controls. Asset bands are below \$10bn, \$10--50bn, \$50--250bn and above \$250bn. The second row excludes the thirty banks in different bands in 2015Q1 and 2018Q1; the third reclassifies the retained banks using 2015Q1 assets. Both the second and third rows contain seventeen larger banks. The third does not reconstruct the full banking population that would have qualified in 2015Q1. $A$ denotes total assets; the 2015Q2 asset-change calculation uses the supplied 2015Q1 level as its lag. Large asset changes are sensitivity criteria, not verified merger classifications. Observation exclusions yield unbalanced panels; all fixed effects are refitted. Standard errors cluster by bank using the same CR1 convention. These restrictions change the sample and potentially its causal target.
\end{minipage}\end{table}

The BB\&T/SunTrust merger, completed in December 2019, also affects the comparison \citep{truist2019merger}. Deleting BB\&T's 2019Q4 observation lowers the benchmark to 0.0178; deleting the bank gives 0.0181. Excluding quarters with absolute log asset changes above 0.25 or 0.15 gives 0.0166 and 0.0123. Large asset changes are sensitivity criteria rather than verified merger classifications, and excluding post-enactment changes may remove economic responses to regulation. These restrictions alter the sample and potentially its target population; they do not select a corrected policy effect. Ending in 2019Q4 gives 0.0174 (Supplementary Table~\ref{tab:linear}). The available foreign-activity controls leave the benchmark close to its original value on a common complete-case sample (Supplementary Section~\ref{app:foreign}).

\subsection{Nonlinear adjustment and the order of transformation}\label{sec:emp:panel}
Supplementary Table~\ref{tab:new-panel} reports the original OLS and Random Forest nuisance fits on the same sample, six contemporaneous controls and five repeated five-fold bank partitions. Forests use 100 trees and minimum leaf size five bank-quarter observations. Both columns use held-out residuals, so even OLS need not equal the full-sample coefficient. The bank-mean and within-bank adjusted rows omit unrestricted quarter indicators; quarter covariate means do not replace them. This table therefore records the original implementations, with their different time adjustments.

On two-way demeaned data, forest and OLS medians are 0.0181 and 0.0185. Together with the full-sample linear benchmark of 0.0193, these positive estimates provide evidence of stability to flexible fitting on the same transformed inputs. The comparison does not establish that a nonlinear function of demeaned controls represents the demeaned level function, which is the separate specification issue addressed by the matched basis exercise below. The intermediate bank-mean and adjusted rows in Supplementary Table~\ref{tab:new-panel} are not matched on time adjustment and remain historical diagnostics. Including both bank and quarter treatment means yields perfect forest treatment predictions and an undefined coefficient, as Corollary~\ref{cor:absorbing} predicts. The additive OLS fit retains variation by excluding their interaction, which requires a corresponding outcome restriction. Full-sample two-way demeaning or covariate quarter means also link banks, so the independent-bank inference result does not directly cover those implementations.

Table~\ref{tab:matched-basis} isolates the order of nonlinear expansion and bank demeaning. Each design uses the same twelve basis terms, nineteen deterministic quarter indicators, training-only scaling, bank partitions and OLS fitting rule. Squaring covariates before bank demeaning gives 0.01535; squaring demeaned covariates and then removing their bank means gives 0.01897. Every training design has full rank. Both rows use bank-only demeaning, which preserves independence across banks under independent bank sampling. The nominal standard errors are 0.00500 and 0.00518. The median paired third-minus-first difference is 0.00236, with nominal standard error 0.00412; it is imprecisely estimated. The comparison isolates a specification choice, while the outcome model determines which function space is appropriate. Supplementary Table~\ref{tab:model-linked-fits} reports the corresponding linear-basis restricted projection.

\begin{table}[htbp]
\centering\small\setlength{\tabcolsep}{4pt}
\caption{Matched nonlinear basis comparisons under bank demeaning}\label{tab:matched-basis}
\begin{tabular}{@{}lrrrr@{}}
\toprule
Covariate basis & Median & Nominal SE & Partition range & Median $\widehat J$\\\midrule
$M b(X)$ & 0.01535 & 0.00500 & [0.01441, 0.01735] & 0.04261\\
$b(MX)$ & 0.01890 & 0.00492 & [0.01687, 0.01997] & 0.04167\\
$M b(MX)$ & 0.01897 & 0.00518 & [0.01598, 0.01976] & 0.04216\\
\bottomrule\end{tabular}
\par\smallskip\begin{minipage}{.98\linewidth}\footnotesize
Notes: $M$ subtracts each bank's time mean, and $b(x)$ contains the six benchmark controls and their six individual squares. Each design includes the same nineteen bank-centred quarter indicators. Outcome and treatment are bank-demeaned. Both nuisance functions use OLS without an intercept, training-only column scaling, and the same five complete-bank partitions with five folds each. Every training design has rank 31. The second row leaves squared demeaned covariates uncentred; the third removes their bank means. $\widehat J$ is the held-out treatment-residual second moment. Standard errors use Supplementary Proposition~\ref{prop:finite-basis-inference} and the median aggregation in equation~\eqref{eq:median-variance}, under the independent analysis-ready-bank working model. They omit upstream preprocessing uncertainty and concern projection coefficients. Ranges describe partition sensitivity.
The median paired third-minus-first difference is 0.00236 (nominal SE 0.00412); it need not equal the difference of the row medians.
\end{minipage}\end{table}

\subsection{First-difference nuisance specifications}\label{sec:emp:fd-nuisances}
Table~\ref{tab:fd-nuisance-comparison} uses the same bank partitions and learner settings on 1,729 adjacent differences. Both nuisance regressions include quarter indicators. One specification uses six differenced controls; the other retains their twelve current and lagged values. The latter preserves the arguments of $g_0(X_{it})-g_0(X_{i,t-1})$, which a function of differences alone generally cannot represent. Current adoption-quarter controls nevertheless need not be pre-treatment variables.

\begin{table}[htbp]
\centering\small\setlength{\tabcolsep}{4pt}
\caption{First-difference estimates under two nuisance specifications}\label{tab:fd-nuisance-comparison}
\begin{tabular}{@{}p{.37\linewidth}rrrr@{}}
\toprule
Covariates in both nuisance regressions & OLS & Forest & Forest $\widehat J_\Delta$ & \shortstack{Adoption\\Brier loss}\\\midrule
Six first differences & 0.0125 & 0.0186 & 0.00931 & 0.1769\\
Six current and six lagged values & 0.0143 & 0.0092 & 0.00648 & 0.1231\\
\bottomrule\end{tabular}
\par\smallskip\begin{minipage}{.97\linewidth}\footnotesize
Notes: Medians over the five complete-bank partitions used in Table~\ref{tab:new-panel}, with five folds per partition and identical forest settings and random seeds. Both nuisance specifications include eighteen fixed quarter indicators, omitting 2015Q3. The sample contains 91 banks and 1,729 adjacent differences, 2015Q3--2020Q1. Forest outcome regressions predict the difference in quarterly growth; forest classifiers predict the zero/one first difference of the coded treatment. OLS includes an intercept in each nuisance regression. The final coefficient in every row is the through-origin regression of held-out outcome residuals on held-out treatment residuals. Thus the OLS column reports cross-fitted medians, not full-sample regression coefficients. $\widehat J_\Delta$ averages squared held-out treatment residuals across all adjacent differences. Adoption Brier loss instead averages $(G_i-\widehat m^\Delta_{it^*})^2$ across the 91 enactment-quarter banks; a constant prediction estimated on each training fold gives 0.1715. These fitted losses do not separate prediction error from population residual variation. Forest partition ranges are $[0.0136,0.0232]$ for differenced controls and $[0.0081,0.0122]$ for current and lagged controls; these are not confidence intervals. Both representations include potentially post-treatment information: current/lagged controls contain $X_{it^*}$, and differenced controls contain it through $\Delta X_{it^*}$. The comparison changes how that information enters as well as the functional representation. Different conditioning choices need not share a causal target.
\end{minipage}\end{table}

The OLS medians are 0.0125 and 0.0143; the forest medians are 0.0186 and 0.0092. Substituting the saved forest nuisance predictions one at a time gives 0.0156 when only the treatment prediction comes from current and lagged controls, and 0.0124 when only the outcome prediction does so (Supplementary Table~\ref{tab:fd-nuisance-swaps}). Both nuisance fits contribute to the difference, although these substitutions do not give a causal decomposition or show that the smaller estimate is less biased.

At adoption, the median held-out Brier loss for predicting $G_i$ is 0.1769 with differenced controls and 0.1231 with current and lagged controls. A constant prediction estimated on each training fold gives 0.1715. The differenced-control forest performs worse than that constant in all five partitions. This is evidence of weak predictive performance in the retained fits, rather than a proof that group membership contains no predictable variation. Proposition~\ref{prop:risk} separates true residual variation from prediction error; the observed loss alone cannot establish the size or sign of the nuisance-error product driving a coefficient discrepancy.

An implementation of Corollary~\ref{cor:fd-collapse} instead fits group membership and the adjacent growth change on the 91 adoption-quarter bank observations, setting treatment predictions to zero elsewhere. Its coefficient is exactly the bank-level residual ratio: median 0.0052 and partition range $[-0.0020,0.0073]$ (Supplementary Table~\ref{tab:model-linked-fits}). Its minimum leaf size is five banks, rather than five differenced observations; both the training sample and prediction restriction change. The influence-based standard-error calculation is 0.0155, subject to the unverified conditions of Corollary~\ref{cor:fd-inference} (Supplementary Table~\ref{tab:audit-uncertainty}).

To isolate the support restriction, Supplementary Table~\ref{tab:fd-support} retains all saved outcome predictions and adoption-quarter treatment predictions, changing only off-adoption treatment predictions to zero. Both principal forest medians remain unchanged: the differenced-control forest already predicts zero off adoption, and the largest coefficient change for the current/lagged forest is 0.000074. The separate estimate of 0.0052 therefore cannot be attributed to the support restriction alone.

\subsection{Pre-enactment conditioning and uncertainty}\label{sec:emp:drdid}
The bank-level outcome is $H_i=H_i^{\mathrm{post}}-H_i^{\mathrm{pre}}$, using twelve pre- and eight post-enactment quarters; $Z_i$ contains six pre-enactment covariate averages. Group prediction now concerns $G_i$ conditional on $Z_i$. Table~\ref{tab:covered-uncertainty} compares the overlap partialling-out and normalised doubly robust (DR) estimators across four learner families using the same ten repeated five-fold bank partitions. Panel A reports point estimates and DR control-weight concentration. Panel B gives nominal intervals for forests with 200 trees and minimum leaf size five banks. At the true group propensity and under their identifying assumptions, their targets are the overlap-weighted treated effect and the ATT, respectively.

\begin{table}[!htbp]
\centering\small\setlength{\tabcolsep}{4pt}
\caption{Bank-level estimates across learners and nominal forest confidence intervals}\label{tab:covered-uncertainty}\label{tab:bank-change-comparison}
\begin{tabular}{@{}llrrrr@{}}
\toprule
\multicolumn{6}{@{}l}{\textit{Panel A. Four-learner comparison and DR control weights}}\\[3pt]
Learner & Features & \shortstack{Overlap\\PLR} & \shortstack{Normalised\\DR-DiD} & \shortstack{Control\\ESS} & \shortstack{Largest\\weight} \\
\midrule
OLS/$L_2$ logit & Six means & 0.0084 & $-0.0048^{\dagger}$ & 1.4 & 0.845 \\
 & $+$ prior growth & $-0.0003$ & $0.0065^{\dagger}$ & 1.4 & 0.850 \\
\addlinespace
Random forest & Six means & 0.0107 & 0.0125 & 17.6 & 0.114 \\
 & $+$ prior growth & 0.0015 & 0.0026 & 20.7 & 0.107 \\
\addlinespace
Polynomial LASSO & Six means & 0.0095 & $-0.0033^{\dagger}$ & 1.3 & 0.873 \\
 & $+$ prior growth & 0.0015 & $-0.0053^{\dagger}$ & 1.3 & 0.866 \\
\addlinespace
XGBoost & Six means & 0.0106 & 0.0117 & 19.8 & 0.129 \\
 & $+$ prior growth & 0.0024 & 0.0046 & 16.1 & 0.189 \\
\bottomrule
\end{tabular}
\par\medskip
\begin{tabular}{@{}p{.49\linewidth}rrr@{}}
\toprule
\multicolumn{4}{@{}l}{\textit{Panel B. Forest estimates and nominal confidence intervals}}\\[3pt]
Specification & Median & $s_{\mathrm{med}}$ & Nominal 95\% interval\\\midrule
\multicolumn{4}{l}{\textit{Overlap PLR}} \\
Six pre-enactment means & 0.0107 & 0.0042 & [0.0024, 0.0189] \\
Add prior growth & 0.0015 & 0.0037 & [-0.0057, 0.0088] \\
Paired change: augmented minus baseline & -0.0103 & 0.0040 & [-0.0182, -0.0024] \\
\midrule
\multicolumn{4}{l}{\textit{DR}} \\
Six pre-enactment means & 0.0125 & 0.0040 & [0.0047, 0.0202] \\
Add prior growth & 0.0026 & 0.0035 & [-0.0042, 0.0094] \\
Paired change: augmented minus baseline & -0.0092 & 0.0040 & [-0.0171, -0.0014] \\
\bottomrule\end{tabular}

\par\smallskip\begin{minipage}{.97\linewidth}\footnotesize
Notes: The outcome is the change between twelve pre- and eight post-enactment averages of quarterly RWA growth. All rows use the same 91 banks and ten saved five-fold bank partitions. Panel A reports partition medians. OLS/$L_2$ logit and forest predictions use their retained settings. LASSO and XGBoost use nested three-fold bank validation. Forests have 200 trees and minimum leaf size five banks. DR uses propensity clipping at $[0.01,0.99]$. The overlap score uses unmodified predictions. Control ESS and largest normalised control weight refer to DR. $\dagger$ marks DR estimates with very concentrated weights and empirically uncalibrated nominal intervals. Supplementary Table~\ref{tab:learner-comparison} reports standard errors and predictive diagnostics for all learners.

For partition estimates $\widehat\theta_r$ and influence-based standard errors $s_r$, $\widehat\theta_{\mathrm{med}}=\operatorname{med}_r\widehat\theta_r$ and $s_{\mathrm{med}}=\{\operatorname{med}_r[s_r^2+(\widehat\theta_r-\widehat\theta_{\mathrm{med}})^2]\}^{1/2}$. Panel B gives the forest intervals $\widehat\theta_{\mathrm{med}}\pm1.96s_{\mathrm{med}}$. Paired changes and influence differences are formed within each partition before aggregation. Their median need not equal the difference of the two displayed medians. A confidence-interval interpretation requires the sampling, support, clipping and nuisance-rate assumptions in Section~\ref{sec:estimation}, jointly for paired comparisons. These conditions are not verified for the banking fits, and the reported calculations do not establish finite-sample coverage.
\end{minipage}
\end{table}

The forest baseline medians are 0.0107 and 0.0125, compared with the unadjusted difference in mean changes of 0.0166. The outcome-regression-only (OR) median is 0.0150. OR averages treated-bank residuals from the comparison-bank outcome regression; DR additionally subtracts the odds-weighted comparison-bank residual mean. Supplementary Tables~\ref{tab:propensity-or-diagnostics} and \ref{tab:audit-uncertainty} report OR estimates, alternative learners and weight concentration. Orthogonality does not make the overlap score doubly robust: its nuisance remainder includes squared propensity error, so the point-estimate gap cannot isolate effect weighting.

Panel B of Table~\ref{tab:covered-uncertainty} uses the point estimates' aggregation rule for nominal influence-based intervals. Section~\ref{sec:estimation} states their sampling, preprocessing, nuisance-rate and clipping qualifications. In the $N=91$ simulations using the same median aggregation, overlap and DR coverage are 92.25\% and 95.50\% under ordinary overlap, and 92.75\% and 89.50\% under weaker overlap (Supplementary Table~\ref{tab:collapsed-inference-calibration}). These results justify treating the banking intervals as nominal, rather than as demonstrated 95\% coverage. Baseline fitted probabilities fall below 0.01 for 11--15 comparison banks per partition. Removing clipping changes the DR median by only 0.000013, but this small numerical difference does not verify the population conditions for inference. The OLS/logit DR fits remain unstable, with median control effective sample size about 1.4; Supplementary Table~\ref{tab:audit-uncertainty} retains their mean--median discrepancies.

The forest group-prediction Brier loss is 0.1242 with $Z_i$ and 0.1215 with prior growth added, compared with 0.1715 for the constant sample share. This loss includes irreducible group variation and does not directly estimate propensity-function error or provide a guaranteed population dispersion bound. Six larger banks fall outside the comparison sample's marginal range on at least one covariate (Supplementary Table~\ref{tab:covariate-support}). These observed ranges raise concerns about extrapolation of the control-outcome regression; they do not establish a failure of population overlap.

\subsection{Prior growth, outcome fitting and earlier comparisons}\label{sec:emp:pre-growth}
Average pre-enactment growth is 0.0100 for larger banks and 0.0306 for comparison banks. Unequal initial growth does not itself violate parallel trends, but may matter through persistence or mean reversion. Adding $H_i^{\mathrm{pre}}$ reduces the overlap median from 0.0107 to 0.0015 and the DR median from 0.0125 to 0.0026. Within-partition changes are negative throughout; their medians are $-0.0103$ and $-0.0092$, with nominal standard-error calculations approximately 0.0040. Medians of paired changes need not equal differences of the separately displayed medians.

This adjustment changes the conditional-trends restriction, and neither restriction generally implies the other. The overlap-weighted target can also change. If both DR specifications identify the same ATT and satisfy the joint inference conditions, a valid paired test can assess their compatibility. The nominal paired DR interval excludes zero, providing evidence against simultaneous validity of that joint set of identifying and estimation assumptions, conditional on the inference approximation. The unverified conditions and finite-sample coverage limit that interpretation; the comparison cannot determine which identifying restriction or nuisance model is valid. Holding the enlarged feature set, fitted propensities, training banks and forest settings fixed, fitting post-enactment growth and subtracting observed prior growth gives a DR median of $-0.0023$, instead of 0.0026 when fitting the change directly. The overlap median remains positive at 0.0003. Both response constructions represent the same population conditional mean, but forests need not deliver identical fitted values; OLS does when prior growth is a regressor (Supplementary Table~\ref{tab:response-check}).

The downward adjustment persists after excluding endpoint-band crossers, BB\&T, or banks with large post-period asset changes, and after ending the post period in 2019Q4 (Supplementary Table~\ref{tab:audit-prepost-samples}). In the first-six versus last-six pre-enactment quarters, however, the median paired adjustments are only $-0.0006$ for overlap and $-0.0022$ for DR, with nominal standard errors 0.0041 and 0.0057. Their intervals include zero (Supplementary Table~\ref{tab:prereform-placebo}). The negative adjusted pre-enactment levels are compatible with the persistent group-mean mechanism discussed in Supplementary Section~\ref{app:prior-growth-interpretation}. However, the paired adjustments are small relative to
their nominal uncertainty and do not distinguish this mechanism from alternative untreated-growth dynamics or reproduce the sizeable post-enactment contraction. The DR level moves from $-0.0076$ to $-0.0105$ after prior-growth adjustment and to $-0.0148$ when later growth is the fitted response; a negative adjusted level does not establish a nonzero paired adjustment. A shorter comparison, 2015Q2--2016Q1 against 2016Q2--2016Q4, gives descriptive DR medians $-0.0009$ and $-0.0103$ before and after adjustment. These windows are not policy-free experiments: the shorter one includes the September 2016 CCAR proposal \citep{fed2016ccar}.

Full-sample linear diagnostics show downward adjustments in all three windows (Supplementary Table~\ref{tab:prior-growth-linear}). In the post-enactment window, the coefficient moves from 0.0110 (HC3 standard error 0.0047) to $-0.0012$ (0.0040), a change of $-0.0122$ with paired standard error 0.0039. In the six-versus-six pre-enactment window, it moves from $-0.0073$ to $-0.0161$, a change of $-0.0088$ with paired standard error 0.0053; its nominal interval includes zero. The corresponding early-growth slopes are $-0.7343$ and $-0.7349$. Thus the linear earlier-window adjustment is sizeable but imprecise, whereas the direct-change forest adjustments are smaller. The qualitative post-enactment finding does not require DML. In the shorter pre-2017 window, the unadjusted-for-growth linear coefficient is 0.0110 (0.0112), compared with a near-zero forest coefficient; no formal between-estimator difference is inferred.

Panel A of Table~\ref{tab:covered-uncertainty} holds the banks, conditioning sets, scores and saved folds fixed while adding polynomial LASSO and XGBoost. Supplementary Table~\ref{tab:learner-comparison} gives the full results and predictive diagnostics. Their overlap medians fall from 0.0095 to 0.0015 and from 0.0106 to 0.0024 after prior-growth adjustment; the paired changes are negative in every partition for all four learner families. XGBoost DR medians also fall, from 0.0117 to 0.0046. Loss-minimising LASSO instead gives negative DR medians and a control effective sample size near 1.3, with roughly 87\% of normalised control weight on one bank. The alternative one-standard-error penalty rule substantially reduces that concentration. These results strengthen the evidence for the overlap contraction while showing why predictive tuning alone does not ensure stable odds weighting; Supplementary Section~\ref{app:learner-comparison} reports the paired uncertainty and diagnostics.

A negative coefficient on prior growth partly reflects the outcome construction: if $E_i$ and $L_i$ are early and late growth averages, then the slope of $L_i-E_i$ on $E_i$ equals the slope of $L_i$ on $E_i$ minus one, holding the other regressors fixed. Two interpretations are possible. With persistent group-specific mean growth, adjusting for a noisy initial growth measure can introduce a group contrast even when untreated average changes are equal. With a common conditional growth relationship and a transitory initial difference, prior-growth adjustment can instead remove an upward discrepancy in the unadjusted comparison. Supplementary Section~\ref{app:prior-growth-interpretation} states simple models for both cases. The slope identity and these models establish competing explanations; the available earlier comparisons do not select one. Numerical similarity between the observed contraction and a calculation under the persistent-mean model would not establish that this mechanism explains the entire contraction or that either estimate is more causally valid. Different significance findings across periods are not a test of a difference between their adjustments. Causal bracketing would additionally require the outcome-dynamics and distributional-ordering conditions of \citet{ding2019bracketing}, applied to a common target.

\subsection{Outcome construction and regulatory timing}\label{sec:emp:outcome-timing}
The statutory-timing proxy in Supplementary Section~\ref{app:timing-outcomes} assigns 2018Q2 to the five larger-group banks below \$100 billion and 2019Q4 to the other fifteen, using the eighteen-month delay. Its conditional coefficient is 0.02285 (CR1 standard error 0.00729); the later cohort has only two post quarters. Separately, splitting the common-enactment coefficient by these size bands gives 0.0258 (0.0074) and 0.0171 (0.0049), with a contrast of 0.0086 (0.0081; $p=0.29$). These retained sensitivities distinguish enactment coding from an effective-date proxy. The latter changes timing and cohort comparisons, so the common-date score reductions do not automatically apply. It does not establish bank-specific relief dates or a cleaner causal interpretation. We retain the common 2018Q2 design.

Raw quarterly growth retains observations affected by the supplied winsorisation. Replacing the outcome and its prior-growth feature with raw values raises the baseline overlap and DR medians to 0.0145 and 0.0167; adding prior growth lowers them to 0.0035 and 0.0050 (Supplementary Table~\ref{tab:final-outcome-timing}). The direction of the adjustment thus survives this change, while the magnitudes and nominal standard errors increase. The six control variables retain their inherited preprocessing, so this is an outcome-construction comparison. Because the pooled cutoffs use post-enactment as well as pre-enactment outcomes, the raw-growth comparison also assesses sensitivity to that transformation. Parallel trends must be justified on the chosen outcome scale.

The last quarter also has a distinct economic interpretation. Pandemic credit-line drawdowns in March 2020 were concentrated at large banks and increased balance-sheet lending and risk-weighted assets \citep{li2020lenders}. Ending in 2019Q4 lowers the supplied-outcome baseline medians to 0.0085 and 0.0101, and the prior-growth-adjusted medians to 0.0000 and 0.0003. The corresponding raw-outcome adjusted medians remain positive at 0.0037 and 0.0036. The comparison identifies sensitivity to the pandemic quarter; attributing its contribution to drawdowns would require bank-specific evidence.

The January 2017 CCAR change is an earlier source of differential regulatory exposure. Restricting the pre-period to 2017Q1--2018Q1 and ending the post-period in 2019Q4 gives baseline medians 0.0130 and 0.0189, falling to $-0.0021$ and $-0.0031$ with prior growth. This exercise recomputes all pre-period features from five quarters and therefore changes both the conditioning information and the comparison window. It does not isolate the 2017 policy effect or eliminate possible anticipation of the 2018 reform. Together, the window comparisons show that the contraction persists, but its size and the adjusted point estimates depend on the economic periods being compared.

\subsection{Empirical interpretation}\label{sec:emp:trends}\label{sec:emp:dynamics}
The supplied linear specification reproduces a positive size-group coefficient, and the demeaned forest estimate is similar. Its magnitude is sensitive to classification date and influential balance-sheet changes. Flexible estimates also depend on the order of transformation, the information retained in first differences, prior-growth conditioning and outcome fitting. Overall, we reproduce the positive conditional
size-group differential under the original specification, but its magnitude varies with sample construction and conditioning choices. The available evidence does not
determine which specification best represents the counterfactual growth path for the banking comparison.

\begin{figure}[htbp]
\centering
\includegraphics[width=.94\linewidth]{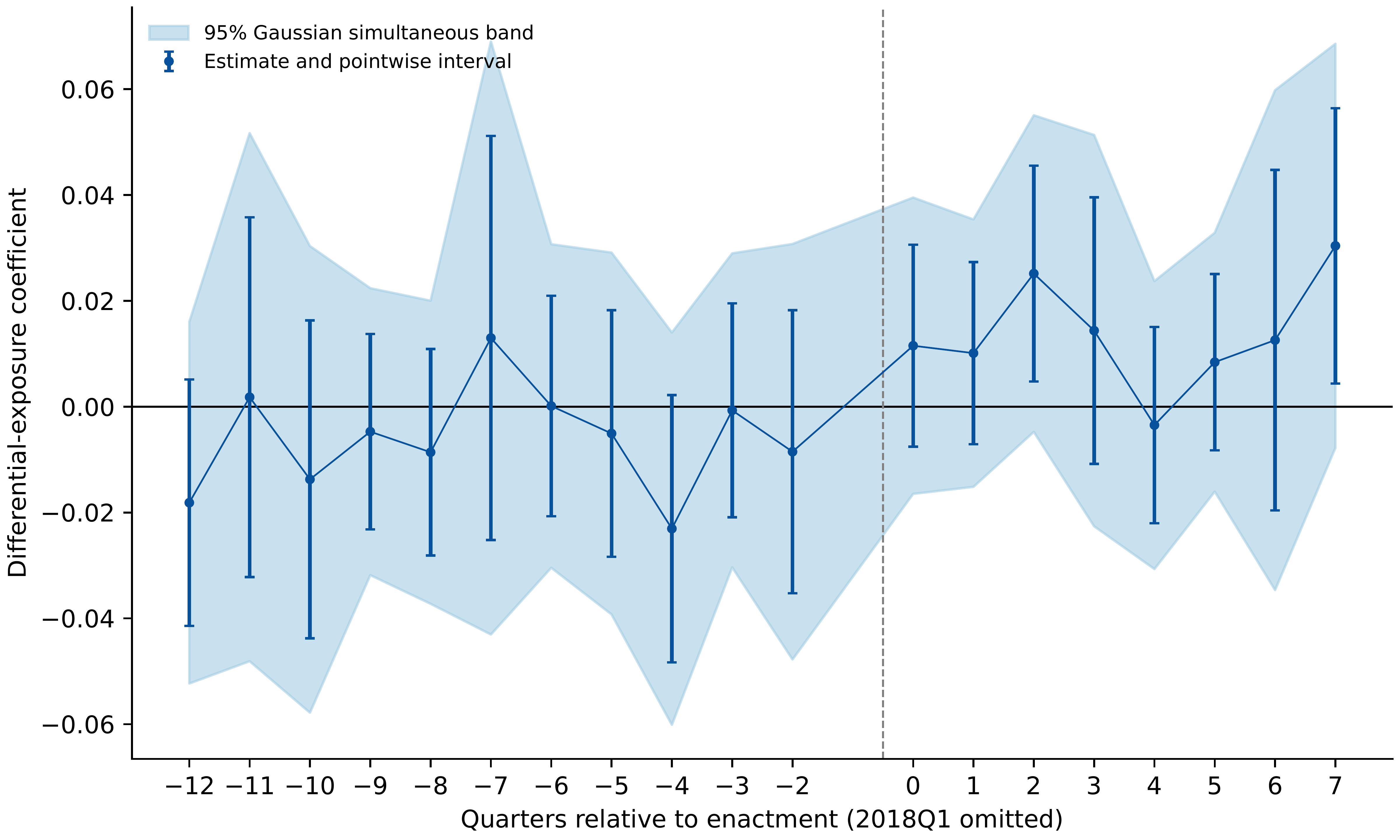}
\caption{Event-study coefficients for winsorised quarterly growth of risk-weighted assets, controlling for bank and quarter fixed effects and the six contemporaneous covariates. The omitted quarter is 2018Q1 ($k=-1$). The sample contains 91 banks over 20 quarters. Pointwise 95\% intervals use finite-sample-corrected (CR1) bank-clustered standard errors and the $t_{90}$ critical value. Simultaneous 95\% bands use the same estimated covariance and a Gaussian maximum-$t$ critical value of 2.9158, computed from 300,000 seeded draws across the nineteen reported coefficients. Both calculations maintain the bank-clustered sampling approximation.}
\label{fig:event}
\end{figure}

The raw-growth accounting decomposition assigns about seventy percent of its RWA coefficient to asset growth and the remainder to regulatory risk density (Supplementary Table~\ref{tab:outcomes}); this is not causal mediation. The event study does not reject zero pre-enactment coefficients, but its imprecision limits that diagnostic (Figure~\ref{fig:event}). Its reference quarter, 2018Q1, includes the Senate vote discussed in Section~\ref{sec:setup:target}. Changing the omitted quarter would renormalise coefficients without changing a fixed pre/post contrast; addressing possible anticipation instead requires a substantive change in the exposure or comparison window. The 2020Q1 coefficient and the sensitivity to ending in 2019Q4 also leave possible differential pandemic exposure within the comparison; those exercises do not identify a separate pandemic effect. Both groups received relief. Consequently, the banking evidence concerns differential exposure under maintained assumptions and does not identify an overall effect of deregulation.

\FloatBarrier
\section{Conclusion}\label{sec:conclusion}
Our bank regulation analysis finds stable results with flexible fitting on matched demeaned inputs, alongside sensitivity to the construction of economic settings. Asset-size classification and influential balance-sheet changes affect the positive effect in the benchmark analysis. Adding prior growth reduces bank-level overlap estimates across four learner families, while DR-DiD estimates depend more strongly on propensity fitting and concentrated control weights. The alternative LASSO penalty rule reduces that concentration, illustrating why predictive performance alone cannot assess the reliability of odds weighting. Pre-enactment comparisons and alternative outcome constructions and windows leave competing explanations for prior-growth sensitivity unresolved. The evidence concerns differential regulatory exposure under maintained assumptions.

For panels with a common treatment date, our results connect panel transformations to nuisance restrictions, identification and treatment-effect targets. Restricting nuisance functions can preserve residual treatment variation, but identification also requires a compatible outcome model and rank condition. Exact score reductions provide bank-level representations for specified implementations. Their causal interpretations require further assumptions and distinguish conditional-variance-weighted post-treatment and adoption-quarter effects from the unweighted ATT. The inference results state the additional conditions for bank-level effect calculations to apply. The coverage experiments document undercoverage for DR-DiD with concentrated control weights, including when the nuisance functions are known.

Applied researchers should carefully define the intended treatment-effect object and justify the identifying restrictions when choosing a panel transformation and nuisance learner. Bank-level cross-fitting, regularisation and predictive validation address estimation problems within each specification. Residual treatment variation, propensity support, weight concentration and partition sensitivity help assess implementation, alongside the conditions governing sampling uncertainty. This makes DML a useful instrument for robustness analysis while making the causal content explicit.
\Needspace{10\baselineskip}
\section*{Acknowledgements}
We thank Dimitris K. Chronopoulos and Muhammed H. Yilmaz for sharing their data, and Annalivia Polselli for clarifications concerning the panel methods in \citet{clarke2026}. We are grateful to participants at New York Camp Econometrics XX, the QFFE 2026 International Conference on Quantitative Finance and Financial Econometrics, the 31st International Panel Data Conference, and the 95th Annual Conference of the Southern Economic Association for helpful comments. All remaining errors are our own.
\section*{Data availability}
The replication package distinguishes reconstruction from retained predictions, fresh fitting and upstream data construction. It contains code, numerical records, software requirements and verification routines; the manuscript and supplement compile without banking data. The complete package used for private verification includes the analysis-ready inputs supplied by Dimitris K. Chronopoulos and Muhammed H. Yilmaz. Permission for public redistribution of those inputs and recoverable bank-level derivatives has not been confirmed; access requests should be directed to the providers. The supplied outcome can be reproduced by the winsorisation rule stated in Section~\ref{sec:emp:data}, but the original collection, complete sample-selection history and preprocessing program remain unavailable. Code and non-disclosive aggregate results can be provided separately.
\section*{Declaration of Generative AI Use}
We used Anthropic Claude Opus 5.5 and OpenAI ChatGPT (GPT-6 Astra)
to proofread author-written passages and improve grammar, clarity,
and flow. We used OpenAI Codex (GPT-6) to check LaTeX formatting,
cross-references, and PDF compilation. We reviewed the tools' terms
of use and confirmed their suitability for these tasks. We reviewed
the suggested changes, verified the originality and accuracy of the
final manuscript, and take full responsibility for its content,
including the references.
\FloatBarrier
\clearpage
\pdfbookmark[0]{References}{main-references}
\setlength{\bibsep}{3pt}
\bibliographystyle{plainnat}
\bibliography{references}

\clearpage
\setcounter{section}{0}
\setcounter{equation}{0}
\setcounter{table}{0}
\setcounter{figure}{0}
\setcounter{theorem}{0}
\setcounter{proposition}{0}
\setcounter{lemma}{0}
\setcounter{corollary}{0}
\setcounter{example}{0}
\setcounter{remark}{0}
\setcounter{assumption}{0}
\renewcommand{\thesection}{S\arabic{section}}
\renewcommand{\theequation}{S\arabic{equation}}
\renewcommand{\thetable}{S\arabic{table}}
\renewcommand{\thefigure}{S\arabic{figure}}
\renewcommand{\thetheorem}{S\arabic{theorem}}
\renewcommand{\theproposition}{S\arabic{proposition}}
\renewcommand{\thelemma}{S\arabic{lemma}}
\renewcommand{\thecorollary}{S\arabic{corollary}}
\renewcommand{\theexample}{S\arabic{example}}
\renewcommand{\theremark}{S\arabic{remark}}
\renewcommand{\theassumption}{S\arabic{assumption}}
\providecommand{\theHsection}{}
\renewcommand{\theHsection}{supp.\arabic{section}}
\providecommand{\theHequation}{}
\renewcommand{\theHequation}{supp.\arabic{equation}}
\providecommand{\theHtable}{}
\renewcommand{\theHtable}{supp.\arabic{table}}
\providecommand{\theHfigure}{}
\renewcommand{\theHfigure}{supp.\arabic{figure}}
\providecommand{\theHtheorem}{}
\renewcommand{\theHtheorem}{supp.\arabic{theorem}}
\providecommand{\theHproposition}{}
\renewcommand{\theHproposition}{supp.\arabic{proposition}}
\providecommand{\theHlemma}{}
\renewcommand{\theHlemma}{supp.\arabic{lemma}}
\providecommand{\theHcorollary}{}
\renewcommand{\theHcorollary}{supp.\arabic{corollary}}
\providecommand{\theHexample}{}
\renewcommand{\theHexample}{supp.\arabic{example}}
\providecommand{\theHremark}{}
\renewcommand{\theHremark}{supp.\arabic{remark}}
\providecommand{\theHassumption}{}
\renewcommand{\theHassumption}{supp.\arabic{assumption}}

\pdfbookmark[0]{Supplementary material}{supplement-title}\begin{center}{\Large\bfseries Supplementary material\par}\vspace{1ex}{\large Double Machine Learning for Panel Data:\\ Identification, Specification, and a Bank Regulation Effect\par}\vspace{1ex}Ali Habibnia \quad Esfandiar Maasoumi \quad Gaurav Shah\par October 2026\end{center}\vspace{1.5ex}

\section{Supporting identification and projection results}\label{app:supporting-results}
This section records results used in the main text. Proofs follow in Section~\ref{app:proofs}. All symbols retain their main-text definitions unless defined below.
\subsection{Fixed effects and the conditional DiD formula}
\begin{proposition}[The no-control DiD and two-way fixed-effects benchmark]\label{prop:did-benchmark}
Consider a balanced panel with common treatment timing and both groups and both period blocks present.
\begin{enumerate}
\item[(i)] The ordinary least squares (OLS) coefficient on $D_{it}$ in a two-way fixed-effects (TWFE) regression with unit and period effects and no other regressors equals the difference in group means of $H_i$.
\item[(ii)] Under Assumptions~\ref{ass:consistency}, \ref{ass:noanticipation} and \ref{ass:sampling}, and Assumption~\ref{ass:pt} with $Z$ empty, the population difference in group means of $H$ equals $\bar\tau$. Effects may vary across treated units and post-treatment periods.
\item[(iii)] Under Assumptions~\ref{ass:consistency} to \ref{ass:sampling},
\begin{equation}
\bar\tau=\E\!\left[\E[H\mid G=1,Z]-\E[H\mid G=0,Z]\,\middle|\,G=1\right].
\end{equation}
\end{enumerate}
\end{proposition}

\subsection{Restricted coefficients and heterogeneous effects}
\begin{corollary}[The linear benchmark and the parallel-trends discrepancy]\label{cor:twfe}
On a balanced panel with common treatment timing and $n=NT$ observations, use the empirical inner product $\langle a,b\rangle_n=n^{-1}a'b$. Let $\mathsf L$ contain a constant, $G_i$, $P_t$, the controls $X_{it}$ and their unit and time means, let $\Pi_{\mathsf L}$ project onto its column space in this inner product, and let $r=(I-\Pi_{\mathsf L})d$. Under the rank condition of Lemma~\ref{lem:mundlak}, the restricted residual-ratio coefficient $\widehat\theta_{\mathsf L}=\langle r,(I-\Pi_{\mathsf L})y\rangle_n/\langle r,r\rangle_n$ equals the two-way fixed-effects coefficient of Lemma~\ref{lem:mundlak}. If $y=\theta_0d+g+u$ with $g$ the untreated conditional-mean vector and $\E[u\mid\mathcal X]=0$ given the design $\mathcal X$, then
\begin{equation}
\E[\widehat\theta_{\mathsf L}\mid\mathcal X]-\theta_0
=\frac{\langle r,g\rangle_n}{\langle r,r\rangle_n},
\label{eq:twfe-conditional}
\end{equation}
which is zero if $g$ lies in the column space of $\mathsf L$ and, more generally, if and only if $r\perp g$. Without controls, with $\bar g_{gp}$ the mean of $g$ in group $g$ and period block $p$, $\widehat\pi=N^{-1}\sum_iG_i$ and $q=T^{-1}\sum_tP_t$,
\begin{equation}
\langle r,r\rangle_n=\widehat\pi(1-\widehat\pi)q(1-q),\qquad
\E[\widehat\theta_{\mathsf L}\mid\mathcal X]-\theta_0
=(\bar g_{11}-\bar g_{10})-(\bar g_{01}-\bar g_{00})=\kappa_{\mathcal X}.
\label{eq:kappa}
\end{equation}
\end{corollary}

\begin{corollary}[Heterogeneous effects under restricted adjustment]\label{cor:heterogeneous}
Keep the projections of Proposition~\ref{prop:restricted} and suppose instead that $Y=\tau(C)D+g_0(C)+U$ with $\E[U\mid D,C]=0$, $D,g_0,U\in L^2$ and bounded $\tau(C)$. Write $s(C)=\Var(D\mid C)$ and $\eta_{\mathcal M}=m_0-m_{\mathcal M}$. If $J_{\mathcal M}>0$,
\begin{equation}
\begin{aligned}
\theta_{\mathcal M}
&=\E[w_{\mathcal M}(C)\tau(C)]
+\frac{\E[\eta_{\mathcal M}(g_0-\Pi_{\mathcal M}g_0)]}{J_{\mathcal M}},\\
w_{\mathcal M}(C)
&=\frac{s(C)+m_0(C)\eta_{\mathcal M}(C)}{J_{\mathcal M}},
\qquad \E[w_{\mathcal M}(C)]=1.
\end{aligned}\label{eq:heterogeneous-restricted}
\end{equation}
The weights need not be positive. If also $J(C)=\E[s(C)]>0$, define $\theta_V=\E[s(C)\tau(C)]/J(C)$; then
\begin{equation}
\theta_{\mathcal M}-\theta_V
=\frac{\E\!\left[\eta_{\mathcal M}
\{g_0-\Pi_{\mathcal M}g_0+m_0(\tau-\theta_V)\}\right]}{J_{\mathcal M}}.
\label{eq:heterogeneous-gap}
\end{equation}
\end{corollary}

\subsection{Algebra of the bank-mean prediction correction}
\begin{proposition}[Within-bank centring and the linear feature comparator]\label{prop:structured}
For a complete held-out unit, write $D_i,Y_i,\widetilde m_i,\widetilde\ell_i\in\mathbb R^T$ for its treatment, outcome and preliminary nuisance predictions, and $Q_T=I_T-\mathbf1_T\mathbf1_T'/T$. Define the adjusted prediction and residual by $\widehat m_i^*=\widetilde m_i+\mathbf1_T(\bar D_i-\bar{\widetilde m}_i)$ and $v_i^*=D_i-\widehat m_i^*$. Then, for any fitted functions,
\begin{align}
 v_i^*&=Q_T(D_i-\widetilde m_i),\label{eq:structured-projection}\\
 \|D_i-\widetilde m_i\|^2
 &=\|v_i^*\|^2+T(\bar D_i-\bar{\widetilde m}_i)^2,\label{eq:structured-loss}\\
 (v_i^*)'(Y_i-\widetilde\ell_i)
 &=(v_i^*)'Q_T(Y_i-\widetilde\ell_i).\label{eq:structured-score}
\end{align}
If every fold contains complete units and both nuisance regressions are OLS with an intercept, the same time-varying covariates $X_{it}$ and all their unit means $\bar X_i$, with the within-covariate design and the between design $(1,\bar X_i,\bar D_i)$ of full column rank in every training fold, then the cluster-summed numerator and denominator equal those obtained by adding $\bar D_i$ as a linear feature to both nuisance regressions, and the two residual-ratio estimates coincide whenever their denominator is positive.
\end{proposition}

\begin{example}[Different targets with oracle nuisances]\label{ex:overlap-att}
Let $Z$ take two equally likely values with $p(0)=1/5$ and $p(1)=4/5$, set every untreated outcome to zero, and let the post-period effects be $\tau(0)=1$ and $\tau(1)=3$, so that Assumptions~\ref{ass:noanticipation} to \ref{ass:support} hold exactly. Then
\[
\tau_{\mathrm{ov}}=\frac{(4/25)\cdot1+(4/25)\cdot3}{(4/25)+(4/25)}=2,
\qquad
\bar\tau=\frac{(1/5)\cdot1+(4/5)\cdot3}{(1/5)+(4/5)}=2.6.
\]
The gap of $-0.6$ remains with the nuisances known and for every $q\in(0,1)$; it is a difference in weighting, not an estimation failure.
\end{example}

\subsection{First differences and dynamic effects}\label{sec:method:fd}
First differencing (FD) removes the unit effect. With common treatment timing, the differenced treatment is nonzero only at adoption, which suggests an impact interpretation. Adjusting for controls changes which event periods the coefficient weights. Let $t^*$ be the common treatment date, so that $t_i^*=t^*$ for every treated unit, let $s=0,\ldots,S$ index post-adoption periods, and suppose the stacked outcome satisfies
\begin{equation}
y=\mathsf F_0\gamma+\sum_{s=0}^{S}\beta_s e_s+u,\qquad
(e_s)_{it}=G_i\ind\{t=t^*+s\},\quad d=\sum_{s=0}^{S}e_s,\label{eq:dynamic}
\end{equation}
where $\mathsf F_0$ holds the level nuisance regressors. A causal interpretation of this event-effect model takes $\beta_s$ to be the effect of path $\mathbf d^1$ relative to $\mathbf d^0$ at duration $s$, with $\beta_0$ the adoption-period impact. Let $B$ difference successive observations within a unit, let $\mathsf F$ hold the differenced nuisance regressors including period effects, let $M_{\mathsf F}=I-\Pi_{\mathsf F}$, and define $v=M_{\mathsf F}Bd$.

\begin{proposition}[Residualised first-difference weights]\label{prop:fd}
If $v'v>0$, $B\mathsf F_0\gamma$ lies in the column space of $\mathsf F$, and $\E[v'Bu\mid\mathsf F_0,\mathsf F,d]=0$, then the OLS coefficient on $Bd$ satisfies
\begin{equation}
\E[\widehat\beta_{\mathrm{FD}}\mid\mathsf F_0,\mathsf F,d]=\sum_{s=0}^{S}\omega_s\beta_s,
\qquad \omega_s=\frac{v'Be_s}{v'v},\qquad \sum_s\omega_s=1.\label{eq:fdweights}
\end{equation}
The weights can be signed and need not vanish after adoption. If the only nuisance regressors in the differenced regression are saturated period effects, on a balanced panel with common treatment timing and both groups present, $\omega_0=1$ and every later weight is zero.
\end{proposition}

The conditional error moment and nuisance-span condition are assumptions of this event-effect regression; they do not follow from the average parallel-trends restriction in Assumption~\ref{ass:pt}. In particular, the DiD assumptions alone do not justify adding arbitrary contemporaneous controls to this regression.

\paragraph{Omitting quarter effects.}
To describe the original FD implementation, let $\mathsf X_\Delta$ contain its included differenced controls, without an intercept or period indicators, and put $v_X=(I-\Pi_{\mathsf X_\Delta})Bd$. Let $L_\Delta$ be the stacked vector with entries $\lambda_t-\lambda_{t-1}$. Suppose the differenced untreated component has the form $\mathsf X_\Delta\gamma+L_\Delta$, $v_X'v_X>0$, and $\E[v_X'Bu\mid\mathcal X]=0$, where $\mathcal X$ contains the treatment and regressor design. The Frisch--Waugh--Lovell formula gives
\begin{equation}
\E[\widehat\beta_{\mathrm{FD},X}\mid\mathcal X]
=\sum_{s=0}^S\omega_{s,X}\beta_s+
\frac{v_X'L_\Delta}{v_X'v_X},
\qquad
\omega_{s,X}=\frac{v_X'Be_s}{v_X'v_X},\quad
\sum_s\omega_{s,X}=1.
\label{eq:fd-omitted-time}
\end{equation}
Indeed, $v_X'\mathsf X_\Delta=0$ and $v_X'Bd=v_X'v_X$, so substitution of the differenced model proves the identity. The second term is omitted-time-effect bias relative to the displayed weighted effect. Under constant effects its first term is $\theta_0$. If $\mathsf X_\Delta$ is empty, then $v_X=Bd$ is one for treated banks at $t^*$ and zero elsewhere: $\omega_{0,X}=1$, later weights vanish, and the time term equals $\lambda_{t^*}-\lambda_{t^*-1}$. In that special regression the sample coefficient is the treated banks' mean $\Delta Y_{it^*}$, without a comparison-bank subtraction. With controls, residualisation can involve comparison banks and other quarters, so neither simplification generally describes the original controlled specification. If an intercept is included, it must also enter the projection design.

\begin{remark}[Distance from the impact effect]\label{rem:fdbound}
Because the weights sum to one,
\begin{equation}
\bigl|\E[\widehat\beta_{\mathrm{FD}}\mid\mathsf F_0,\mathsf F,d]-\beta_0\bigr|
\leq\Bigl(\sum_{s>0}|\omega_s|\Bigr)\max_{s>0}|\beta_s-\beta_0|.\label{eq:fdbound}
\end{equation}
The first factor is computable from the design alone and is the quantity to report when a first-difference coefficient is offered as an impact effect. It is zero in the saturated case, one in Example~\ref{ex:fd}, and approximately $0.00765$ in the application.
\end{remark}

Negative weights and the resulting non-convex treatment-effect averages are studied by \citet{dechaisemartin2020} and \citet{borusyak2024}; the proposition is the corresponding statement for a residualised first-difference regression, and the distance bound uses a multiplier computable from the design.

\begin{example}[Control adjustment changes the impact interpretation]\label{ex:fd}
Take one treated and one control unit, three periods, adoption in period 2, effects $(\beta_0,\beta_1)=(1,2)$, and all untreated outcomes and errors zero. Let an exogenous control have levels $(0,1,2)$ for the treated unit and $(0,0,0)$ for the control unit, with true coefficient zero. Ordering differenced observations as control periods 2, 3 and treated periods 2, 3, $Bd=(0,0,1,0)'$, the differenced control is $(0,0,1,1)'$, and after removing it and the period effects $v=(-1,1,1,-1)'/4$. The weights are $(2,-1)$ and the coefficient is zero despite a unit impact effect.
\end{example}

Constant effects make every weighted target equal because the weights sum to one; the converse fails, so agreement between first-difference and two-way fixed-effects coefficients does not indicate a flat effect profile. In the application, $\beta_0$ is an adoption-period effect on quarterly log growth. Differencing that growth outcome does not itself rule out an impact interpretation: the limitations are the event-effect model and the later-period weights in~\eqref{eq:fdweights}.

\subsection{Prediction error and fitted treatment residuals}\label{sec:method:fitted}
Identification concerns the population model; the empirical score uses fitted nuisances. Error in a fitted treatment nuisance leaves treatment residuals even when the features determine treatment exactly, so a positive fitted denominator is not evidence of identifying variation. For held-out predictions we report
\begin{equation}
\widehat J=\frac1{NT}\sum_{it}(D_{it}-\widehat m_{-k(i)}(C_{it}))^2,
\qquad \widehat R=\widehat J/\widehat{\Var}(D),\label{eq:diag}
\end{equation}
the mean squared held-out treatment residual and its ratio to the raw treatment variance. The following identities separate population treatment variation from prediction error.

\begin{proposition}[Prediction error and fitted residual variation]\label{prop:risk}
Condition on a training sample independent of a validation unit and regard $\widehat m,\widehat\ell$ as fixed functions of $C$. Under square integrability and the conditional mean restriction of~\eqref{eq:plr}, with $\delta_m=\widehat m-m_0$ and $\delta_\ell=\widehat\ell-\ell_0$,
\begin{align}
\E[(D-\widehat m(C))^2\mid\text{training}]
&=J+\E[\delta_m^2\mid\text{training}],\label{eq:risk}\\
\E[\psi(W;\theta_0,\widehat\ell,\widehat m)\mid\text{training}]
&=\E[\delta_m\delta_\ell-\theta_0\delta_m^2\mid\text{training}].\label{eq:remainder}
\end{align}
If the left side of~\eqref{eq:risk} is positive, the root $\theta(\widehat\ell,\widehat m)$ of the population moment conditional on the training sample satisfies
\begin{equation}
\theta(\widehat\ell,\widehat m)-\theta_0=
\frac{\E[\delta_m\delta_\ell-\theta_0\delta_m^2\mid\text{training}]}
{J+\E[\delta_m^2\mid\text{training}]}
=\frac{\E[\delta_m e_g\mid\text{training}]}{J+\E[\delta_m^2\mid\text{training}]},
\qquad e_g=\delta_\ell-\theta_0\delta_m.\label{eq:rootbias}
\end{equation}
\end{proposition}

Writing $V=D-m_0$ and $U=Y-\ell_0-\theta_0V$, the score at $(\theta_0,\ell_0+\delta_\ell,m_0+\delta_m)$ expands as $(V-\delta_m)(U-\delta_\ell+\theta_0\delta_m)$. The conditional mean restrictions remove the linear terms in expectation, giving Neyman orthogonality \citep{chernozhukov2018,chernozhukov2022locally}; equation~\eqref{eq:remainder} is the remaining quadratic term. The nuisance-rate conditions in Theorem~\ref{thm:collapsed-inference} control this remainder.

When $J=0$, the denominator in~\eqref{eq:risk} consists entirely of prediction error. The root discrepancy has no general sign because its error product can offset its squared treatment-error term. When that product is small relative to $\theta_0\E[\delta_m^2]$, the root is attenuated towards zero; the sign does not identify a smoothing mechanism. Realised validation losses also contain sampling variation: $\widehat R$ can exceed one and need not fall when features are added. Supplementary Section~\ref{app:oracle-errors} gives an oracle substitution that worsens the root by removing error cancellation (Corollary~\ref{cor:oracle-swap}) and vanishing nuisance errors with any prescribed residual ratio when $J=0$ (Example~\ref{ex:arbitrary}). Supplementary Section~\ref{app:classes} gives the exact additive and unit-mean oracle risks on the four adoption cells.
\section{Additional results on approximation and nuisance error}\label{app:extended}
The main text separates identification and target weighting from errors introduced by fitted nuisance functions. This section retains the detailed learner-class calculations and exact examples supporting that distinction. Oracle risk is the infimum of population prediction loss over a specified class. When the infimum is attained, a minimiser is an oracle predictor; for unrestricted squared-error prediction, it is the true conditional mean.

\subsection{Oracle approximation risks by learner class}\label{app:classes}
\begin{proposition}[Oracle approximation risks for assignment classes]\label{prop:classes}
In the setting with common treatment timing of Corollary~\ref{cor:absorbing}, sample a
unit and a period independently and uniformly from the balanced panel.
Under this fixed-panel sampling law, $\pi=\widehat\pi$ is the realised treated share,
$G\sim\mathrm{Bernoulli}(\pi)$ and
$P\sim\mathrm{Bernoulli}(q)$ are independent, $D=GP$, and
$\Var(D)=\pi q(1-\pi q)$. For a prediction class $\mathcal F$ on the
treatment means, define its oracle squared prediction risk
\begin{equation}
L_{\mathcal F}=\inf_{f\in\mathcal F}
\E[(D-f(\bar D_i,\bar D_t))^2],\qquad
R_{\mathcal F}=L_{\mathcal F}/\Var(D).
\end{equation}
The unrestricted conditional variance $J(C)$ and the approximation risk
$L_{\mathcal F}$ are different objects.
\begin{enumerate}
\item[(i)] If $\mathcal F$ contains the reconstruction map
$f(a,b)=ab/(q\pi)$, or an equivalent representation of $GP$, then
$L_{\mathcal F}=0$.
\item[(ii)] For the additive class
$\mathcal F_{\mathrm{add}}=\{\alpha+g_1(\bar D_i)+g_2(\bar D_t)\}$,
the minimising predictor is $qG+\pi P-\pi q$, and
\begin{equation}
L_{\mathrm{add}}=\pi(1-\pi)q(1-q),\qquad
R_{\mathrm{add}}=\frac{(1-\pi)(1-q)}{1-\pi q}.
\label{eq:Radd}
\end{equation}
\item[(iii)] For all square-integrable functions of $\bar D_i$ alone,
the minimising predictor is $qG$, and
\begin{equation}
L_{\mathrm{unit}}=\pi q(1-q),\qquad
R_{\mathrm{unit}}=\frac{1-q}{1-\pi q}.
\end{equation}
\item[(iv)] For the logistic link $\Lambda(x)=(1+e^{-x})^{-1}$ and class
$\{\Lambda(\alpha+\beta\bar D_i+\gamma\bar D_t)\}$, the infimum
of squared prediction loss is zero and is not attained at finite
coefficients. If coefficients are instead fitted by expected Bernoulli
negative log-likelihood plus $\lambda(\beta^2+\gamma^2)$ with
$\lambda>0$, a finite minimiser exists. Its squared prediction loss is
strictly positive and converges to zero as $\lambda\downarrow0$.
\end{enumerate}
Enlarging a prediction class can only lower its oracle risk. The statement
does not imply a corresponding inequality for fitted validation losses.
\end{proposition}

A depth-two tree can represent the four-cell assignment rule. Forest or
boosting classes that contain this tree and polynomial dictionaries that
contain its scaled interaction also have zero oracle risk. Fixed
implementation constraints, penalisation, finite training samples and the
fitting algorithm still determine whether a fitted learner attains that
risk. Additive least squares omits the interaction, whereas a linear-index
logit can separate the cells because its output link is nonlinear. Positive
loss from a penalised logit in this deterministic design does not restore
unrestricted identification.

At $\pi=20/91$ and $q=8/20$, the reference ratios are
$R_{\mathrm{unit}}=0.658$ and $R_{\mathrm{add}}=0.513$.
They describe idealised fits using only the stated treatment means.
Additional features can lower the best achievable risk if the enlarged
class nests the original one. A held-out empirical ratio can be above or
below these references, owing to estimation error and validation-sample
variation. In particular, numerical proximity to a reference is not a
test that the other covariates carry little assignment information.
When $\mathcal F$ is a closed linear space, Proposition~\ref{prop:restricted}
states what a positive oracle risk can identify under a maintained
restriction on the outcome nuisance. Algorithmic inability to approximate
treatment is not a substitute for that restriction.

Conditioning on $\bar D_i$ alone gives $\E[D\mid\bar D_i]=qG_i$ under uniform sampling over the balanced periods, as in Proposition~\ref{prop:classes}(iii). This equality need not survive the addition of other informative features. Nor does identifying each observed bank by a distinct covariate vector prove that assignment is determined by that vector in a superpopulation. Finite-sample memorisation and population support are separate questions.

The oracle-risk qualification also matters without any covariates. In Theorem~\ref{thm:design}, at $\pi=20/91$ and $q=0.4$, an assignment prediction fixed at $a=0.01$ gives a normalised adjusted treatment-residual loss of $0.645$, exceeding the oracle bound $R_{\mathrm{add}}=0.513$. The prediction lies inside $[0,1]$; the discrepancy is ordinary assignment-prediction error.

\subsection{Oracle substitution and vanishing-error examples}\label{app:oracle-errors}
The fitted-root identity in Proposition~\ref{prop:risk} contains both a product of outcome and treatment errors and a squared treatment error. The two can offset. Holding one nuisance fixed while replacing the other by its truth makes that mechanism explicit.

\begin{corollary}[Oracle substitution and error cancellation]\label{cor:oracle-swap}
Under Proposition~\ref{prop:risk}, let $J>0$ and hold a fitted treatment nuisance $\widehat m$ fixed. Replacing the outcome nuisance by $\ell_0$ gives the validation-population root
\begin{equation}
\theta(\ell_0,\widehat m)
=\theta_0\frac{J}{J+\E[\delta_m^2\mid\mathrm{training}]}.
\end{equation}
Replacing the treatment nuisance by $m_0$, with a fixed square-integrable outcome nuisance, instead gives population root $\theta_0$. Under the heterogeneous model of Corollary~\ref{cor:heterogeneous} with $J(C)>0$, both statements hold with $\theta_V$ in place of $\theta_0$ (Section~\ref{app:proofs}). These statements concern roots of expected validation moments, not unbiasedness of finite-sample ratios.
\end{corollary}
The result follows by setting the corresponding error to zero in equation~\eqref{eq:rootbias}. For positive $\theta_0$, an oracle outcome regression can increase downward discrepancy if it removes an offsetting outcome-error product. Lower outcome prediction error therefore need not monotonically improve the coefficient when treatment prediction remains imperfect. The simulations use oracle substitutions only for mechanism diagnosis; nuisance choices are selected by training-sample prediction criteria.

\begin{example}[Vanishing nuisance errors and an arbitrary residual ratio]\label{ex:arbitrary}
Let $D=h(C)\in\{0,1\}$, $p=\E[D]\in(0,1)$, $r(C)=h(C)-p$, and $Y=\beta h(C)$ without noise. Then $m_0=h$ and $\ell_0=\beta h$. For any fixed real $b$ and $0<\varepsilon<1$, set
\begin{equation}
\widehat m_\varepsilon=h-\varepsilon r,
\qquad \widehat\ell_\varepsilon=\beta h-b\varepsilon r.
\end{equation}
The treatment prediction lies strictly between zero and one. Both nuisance errors converge to zero in $L^2$ as $\varepsilon\downarrow0$, but
\begin{equation}
\frac{\E[(D-\widehat m_\varepsilon)(Y-\widehat\ell_\varepsilon)]}
{\E[(D-\widehat m_\varepsilon)^2]}=b,
\qquad
\frac{\E[(D-\widehat m_\varepsilon)^2]}{\Var(D)}=\varepsilon^2.\label{eq:arbitrary}
\end{equation}
Indeed, the two residuals are $\varepsilon r$ and $b\varepsilon r$.
\end{example}

The coefficient $\beta$ is a designated coefficient in this data-generating representation, not an identified parameter of the unrestricted model: any $\theta$ can be offset by $g(C)=(\beta-\theta)h(C)$. Thus, Example~\ref{ex:arbitrary} is not a failure of DML under its standard identification conditions. It demonstrates why even increasingly accurate nuisance fits need not select a unique coefficient when those conditions fail. Choosing $\varepsilon_N=o(N^{-1/4})$ also satisfies the usual numerical nuisance-rate bound while leaving the indispensable requirement $J>0$ violated. No direction of the resulting discrepancy is implied.

\subsection{Interactive regression models and counterfactual support}\label{app:interactive}
Interactive outcome regressions after adoption require counterfactual support and a conditional independence assumption. If the features determine treatment, unobserved counterfactual outcomes can vary without changing the observed distribution; the unrestricted model does not identify the ATT. DiD instead imposes a restriction on untreated changes. Exact support failure also differs from propensities approaching the boundaries while remaining in $(0,1)$, where the efficiency bound depends on outcome-variance-weighted inverse-propensity moments \citep{khan2010irregular}.
\section{Sufficient conditions for pre/post-score inference}\label{app:inference-statement}
The following results specialise the partially linear DML argument to one observation per bank. Proposition~\ref{prop:dr-inference} treats the normalised doubly robust estimator separately; Section~\ref{app:collapsed-inference} gives the proofs. Nuisance convergence rates are maintained assumptions and are not established for the empirical forests.

\begin{theorem}[Independent-bank inference for the overlap score]\label{thm:collapsed-inference}
Under Assumptions~\ref{ass:consistency} to \ref{ass:sampling}, let $W_i=(H_i,G_i,Z_i)$ be i.i.d., with $\E[H^2]<\infty$. Here $H_i$ and $Z_i$ are fixed measurable functions of bank $i$'s own record. Suppose $\pi=\Pr(G=1)\in(0,1)$ and $J_G=\E[p(Z)\{1-p(Z)\}]>0$ are fixed as $N\to\infty$, with $T$ fixed. Divide complete banks into a fixed number $K\geq2$ of folds using either of two schemes: (a) allocation independent of the records, with fold proportions tending to positive constants; or (b) allocation depending only on $(G_1,\ldots,G_N)$ and external randomisation independent of the records, with each fold containing $n_{gk}=N_g/K+O(1)$ banks from group $g\in\{0,1\}$, where $N_g=\sum_i\ind\{G_i=g\}$ and the count discrepancy is uniformly bounded over folds. Scheme (b) includes balanced random allocation within each treatment group and excludes stratification or fold selection using outcomes or other covariates. All nuisance fitting, tuning and fitted preprocessing use only the complementary training folds; any fitted preprocessing is part of the fold-specific prediction rule.

Use $\widehat p\in[0,1]$ and $\widehat\ell_H$ in equation~\eqref{eq:design-collapse}, pooling held-out observations before taking the ratio. For a nuisance $f$ with truth $f_0$, define
\[
r_f=\max_{k\leq K}\left\{\int
   \bigl(\widehat f_{-k}(z)-f_0(z)\bigr)^2\,d\mathbb P_Z(z)
   \right\}^{1/2}.
\]
The integral holds each trained prediction rule fixed and uses the covariate law of a fresh independent bank. The norm is random through the training sample and learner randomisation; it is not an empirical validation-fold norm. Treatment-stratified allocation changes the training-sample law conditional on group counts. The rate assumptions concern that actual training sample; an i.i.d.-training rate theorem requires adaptation. The proof handles validation-fold averaging conditional on groups, but does not establish learner convergence. Write $r_p$ and $r_\ell$ for the propensity and all-bank outcome errors. Assume $r_p+r_\ell=o_p(1)$ and
\begin{equation}
r_p(r_\ell+r_p)=o_p(N^{-1/2}).
\label{eq:collapsed-rates}
\end{equation}
Then the fitted denominator is positive with probability tending to one. Put $V=G-p(Z)$. The influence function is
\begin{equation}
\varphi_{\mathrm{ov}}(W)
=\frac{V\{H-\ell_H(Z)-\tau_{\mathrm{ov}}V\}}{J_G},
\label{eq:collapsed-if-ov}
\end{equation}
and, if $\sigma_{\mathrm{ov}}^2=\E[\varphi_{\mathrm{ov}}^2]>0$,
\begin{equation}
\sqrt N(\widehat\tau_{\mathrm{ov}}-\tau_{\mathrm{ov}})
=N^{-1/2}\sum_i\varphi_{\mathrm{ov}}(W_i)+o_p(1)
\ \Longrightarrow\ N(0,\sigma_{\mathrm{ov}}^2).
\label{eq:collapsed-alr}
\end{equation}
For $v_i=G_i-\widehat p_{-k(i)}(Z_i)$ and $\widehat J_G=N^{-1}\sum_i v_i^2$, the fitted influence value is
\[
\widehat\varphi_{\mathrm{ov},i}
=\frac{v_i\{H_i-\widehat\ell_{H,-k(i)}(Z_i)-\widehat\tau_{\mathrm{ov}}v_i\}}{\widehat J_G}.
\]
Its sample variance consistently estimates $\sigma_{\mathrm{ov}}^2$; dividing by $N$ estimates the leading sampling variance of $\widehat\tau_{\mathrm{ov}}$.

With a fixed number $R$ of repeated partitions, assume the same conditions with each nuisance error maximised over both folds and repetitions. No independence between repetitions is required. Their mean estimate has the same expansion~\eqref{eq:collapsed-alr}. Averaging each bank's fitted influence values across repetitions before computing their sample variance consistently estimates $\sigma_{\mathrm{ov}}^2$. The median estimate $\widetilde\tau_{\mathrm{ov}}$ also has expansion~\eqref{eq:collapsed-alr}. If $\widehat s_r^2$ is the influence-based sampling variance estimate for repetition $r$, the median convention
\begin{equation}
\widehat s_{\mathrm{med}}^2
=\operatorname{med}_{r=1,\ldots,R}
 \{\widehat s_r^2+(\widehat\tau_{\mathrm{ov},r}-\widetilde\tau_{\mathrm{ov}})^2\}
\label{eq:median-variance}
\end{equation}
satisfies $N\widehat s_{\mathrm{med}}^2\to_p\sigma_{\mathrm{ov}}^2$. For even $R$, each median is the arithmetic mean of the two central order statistics. The median is taken after adding the squared quantities. Neither aggregation rule supplies a finite-sample coverage guarantee.
\end{theorem}

No uniform propensity bound is needed: fitted propensities in $[0,1]$ bound treatment residuals, and the displayed moments and rates control the remainder. Fixed positive $J_G$ excludes vanishing residual variation. Causal assumptions enter only through the target's interpretation. Without them, the same expansion holds after replacing $\tau_{\mathrm{ov}}$ by the statistical projection target
\[
\frac{\E[(G-p(Z))\{H-\ell_H(Z)\}]}{J_G}.
\]
The proof uses the conditional-mean identities and zero expected score at this root. The fixed-$R$ median statement follows because all repetitions share the same leading empirical term; their remainders alone differ. Increasing $R$ with $N$ requires further analysis.

\paragraph{Paired specification comparisons.}
Let $\widehat\theta^{(j)}$, $j=0,1$, be two estimators computed on the same banks, with statistical targets $\theta^{(j)}$ and influence functions $\varphi^{(j)}$. Suppose each satisfies its stated expansion, with all moment and nuisance-rate conditions holding jointly and uniformly over the fixed number of repetitions. Subtraction gives
\[
\sqrt N\bigl[\{\widehat\theta^{(1)}-\widehat\theta^{(0)}\}
-\{\theta^{(1)}-\theta^{(0)}\}\bigr]
=N^{-1/2}\sum_i\{\varphi^{(1)}(W_i)-\varphi^{(0)}(W_i)\}+o_p(1).
\]
When the variance of the influence difference is positive, its fitted sample variance divided by $N$ consistently estimates the sampling variance of the contrast. This calculation retains the covariance between the two estimators. The same fixed-$R$ argument applies to the median of paired changes, using equation~\eqref{eq:median-variance} with each repetition's paired estimate and variance. That median need not equal the difference of the two separately reported medians. Conditioning on different covariates can change the overlap target; causal interpretation of either target requires its own identifying assumptions. These joint conditions provide an asymptotic justification, not a finite-sample coverage guarantee for the banking comparison.

\begin{corollary}[Inference for the first-difference representation]\label{cor:fd-inference}
Let $R_i=\Delta Y_{it^*}$ and let $B_i$ be a fixed measurable function of bank $i$'s own record. Suppose $(R_i,G_i,B_i)$ are i.i.d., $\E[R^2]<\infty$, $\pi\in(0,1)$, and
\[
p_R(B)=\Pr(G=1\mid B),\qquad
\ell_R(B)=\E[R\mid B],\qquad
J_R=\E[p_R(B)\{1-p_R(B)\}]>0
\]
with $J_R$ fixed as $N\to\infty$. Define the statistical target
\[
\theta_R=\frac{\E[(G-p_R(B))\{R-\ell_R(B)\}]}{J_R}.
\]
Use the admissible folds and training-only fitting of Theorem~\ref{thm:collapsed-inference}, fitted $a_i=\widehat p_{R,-k(i)}(B_i)\in[0,1]$ and $b_i=\widehat\ell_{R,-k(i)}(B_i)$, and impose the treatment fit of Corollary~\ref{cor:fd-collapse}. Suppose the fresh-bank $L^2$ errors satisfy $r_{p_R}+r_{\ell_R}=o_p(1)$ and $r_{p_R}(r_{\ell_R}+r_{p_R})=o_p(N^{-1/2})$. If the variance of
\begin{equation}
\varphi_R=
\frac{(G-p_R(B))\{R-\ell_R(B)-\theta_R(G-p_R(B))\}}{J_R}
\label{eq:fd-if}
\end{equation}
is positive, then $\sqrt N(\widehat\theta_\Delta-\theta_R)$ has the asymptotically linear and normal representation of Theorem~\ref{thm:collapsed-inference}, with $\varphi_R$ as its influence function. Its fitted influence variance and the theorem's fixed-number-of-partitions aggregation rules are consistent under the corresponding uniform rate conditions.
\end{corollary}

\begin{proof}
Corollary~\ref{cor:fd-collapse} makes the panel ratio exactly the bank-level ratio with outcome $R$. Conditional means give $\E[G-p_R\mid B]=0$, $\E[R-\ell_R\mid B]=0$ and the zero expected score at $\theta_R$. The proof of Theorem~\ref{thm:collapsed-inference} uses precisely these identities, bounded treatment residuals, the displayed second moments and the nuisance rates. Replacing $(H,Z,p,\ell_H,J_G)$ by $(R,B,p_R,\ell_R,J_R)$ in that proof establishes the conclusion; no causal assumption is needed for this statistical target.
\end{proof}

With $B=A_i=(X_{it^*},X_{i,t^*-1})$ and the invariant-covariate untreated model in Corollary~\ref{cor:dynamic-panel}, $\theta_R=\theta_\Delta$ in equation~\eqref{eq:dynamic-panel-targets}. Alternatively, consistency, no anticipation and conditional parallel trends for the untreated change between $t^*-1$ and $t^*$ given pre-treatment $B$ yield the corresponding overlap-weighted impact effect. Parallel trends for the average change $H$ alone does not supply that single-quarter restriction. In the banking comparison, contemporaneous elements of $A_i$ may respond to the reform, so the fitted FD comparator is interpreted as a specification check.

The sample identities transfer these results to specified panel scores. Honest-forest rates depend on regularity, dimension and fitting conditions not verified for the empirical forests \citep{wager2018}. The following standard finite-dimensional projection argument covers the bank-demeaned OLS comparisons.

\begin{proposition}[Fixed-basis projection inference]\label{prop:finite-basis-inference}
Let complete bank records be i.i.d., with fixed $T$, and let $Q_{it}\in\mathbb R^d$ be a fixed basis of that bank's covariates and calendar indicators, with fixed $d$. Write $\langle a\rangle_i=T^{-1}\sum_t a_{it}$ and let $Y^w,D^w$ denote bank-demeaned outcome and treatment. Suppose
$\E\sum_t(|Y^w_{it}|+|D^w_{it}|+\|Q_{it}\|)^4<\infty$ and $\Gamma=\E\langle QQ'\rangle$ is positive definite. Define
\[
\gamma_\ell=\Gamma^{-1}\E\langle QY^w\rangle,\quad
\gamma_m=\Gamma^{-1}\E\langle QD^w\rangle,\quad
v_{it}=D^w_{it}-Q_{it}'\gamma_m,
\]
\[
J_B=\E\langle v^2\rangle>0,\qquad
\theta_B=\frac{\E\langle v(Y^w-Q'\gamma_\ell)\rangle}{J_B}.
\]
Fit both nuisance projections by OLS on each training sample, using the folds of Theorem~\ref{thm:collapsed-inference}, and pool held-out scores. If $\sigma_B^2=\E[\varphi_{B,i}^2]>0$, then
\begin{equation}
\sqrt N(\widehat\theta_B-\theta_B)
=N^{-1/2}\sum_i\varphi_{B,i}+o_p(1)
\ \Longrightarrow\ N(0,\sigma_B^2),\qquad
\varphi_{B,i}=\frac{\langle v(Y^w-Q'\gamma_\ell-\theta_B v)\rangle_i}{J_B}.
\label{eq:finite-basis-if}
\end{equation}
The sample variance of the fitted bank influence values divided by $N$ consistently estimates sampling variance. The fixed-repetition median and paired-comparison rules above also apply. The target $\theta_B$ is a statistical projection coefficient; a causal interpretation additionally requires the corresponding outcome and identifying restrictions.
\end{proposition}

Here the basis and transformed outcomes must be functions of each bank's own record. Full-sample two-way demeaning and sample-estimated winsorisation cutoffs require additional arguments. The empirical fixed-basis standard errors use the independent-record working model for the supplied analysis-ready data and omit upstream preprocessing uncertainty.
\section{The doubly robust ATT comparison}\label{app:dr}
The established doubly robust DiD score~\eqref{eq:drdid} and normalised estimator~\eqref{eq:normalized} target $\bar\tau$ under the causal assumptions \citep{santanna2020,chang2020}. Their outcome nuisance is $\mu_0(Z)=\E[H\mid G=0,Z]$. The overlap score instead uses $\ell_H(Z)=\E[H\mid Z]$ and targets $\tau_{\mathrm{ov}}$; matching covariates and folds does not equate the targets.

\subsection{Population root and normalisation}
Let $a(Z)\in[0,1)$ and $b(Z)$ be working propensity and control-outcome regressions, and put $w_a=(1-G)a(Z)/\{1-a(Z)\}$. Provided the relevant expectations exist and $0<\E[w_a]<\infty$, the population counterpart of equation~\eqref{eq:normalized} is
\begin{equation}
\mathcal T(b,a)=\frac{\E[G\{H-b(Z)\}]}{\pi}
-\frac{\E[w_a\{H-b(Z)\}]}{\E[w_a]}.
\label{eq:dr-normalized-functional}
\end{equation}
If $b=\mu_0$, the control term is zero by iterated expectations and the treated term is $\bar\tau$ under the causal assumptions. If $a=p$, then $\E[w_a\mid Z]=p(Z)$ and $\E[w_a]=\pi$, so the contribution of $b-\mu_0$ cancels between the two terms. Thus normalisation preserves the population double-robustness property. The sample normalised estimator need not equal the root obtained by replacing the single common denominator of the population score with a treated-share estimate.

Normalising the treated and control contributions separately also changes the finite-sample influence calculation. Equation~\eqref{eq:normalizedif} centres each residual at its own group-weighted residual mean and uses its own estimated denominator. At the true nuisance functions these centres converge to $\bar\tau$ and zero, respectively, yielding the influence function below. Section~\ref{app:collapsed-inference} verifies that nuisance estimation and both random denominators leave only a negligible remainder under the stated rates.

\subsection{Sufficient conditions for inference}
\begin{proposition}[Independent-bank inference for normalised DR]\label{prop:dr-inference}
Maintain the causal assumptions, i.i.d.\ bank records, finite $\E[H^2]$, fixed $\pi\in(0,1)$, fixed $T$ and $K$, admissible folds and training-only fitting of Theorem~\ref{thm:collapsed-inference}. Suppose in addition that $p(Z)\leq1-\varepsilon$ almost surely for a fixed $\varepsilon>0$. Use $\widehat p\in[0,1-c]$, with $0<c<\varepsilon$, and $\widehat\mu_0$ in equation~\eqref{eq:normalized}, pooling held-out observations before taking either ratio. Define $r_p$ and $r_\mu$ by the fresh-bank $L^2(\mathbb P_Z)$ norms in that theorem, and assume $r_p+r_\mu=o_p(1)$ and
\begin{equation}
r_p r_\mu=o_p(N^{-1/2}).
\label{eq:dr-rates}
\end{equation}
The overlap estimator's outcome nuisance and rate condition are not required here. Both fitted denominators are positive with probability tending to one. With
\begin{equation}
\varphi_{\mathrm{DR}}(W)
=\frac{G}{\pi}\{H-\mu_0(Z)-\bar\tau\}
-\frac{1-G}{\pi}\frac{p(Z)}{1-p(Z)}\{H-\mu_0(Z)\},
\label{eq:collapsed-if-dr}
\end{equation}
and $\sigma_{\mathrm{DR}}^2=\E[\varphi_{\mathrm{DR}}^2]>0$,
\begin{equation}
\sqrt N(\widehat\tau_{\mathrm{DR}}-\bar\tau)
=N^{-1/2}\sum_i\varphi_{\mathrm{DR}}(W_i)+o_p(1)
\ \Longrightarrow\ N(0,\sigma_{\mathrm{DR}}^2).
\label{eq:dr-alr}
\end{equation}
The sample variance of the fitted influence values in equation~\eqref{eq:normalizedif} consistently estimates $\sigma_{\mathrm{DR}}^2$; dividing by $N$ estimates the leading sampling variance of $\widehat\tau_{\mathrm{DR}}$. The fixed-$R$ mean, median and variance conclusions of Theorem~\ref{thm:collapsed-inference} also hold for DR, with nuisance errors maximised over both folds and repetitions and with $\varphi_{\mathrm{DR}}$, $\bar\tau$ and $\widehat\tau_{\mathrm{DR}}$ replacing the overlap quantities.
\end{proposition}

The causal assumptions give the ATT interpretation. Without them, the same expansion applies to the statistical target $\E[H-\mu_0(Z)\mid G=1]$. The upper propensity bound controls the odds weights and is stronger than the point-support requirement for ATT identification; no lower bound is needed. Upper truncation at $1-c$ preserves the propensity rate because it fixes the true $p$. A fixed lower clip is covered only if the true propensity lies at or above it almost surely. Otherwise the clipped fit cannot converge to $p$, so the proposition does not apply. Equation~\eqref{eq:clipped-dr-root} gives the resulting population discrepancy when the outcome limit is misspecified. A correct control-outcome regression still gives the ATT root, but does not by itself restore orthogonality or this inference result at a clipped propensity limit. In particular, for a working propensity $a$, the outcome derivative in direction $h$ is $-\E[p h]/\pi+\E[\rho_a h]/\E[\rho_a]$, where $\rho_a=(1-p)a/(1-a)$. It vanishes in every direction exactly when these two normalised covariate weights agree almost surely. Binding clipping generally changes those weights, although special cases can preserve them. Observed threshold crossings alone do not establish whether clipping binds in population.

Concentrated normalised weights can make finite-sample behaviour unstable even when they are bounded in the population. The effective sample size $\mathrm{ESS}_0=(\sum_iw_i)^2/\sum_iw_i^2$ describes control-weight concentration; it is not a test of population support. Lemma~\ref{lem:weight-centring} isolates a source of downward distortion in a centred variance component under concentrated control weights. Neither that calculation nor the stated asymptotic result supplies a finite-sample correction or establishes coverage for the banking data.
\section{Panel transformations, restricted classes and dynamic targets}\label{app:panel-transformations}
This section proves the common-timing results in the main text. The transformation constructions build on \citet[Section~5.1.2]{clarke2026}; partialling out and orthogonal estimation follow \citet{robinson1988} and \citet{chernozhukov2018}. The additive CRE comparison uses the distinction between a conditional-effect restriction and the linear coefficient identity associated with \citet{mundlak1978}. The calculations below impose the stated common-timing treatment path and do not infer those assumptions from an empirical learner.

\subsection{Population averaging and transformed conditional means}
Take a complete bank record and an independent index $\mathsf t$ drawn uniformly from $\{1,\ldots,T\}$ for levels and within transformations, or from $\{2,\ldots,T\}$ for first differences. An expectation of a panel variable means expectation over both this record and $\mathsf t$. This convention does not assume independence of observations from the same bank. The calendar is fixed, $P_t=\ind\{t\geq t^*\}$, $2\leq t^*\leq T$, $q=T^{-1}\sum_tP_t\in(0,1)$ and $D_{it}=G_iP_t$. Let $X_i=(X_{i1},\ldots,X_{iT})$ and $A_i=(X_{it^*},X_{i,t^*-1})$. Denote the assignment probabilities by $p_W(X_i)=\E[G_i\mid X_i]$ and $p_\Delta(A_i)=\E[G_i\mid A_i]$.

For this subsection assume the square-integrable static model
\[
Y_{it}=\alpha_i+\lambda_t+\theta_0G_iP_t+g_0(X_{it})+U_{it},\qquad
\E[U_{it}\mid G_i,X_i,\alpha_i]=0\quad(t=1,\ldots,T).
\]
Because the common-timing treatment history contains a post-period entry equal to $G_i$, this error condition is equivalent to the history formulation in equation~\eqref{eq:panel-strict-exogeneity}. Iterated expectations give $\E[U_{it}\mid G_i,X_i]=0$ for every $t$.

\paragraph{Proof of Proposition~\ref{prop:panel-transformations}(i).}
Differencing eliminates $\alpha_i$ and gives
\[
\Delta Y_{it}=\theta_0\ind\{t=t^*\}G_i+
\lambda_t-\lambda_{t-1}+g_0(X_{it})-g_0(X_{i,t-1})+\Delta U_{it}.
\]
For $C^\Delta_{it}=(X_{it},X_{i,t-1},t)$, both the deterministic nuisance term and $\Delta D_{it}$ are measurable functions of the displayed arguments and $G_i$. Strict exogeneity at $t$ and $t-1$ implies
\[
\E[\Delta U_{i\mathsf t}\mid G_i,X_i,\mathsf t]=0,
\qquad
\E[\Delta U_{i\mathsf t}\mid\Delta D_{i\mathsf t},C^\Delta_{i\mathsf t}]=0.
\]
At $\mathsf t=t^*$ the covariate pair is $A_i$; elsewhere $\Delta D$ is zero. Hence
\[
\E[\Delta D\mid C^\Delta]=\ind\{\mathsf t=t^*\}p_\Delta(A_i),\qquad
\E[\Delta Y\mid C^\Delta]=\theta_0\E[\Delta D\mid C^\Delta]+g_0^\Delta(C^\Delta).
\]
Although $A_i$ need not be included in $C^\Delta$ away from adoption, the right-hand side of the treatment equation is identically zero there, so it is a function of $C^\Delta$ throughout. The residual $r_\Delta=\ind\{\mathsf t=t^*\}(G_i-p_\Delta)$ has squared expectation
\[
\E[r_\Delta^2]=\frac{1}{T-1}\E[(G_i-p_\Delta)^2]
=\frac{\E[p_\Delta(1-p_\Delta)]}{T-1}.
\]

\paragraph{Proof of Proposition~\ref{prop:panel-transformations}(ii).}
Subtracting bank means gives $D^w_{it}=G_i(P_t-q)$ and
\[
Y^w_{it}=\theta_0D^w_{it}+\lambda_t-\bar\lambda+
 g_0(X_{it})-T^{-1}\sum_sg_0(X_{is})+U_{it}-\bar U_i.
\]
The nuisance is a function of $C^W_{it}=(X_i,t)$. Strict exogeneity implies
$\E[U_{it}-\bar U_i\mid G_i,X_i]=0$, and then
$\E[U_{i\mathsf t}-\bar U_i\mid D^w_{i\mathsf t},C^W_{i\mathsf t}]=0$.
The independence of $\mathsf t$ gives
\[
\E[D^w\mid C^W]=(P_{\mathsf t}-q)p_W(X_i),\qquad
\E[Y^w\mid C^W]=\theta_0\E[D^w\mid C^W]+g_0^W(C^W).
\]
Thus $r_W=(P_{\mathsf t}-q)(G_i-p_W)$ and
\[
\E[r_W^2]=\left\{T^{-1}\sum_t(P_t-q)^2\right\}
\E[(G_i-p_W)^2]=q(1-q)\E[p_W(1-p_W)].
\]
In either transformed PLR, the expected score at the true nuisance functions is $(\theta_0-\theta)\E[r^2]$. Positivity of the relevant second moment therefore gives its unique root.

\paragraph{Proof of Proposition~\ref{prop:panel-transformations}(iii).}
Under $\E[\alpha_i\mid G_i,X_i]=r_0(qG_i,\bar X_i)$, put
$e_{it}=\alpha_i-r_0(qG_i,\bar X_i)+U_{it}$. Then
$\E[e_{it}\mid G_i,X_i]=0$. Conditional expectation with respect to
$(D_{it},C^{\mathrm{CRE}}_{it})$, where
$C^{\mathrm{CRE}}_{it}=(X_{it},\bar X_i,G_i,t)$, gives the levels PLR with nuisance
$\lambda_t+g_0(X_{it})+r_0(qG_i,\bar X_i)$. Since $G_iP_t$ is measurable with respect to $(G_i,t)$, unrestricted conditional-mean adjustment has zero treatment residual. It permits unrestricted direct group-by-time interactions and does not impose the stated additive separation. The result therefore does not establish nonidentification in that additive model. If the bank-effect conditional mean does not depend on $G_i$, the smaller information set $(X_{it},\bar X_i,t)$ suffices for the outcome representation, but its treatment residual variance can still be zero if these remaining variables determine $D_{it}$.

\subsection{Calculation for Example~\ref{ex:quadratic-trend}}
Keep the static model and suppose the bank records are independent and identically distributed with $T$ fixed. Let $X_{it}=\nu_i+ct+\varepsilon_{it}$, $c\ne0$, with $\E[\nu_i^4]<\infty$, independent and identically distributed disturbances $\varepsilon_{it}$ having mean zero, variance $\sigma_\varepsilon^2>0$ and finite fourth moment, and the disturbance vector independent of $(G_i,\nu_i)$. These conditions and square integrability of the model components ensure the moments below exist. Write $\pi=\Pr(G_i=1)\in(0,1)$ and $a_t=P_t-q$.

For any integrable trajectory $R_{it}$, a direct sum gives
\begin{equation}
T^{-1}\sum_t a_tR_{it}
=q(1-q)\left\{|\mathcal T_1|^{-1}\sum_{t\in\mathcal T_1}R_{it}
-|\mathcal T_0|^{-1}\sum_{t\in\mathcal T_0}R_{it}\right\}.
\label{eq:panel-block-identity}
\end{equation}
Define population two-way demeaned variables by
$\ddot R_{it}^{\,0}=R_{it}-\bar R_i-\E[R_{it}]+\E[\bar R_i]$.
For treatment and the scalar covariate they are
\[
\ddot D_{it}^{\,0}=(G_i-\pi)a_t,\qquad
\ddot X_{it}^{\,0}=\varepsilon_{it}-\bar\varepsilon_i.
\]
The disturbance-vector independence implies
$\E[\ddot D^{\,0}\mid\ddot X^{\,0},\mathsf t]=0$.
The population Gram matrix for these two regressors is diagonal, with entries
$\pi(1-\pi)q(1-q)$ and $(1-T^{-1})\sigma_\varepsilon^2$, both positive.
Consequently linear adjustment for $\ddot X^{\,0}$ leaves the treatment residual unchanged. Orthogonality to removed bank and time components and equation~\eqref{eq:panel-block-identity} give the population treatment coefficient
\[
\frac{\E[(G_i-\pi)a_{\mathsf t}Y_{i\mathsf t}]}{\pi(1-\pi)q(1-q)}
=\frac{\E[(G_i-\pi)H_i]}{\pi(1-\pi)}
=\E[H_i\mid G_i=1]-\E[H_i\mid G_i=0].
\]
Sample quarter means converge to their population counterparts. The law of large numbers for complete bank records and the positive Gram matrix therefore yield the same probability limit for sample TWFE as $N\to\infty$.

For within-transformed observations conditioned on $C^a_{it}=(X^w_{it},t)$, note that
$X^w_{it}=c(t-\bar t)+\varepsilon_{it}-\bar\varepsilon_i$.
Conditional on $t$, these variables give no information about $G_i$, so
$\E[D^w\mid C^a]=\pi a_{\mathsf t}$ and the residual is again $(G_i-\pi)a_{\mathsf t}$.
It is orthogonal to every square-integrable function of $C^a$, including
$\E[Y^w\mid C^a]$. The population residual ratio therefore has the same value as the linear coefficient. The same argument applies if population quarter means are also removed from the input and outcome. This calculation concerns a population conditional-mean score; convergence of a fitted learner is an additional condition.

For $g_0(x)=x^2$, independence and zero disturbance means imply
\[
\E[X_{it}^2\mid G_i=g]
=\E[\nu_i^2\mid G_i=g]+2ct\E[\nu_i\mid G_i=g]+c^2t^2+\sigma_\varepsilon^2.
\]
The $\nu_i^2$ term disappears in the pre/post contrast; the $c^2t^2$ and disturbance-variance terms disappear in the difference between groups. Strict exogeneity removes the corresponding difference in errors, and quarter effects are common to both groups. The resulting coefficient is exactly equation~\eqref{eq:quadratic-trend-bias}.

The full-history coefficient follows from Proposition~\ref{prop:panel-transformations}(ii) when $J_W>0$. Including $X_{it}^2$ in a linear FE regression makes the untreated covariate term part of its regressor span, but recovery of $\theta_0$ also requires positive treatment residual variance after this expanded adjustment. That rank condition is not implied by the rank of the regression using $X_{it}$ alone.

Nonlinearity by itself does not imply failure of demeaned-input adjustment. For example, with $T=2$, $X_i=(B_i,-B_i)$ and $g_0(x)=x^3$, both $\bar X_i$ and the time average of $g_0(X_{it})$ are zero. Thus the exact transformed nuisance equals $g_0(X^w_{it})$. An assignment probability strictly between zero and one given $B_i$ can retain positive treatment variation. This example only rules out a necessity claim that linearity is the sole valid case.

\subsection{Closed projection spaces and proof of Corollary~\ref{cor:panel-restricted}}
The projection result applies to any closed linear subspace $\mathcal M$ of $L^2(C)$; it does not require a constant. To see this directly, suppose
$Y=\theta_0D+g_0(C)+U$ with $\E[U\mid D,C]=0$ and square-integrable terms.
Let $m_0=\E[D\mid C]$, $m_{\mathcal M}=\Pi_{\mathcal M}D$, and
$r_{\mathcal M}=D-m_{\mathcal M}$. Since $D-m_0$ is orthogonal to all functions of $C$,
$m_{\mathcal M}=\Pi_{\mathcal M}m_0$ and
\[
\E[r_{\mathcal M}^2]
=\E[(D-m_0)^2]+\|m_0-m_{\mathcal M}\|_2^2.
\]
Projection orthogonality and the error restriction give
\[
\E[r_{\mathcal M}(Y-\Pi_{\mathcal M}Y)]
=\theta_0\E[r_{\mathcal M}^2]
+\E[(m_0-m_{\mathcal M})(g_0-\Pi_{\mathcal M}g_0)].
\]
If $g_0\in\mathcal M$, the last term vanishes. If instead $J_{\mathcal M}=0$, then $D=m_{\mathcal M}\in\mathcal M$ almost surely, and replacing $g_0$ by $g_0+(\theta_0-\theta)D$ preserves the same observational model for every $\theta$. These arguments prove the identification and bias statements without using constants. For any directions $h_\ell,h_m\in\mathcal M$, the score derivatives at $(\Pi_{\mathcal M}Y,\Pi_{\mathcal M}D)$ are
\[
-\E[r_{\mathcal M}h_\ell]=0,\qquad
-\E[h_m(Y-\Pi_{\mathcal M}Y)]+2\theta\E[r_{\mathcal M}h_m]=0,
\]
again by projection orthogonality.

For the within model take $C=(X_i,t)$, $Y=Y^w_{it}$ and $D=D^w_{it}$. Its nuisance is
$g_0(X_{it})-T^{-1}\sum_sg_0(X_{is})+\lambda_t-\bar\lambda$,
which belongs to $\mathcal M_W$ by choosing $f=g_0$ and $a_t=\lambda_t-\bar\lambda$.
Every generator of $\mathcal M_W$ has zero time mean for each bank; this remains true in its closure because temporal averaging is a bounded operator in $L^2$. Thus $\mathcal M_W$ excludes nonzero constants, as intended.
For the CRE model, choose $f=g_0$, $h(G_i,\bar X_i)=r_0(qG_i,\bar X_i)$ and $a_t=\lambda_t$. This places the nuisance in $\mathcal M_{\mathrm{CRE}}$.
The conditional errors established above and the respective positive $J_{\mathcal M}$ then give the corollary. Using closures ensures the existence of projections; it does not guarantee that every limit admits a finite additive decomposition or that the residual variance is positive.

\subsection{Within scores and the bank-mean adjustment}
For arbitrary bank-level fitted values $a_i$ and panel fitted values $\widetilde\ell_{it}$, define
$\widehat m^W_{it}=(P_t-q)a_i$ and
$\widehat\ell^W_{it}=\widetilde\ell_{it}-\overline{\widetilde\ell}_i$.
Then
\[
D^w_{it}-\widehat m^W_{it}=(P_t-q)(G_i-a_i)=v^*_{it},\qquad
Y^w_{it}-\widehat\ell^W_{it}=Y_{it}-\widetilde\ell_{it}-(\bar Y_i-\overline{\widetilde\ell}_i).
\]
Since $\sum_tv^*_{it}=0$, the bank-constant term in the second expression contributes zero to its product with $v^*$. The score numerator and denominator therefore coincide exactly with those of Theorem~\ref{thm:design}(i). Applying equation~\eqref{eq:panel-block-identity} to $Y_{it}-\widetilde\ell_{it}$ gives that theorem's pre/post residual ratio. This proof uses no consistency of fitted values and no causal assumptions. Replacing the full-history assignment probability $p_W(X_i)$ by $p(Z_i)$ changes the population conditioning problem; conditional parallel trends given $Z_i$ supplies the separate interpretation used in Theorem~\ref{thm:design}(iii).

\subsection{Proof of Corollary~\ref{cor:dynamic-panel}}
Write $Y_{it}(d)=Y_{it}(\mathbf d^d)$ and $X_{it}(d)=X_{it}(\mathbf d^d)$, $d\in\{0,1\}$, for outcomes and covariates under the two exposure paths. Replace the static outcome equation by the reference-path model
\[
Y_{it}(0)=\alpha_i+\lambda_t+g_0(X_{it})+U_{it},\qquad
\E[U_{it}\mid G_i,X_i,\alpha_i]=0.
\]
The covariates are invariant across the two exposure paths:
$X_{it}(1)=X_{it}(0)=X_{it}$ for all $t$. Consistency and no anticipation imply
$Y_{it}=Y_{it}(0)+G_iP_t\beta_{it}$, where the value of $\beta_{it}$ before adoption can be set to zero and post-period values equal the potential-outcome contrast. Assume all these variables are square integrable. The treatment regressions and residuals proved above depend only on treatment and covariates and therefore remain unchanged. The outcome conditional means are now the true conditional means under this dynamic model; they need not equal the static formulas involving a constant $\theta_0$.

For each transformation, its residual $r$ satisfies $\E[r\mid C]=0$. Subtracting the true outcome conditional mean has no effect on $\E[rY^{\mathrm{transformed}}]$. The transformed untreated nuisance is a function of the specified $C$, and its error has conditional mean zero, so its contribution is zero as well.

For first differences, $r_\Delta$ vanishes outside adoption. At $t^*$, no anticipation gives
\[
\Delta Y_{it^*}=\Delta Y_{it^*}(0)+G_i\beta_{it^*}.
\]
Conditional on $A_i$, the remaining contribution is
\[
\E[(G_i-p_\Delta)G_i\beta_{it^*}\mid A_i]
=p_\Delta(1-p_\Delta)\E[\beta_{it^*}\mid G_i=1,A_i]
=p_\Delta(1-p_\Delta)\tau_\Delta(A_i).
\]
The pooled numerator divides this expectation by $T-1$. Dividing in turn by $J_\Delta$ proves the first formula in~\eqref{eq:dynamic-panel-targets}, independently of the later effect path.

For the within transformation, $\sum_t(P_t-q)=0$ eliminates the bank mean of the treatment-effect path in the numerator. Its time average is
\[
T^{-1}\sum_t(P_t-q)(G_i-p_W)G_iP_t\beta_{it}
=q(1-q)(G_i-p_W)G_i\bar\beta_i.
\]
Conditioning on $X_i$ gives
$q(1-q)p_W(1-p_W)\E[\bar\beta_i\mid G_i=1,X_i]$.
Dividing its expectation by $J_W$ proves the second formula.
Conditional effects can be assigned arbitrary values where their conditioning event has zero treated probability, since the associated weight is zero. The positive-denominator assumptions are sufficient for these weighted averages; no uniform lower overlap bound is needed for this population calculation. Inference for a fitted score additionally requires the sampling, moment and nuisance-rate conditions stated in Theorem~\ref{thm:collapsed-inference} or Corollary~\ref{cor:fd-inference}, together with the relevant sample representation.

The relation to full-history parallel trends also follows directly. Define
\[
c(X_i)=\bar\lambda_{\mathcal T_1}-\bar\lambda_{\mathcal T_0}
 +|\mathcal T_1|^{-1}\sum_{t\in\mathcal T_1}g_0(X_{it})
 -|\mathcal T_0|^{-1}\sum_{t\in\mathcal T_0}g_0(X_{it}).
\]
Here $\bar\lambda_{\mathcal T_j}$ denotes the average quarter effect over block $\mathcal T_j$. The bank effect cancels in $H_i(0)$ and strict exogeneity yields
$\E[H_i(0)\mid G_i,X_i]=c(X_i)$ in both groups. This proves conditional parallel trends given the invariant history wherever the group comparison is defined. For a coarser vector $Z_i$, however, $\E[c(X_i)\mid G_i,Z_i]$ can depend on $G_i$; the full-history statement therefore need not imply the application's restriction given $Z_i$.

\subsection{Why positive residual variance is insufficient for an unweighted ATT}\label{app:dynamic-att-support}
Consider two periods, with treatment beginning in period two. With probability $1/2$ a bank is in stratum A, has covariate history $(0,0)$ and is treated with probability $1/2$. Otherwise it is in stratum B, has history $(0,1)$ and is always treated. Set all bank effects, quarter effects and errors to zero. For any real number $c$, define $g_c(0)=0$, $g_c(1)=c$, with treatment effect one in A and $1-c$ in B. The covariates are invariant to exposure, and all moments and strict-exogeneity restrictions of Corollary~\ref{cor:dynamic-panel} hold.

For every $c$, the observed outcomes are $Y_{i1}=0$ and $Y_{i2}=G_i$. Thus the observed distribution is unchanged. The residual group variance is $1/8$, and the FD and within residual variances are $J_\Delta=1/8$ and $J_W=1/32$. Both weighted roots equal one because stratum B receives zero conditional-variance weight. Among treated banks, however, one third are in A and two thirds in B, so the ATT is $1-2c/3$. It is not identified in this model class. The example separates the corollary's identified weighted coefficients from an unweighted ATT and shows exactly where treated-support overlap or an extrapolation restriction is needed.
\section{Proofs and numerical verification}\label{app:proofs}
\subsection{Proof of Proposition~\ref{prop:did-benchmark}}
Let $\widehat\pi=N^{-1}\sum_iG_i$. On the realised balanced design, the treatment residual after removing unit and period fixed effects is
\[
\ddot D_{it}=(G_i-\widehat\pi)(P_t-q).
\]
It is orthogonal to the removed components. For any complete outcome trajectory,
\[
T^{-1}\sum_t(P_t-q)Y_{it}=q(1-q)H_i.
\]
The Frisch--Waugh--Lovell ratio therefore reduces to
\[
\widehat\theta_{\mathrm{TWFE}}
=\frac{\sum_i(G_i-\widehat\pi)H_i}{N\widehat\pi(1-\widehat\pi)}
=\frac{\sum_iG_iH_i}{\sum_iG_i}
-\frac{\sum_i(1-G_i)H_i}{\sum_i(1-G_i)}.
\]
The assumed presence of both groups and period blocks makes the denominator positive. This is an algebraic equality and requires no outcome model.

For the population counterpart, put $\pi=\Pr(G=1)$ and average uniformly over the fixed periods. The population treatment residual is $(G_i-\pi)(P_t-q)$; its second moment is $\pi(1-\pi)q(1-q)$. Its covariance with the observed outcome, divided by that moment, is
\[
\theta_{\mathrm{TWFE}}
=\frac{\E[(G-\pi)H]}{\pi(1-\pi)}
=\E[H\mid G=1]-\E[H\mid G=0].
\]
For the two exposure paths defined in Section~\ref{sec:setup:target}, write $H_i(0)=\bar Y_{i,\mathcal T_1}(\mathbf d^0)-\bar Y_{i,\mathcal T_0}(\mathbf d^0)$ and $\bar\Delta_i=|\mathcal T_1|^{-1}\sum_{t\in\mathcal T_1}\{Y_{it}(\mathbf d^1)-Y_{it}(\mathbf d^0)\}$. Consistency and no anticipation give $H_i=H_i(0)+G_i\bar\Delta_i$. Thus
\[
\theta_{\mathrm{TWFE}}=\bar\tau+
\E[H(0)\mid G=1]-\E[H(0)\mid G=0].
\]
Equation~\eqref{eq:unconditional-pt} removes the final discrepancy. No restriction on the variation of $\bar\Delta_i$ across units, or on its constituent effects across post-treatment periods, was used. Population parallel trends need not make the corresponding realised-sample discrepancy zero. Under independent sampling of complete units and the stated moments, the two sample group means converge to their population counterparts.

For part (iii), the same decomposition $H_i=H_i(0)+G_i\bar\Delta_i$ gives, conditional on $Z$,
\[
\begin{aligned}
\E[H\mid G=1,Z]-\E[H\mid G=0,Z]
&=\E[\bar\Delta\mid G=1,Z]\\
&\quad+\E[H(0)\mid G=1,Z]-\E[H(0)\mid G=0,Z]\\
&=\tau(Z),
\end{aligned}
\]
where the last equality is equation~\eqref{eq:cpt}, and Assumption~\ref{ass:support} makes the control conditional mean well defined on the treated support. Averaging $\tau(Z)$ over the distribution of $Z$ given $G=1$ gives $\bar\tau$, which is equation~\eqref{eq:att-functional}.

\subsection{Proof of Lemma~\ref{lem:mundlak}}
Let $\mathsf B$ collect the unit and time means of each column of $\mathsf R$ and a constant. The orthogonal projection onto its column space is well defined even if $\mathsf B$ has redundant columns. Decompose $\mathsf R=Q\mathsf R+(I-Q)\mathsf R$. The second term lies in the column space of $\mathsf B$, and $Q\mathsf R$ is orthogonal to $\mathsf B$ because it sums to zero within each unit and time period. Hence $M_{\mathsf B}\mathsf R=Q\mathsf R$. The Frisch--Waugh--Lovell coefficient is $(\mathsf R'M_{\mathsf B}\mathsf R)^{-1}\mathsf R'M_{\mathsf B}Y=(\mathsf R'Q\mathsf R)^{-1}\mathsf R'QY$. Symmetry and idempotence of $Q$ yield the transformed-data coefficient. The assumed within rank makes the inverse well defined.

\subsection{Proof of Proposition~\ref{prop:support} and Corollary~\ref{cor:absorbing}}
Write $m_j=\E[D\mid\mathcal C_j]$. The residual decomposition $D-m_1=(D-m_2)+(m_2-m_1)$ has orthogonal components by iterated expectations. Taking expectations of the squared residual decomposition proves~\eqref{eq:nested}. Nonnegative conditional variance has zero expectation exactly when $D=\E[D\mid\mathcal C]$ almost surely. If $D=h(C)$, replacing $g_0$ as in the proposition preserves $\theta D+g(C)$ at every observation and preserves the conditional error restriction. Therefore the observed law cannot distinguish values of $\theta$ within the unrestricted nuisance model.

The treatment means under common treatment timing follow by summing $G_iP_t$ over periods and units, with $\widehat\pi=N^{-1}\sum_iG_i$ the realised treated share. For absorbing treatment at $t_i^*$, there are $T-t_i^*+1$ treated periods, including zero when $t_i^*=T+1$. Dividing by $T$ and solving for $t_i^*$ proves the staggered-adoption identity.

\subsection{Proof of Proposition~\ref{prop:restricted}}
Every element of $\mathcal M$ is a function of $C$, so $D-m_0$ is orthogonal
to $\mathcal M$ and to $m_0-m_{\mathcal M}$. Hence
$\Pi_{\mathcal M}D=\Pi_{\mathcal M}m_0$ and the Pythagorean identity for
$D-m_{\mathcal M}=(D-m_0)+(m_0-m_{\mathcal M})$ proves
\eqref{eq:restricted-info}.

Write $r_{\mathcal M}=D-m_{\mathcal M}$ and
$g_{\mathcal M}=\Pi_{\mathcal M}g_0$. The conditional mean restriction implies
$\E[Ur_{\mathcal M}]=0$ and $\Pi_{\mathcal M}U=0$. Because
$r_{\mathcal M}$ is orthogonal to $\mathcal M$,
$\E[r_{\mathcal M}\ell_{\mathcal M}]=0$,
$\E[r_{\mathcal M}D]=J_{\mathcal M}$, and
\begin{align*}
\E[r_{\mathcal M}Y]
&=\theta_0J_{\mathcal M}+\E[r_{\mathcal M}(g_0-g_{\mathcal M})]\\
&=\theta_0J_{\mathcal M}
+\E[(m_0-m_{\mathcal M})(g_0-g_{\mathcal M})].
\end{align*}
The second equality uses $\E[D-m_0\mid C]=0$. Dividing by $J_{\mathcal M}$
proves~\eqref{eq:restricted-bias}; Cauchy--Schwarz gives the displayed bound.

If $g_0\in\mathcal M$ and $J_{\mathcal M}>0$, the ratio uniquely recovers
$\theta_0$. Conversely, if $J_{\mathcal M}=0$, then $D=m_{\mathcal M}\in
\mathcal M$ almost surely. For every alternative coefficient $\theta$,
the function $g_0+(\theta_0-\theta)m_{\mathcal M}$ remains in $\mathcal M$
and generates the same conditional outcome mean. This proves the
identification equivalence within the stated restricted model.

For any $\theta\in\mathbb R$ and $h_\ell,h_m\in\mathcal M$, the directional derivatives of the expected
score with respect to $\ell$ and $m$, respectively, are
\begin{align*}
-\E[r_{\mathcal M}h_\ell]&=0,\\
-\E[h_m(Y-\ell_{\mathcal M})]
+2\theta\E[r_{\mathcal M}h_m]&=0.
\end{align*}
Both equalities follow from the defining projection orthogonality. These
zero derivatives hold for every coefficient value in the allowed nuisance
directions; it does not turn an incorrect outcome restriction into a correct
causal model.

\subsection{Proof of Corollary~\ref{cor:twfe}}
All projections in the first part of the corollary use the empirical inner product on the realised balanced design. By definition, $r=(I-\Pi_{\mathsf L})d$ is orthogonal to $\mathsf L$ and satisfies $r'd=r'r$. Thus $r'(I-\Pi_{\mathsf L})y=r'y$, and the empirical residual ratio is the ordinary least-squares coefficient on $d$ in the regression on $(d,\mathsf L)$. Since $\bar D_i=qG_i$ and $\bar D_t=\widehat\pi P_t$, the column space of $(d,\mathsf L)$ is the two-way Mundlak design of Lemma~\ref{lem:mundlak}. That lemma establishes equality with the realised two-way within coefficient without taking expectations.

Conditional on $\mathcal X$, $r$, $d$ and $g$ are fixed. Substitution of $y=\theta_0d+g+u$ gives
\[
\widehat\theta_{\mathsf L}-\theta_0
=\frac{r'g}{r'r}+\frac{r'u}{r'r}.
\]
The conditional mean of the second term is zero. This proves equation~\eqref{eq:twfe-conditional}; it establishes conditional unbiasedness under $r'g=0$, not equality of a realised estimate and $\theta_0$.

Without controls, uniform averaging of every unit-period observation on a balanced panel makes the empirical group and post indicators independent. Their empirical means are $\widehat\pi$ and $q$. Consequently $\Pi_{\mathsf L}d=qG+\widehat\pi P-\widehat\pi q$ and $r_{it}=(G_i-\widehat\pi)(P_t-q)$. Its empirical second moment is $\widehat\pi(1-\widehat\pi)q(1-q)$. Partitioning $\langle r,g\rangle_n$ into the four group-period cells gives
\[
\langle r,g\rangle_n
=\widehat\pi(1-\widehat\pi)q(1-q)
\{\bar g_{11}-\bar g_{10}-\bar g_{01}+\bar g_{00}\},
\]
which proves equation~\eqref{eq:kappa} and defines the finite-design discrepancy $\kappa_{\mathcal X}$. This argument permits arbitrary variation of the untreated conditional mean inside each group-period cell; only the four cell averages enter the coefficient.

For the distinct population statement, use the population inner product of Proposition~\ref{prop:restricted}, with independent Bernoulli indicators $(G,P)$ and $\mathcal M=\operatorname{span}(1,G,P)$. The population projection satisfies $\Pi_{\mathcal M}D=qG+\pi P-\pi q$, so $D-\Pi_{\mathcal M}D=(G-\pi)(P-q)$ and $J_{\mathcal M}=\pi(1-\pi)q(1-q)$. If $g_0(G,P)=a+bG+cP+\kappa GP$, then
\[
g_0-\Pi_{\mathcal M}g_0=\kappa(G-\pi)(P-q).
\]
Equation~\eqref{eq:restricted-bias} therefore gives $\theta_{\mathcal M}-\theta_0=\kappa$. Here $\pi$ and $q$ are probabilities under that population law. If the law instead samples a unit and a period independently and uniformly from the fixed balanced design and averages over its outcome noise, they equal the design shares, and the population projection coefficient is the conditional mean in equation~\eqref{eq:kappa}; it is still distinct from the noisy realised coefficient.

\subsection{Proof of Corollary~\ref{cor:heterogeneous}}
Set $r_{\mathcal M}=D-m_{\mathcal M}$. Projection orthogonality gives
$\E[r_{\mathcal M}\ell_{\mathcal M}]=0$,
$\E[r_{\mathcal M}\Pi_{\mathcal M}g_0]=0$, and
$\E[r_{\mathcal M}D]=J_{\mathcal M}$. Also,
\[
\E[r_{\mathcal M}\mid C]=\eta_{\mathcal M}(C),\qquad
\E[r_{\mathcal M}D\mid C]
=\E[D^2\mid C]-m_{\mathcal M}m_0
=s(C)+m_0(C)\eta_{\mathcal M}(C).
\]
The conditional error restriction implies $\E[r_{\mathcal M}U]=0$. Taking expectations after substituting the heterogeneous-effect model into $\E[r_{\mathcal M}Y]$ therefore yields
\[
\E[r_{\mathcal M}Y]
=\E[\{s+m_0\eta_{\mathcal M}\}\tau]
+\E[\eta_{\mathcal M}(g_0-\Pi_{\mathcal M}g_0)].
\]
Division by $J_{\mathcal M}$ proves equation~\eqref{eq:heterogeneous-restricted}. The weights integrate to one because their numerator has expectation $\E[r_{\mathcal M}D]=J_{\mathcal M}$. If $J(C)>0$, subtracting $\theta_V$ and using $\E[s(\tau-\theta_V)]=0$ proves equation~\eqref{eq:heterogeneous-gap}. When $D$ is binary, $s=m_0(1-m_0)$, so the weight numerator simplifies to $m_0(1-m_{\mathcal M})$.

For a concrete illustration, use the two covariate strata in Example~\ref{ex:overlap-att}, now denoting assignment by $D$, and take $\mathcal M$ to contain only constants. Then $m_{\mathcal M}=1/2$, $J_{\mathcal M}=1/4$, $J(C)=4/25$, $g_0=0\in\mathcal M$, and
\[
\theta_{\mathcal M}=2.6,\qquad \theta_V=2,\qquad
\frac{\E[\eta_{\mathcal M}m_0(\tau-\theta_V)]}{J_{\mathcal M}}=0.6.
\]
Thus a correct untreated-outcome restriction need not make the two projection targets equal.

Signed weights can arise in the same static setting. Let $C=-1,0,1$ with equal probabilities, and let $m_0(C)$ equal $0.05,0.95,0.95$. For $\mathcal M=\operatorname{span}(1,C)$, the projection is $m_{\mathcal M}(C)=0.65+0.45C$. Hence $J_{\mathcal M}=0.0925$ and the integrated weights on the three strata are $16/111$, $133/111$ and $-38/111$. With $g_0=0$ and effects $0,0,1$, the restricted coefficient is $-38/111$ despite nonnegative conditional effects. These calculations concern a restricted linear projection, not a propensity constrained to the unit interval.

Under Corollary~\ref{cor:heterogeneous} with $J(C)>0$, write $V=D-m_0$ and $\ell_0=\E[Y\mid C]$, so that $Y-\ell_0=\tau(C)V+U$. Conditional on the independent training sample,
\begin{align*}
\E[(D-\widehat m)(Y-\widehat\ell)\mid\mathrm{training}]
&=\theta_VJ+\E[\delta_m\delta_\ell\mid\mathrm{training}],\\
\E[(D-\widehat m)^2\mid\mathrm{training}]
&=J+\E[\delta_m^2\mid\mathrm{training}].
\end{align*}
Thus equation~\eqref{eq:rootbias} and Corollary~\ref{cor:oracle-swap} hold with $\theta_V$ replacing $\theta_0$. Both nuisance derivatives still vanish because $\E[V\mid C]=\E[Y-\ell_0\mid C]=0$; this does not require the score itself to have zero conditional mean. Under heterogeneity, however, $\ell_0-\theta_Vm_0=g_0+(\tau-\theta_V)m_0$, so the error $\delta_\ell-\theta_V\delta_m$ is relative to this composite function rather than to $g_0$ alone.

\subsection{Proof of Proposition~\ref{prop:classes}}
Uniform independent sampling of a unit and a period makes $G$ and $P$
independent Bernoulli variables with means $\pi=\widehat\pi$ and $q$ under
this fixed-panel law. The treatment
means $qG$ and $\pi P$ are invertible transformations of the two indicators.
If the reconstruction map belongs to the prediction class, it predicts
$D=GP$ exactly and proves (i).

The centred functions of $G$ and the centred functions of $P$ are
orthogonal in $L^2$ by independence. Thus the projection of $GP$ onto the
additive space is
$\E[GP\mid G]+\E[GP\mid P]-\E[GP]=qG+\pi P-\pi q$.
Its residual is $(G-\pi)(P-q)$, with second moment
$\pi(1-\pi)q(1-q)$. Every function of a binary argument is affine in that
argument, so the additive class on the four cells is exactly the linear
span of $1,G,P$. This proves (ii). For (iii), condition on $G$ to obtain
$\E[D\mid G]=qG$ and risk $\pi q(1-q)$.

For (iv), write the logistic index in the equivalent coordinates
$\alpha+bG+cP$. The sequence $s(G+P-3/2)$ is positive exactly when
$(G,P)=(1,1)$ and converges in magnitude to infinity on every cell as
$s\to\infty$. The associated probabilities tend to $D$ and their squared
prediction loss tends to zero. Finite coefficients give probabilities in
$(0,1)$, so zero squared loss is unattainable. In the original feature
coordinates the same sequence has $\alpha=-3s/2$, $\beta=s/q$ and
$\gamma=s/\pi$.

For positive $\lambda$, consider expected Bernoulli negative
log-likelihood plus $\lambda(\beta^2+\gamma^2)$. On a bounded sublevel
set the two penalised slopes are bounded. Since both outcomes have positive
probability, the negative log-likelihood diverges when the unpenalised
intercept tends to either infinity with bounded slopes. Hence sublevel
sets are compact and continuity gives a finite minimiser. Its squared
prediction loss is strictly positive. For any $\epsilon>0$, choose fixed
finite separating coefficients with negative log-likelihood below
$\epsilon/2$, then choose $\lambda$ sufficiently small that their penalty
is below $\epsilon/2$. The penalised optimum therefore tends to zero as
$\lambda\downarrow0$. On each Bernoulli outcome, squared probability error
is bounded by the negative log-likelihood contribution, so the squared
prediction loss at the optimum also tends to zero. This proves the stated
claim for the usual logistic fitting objective. Finally, an infimum over a
superset cannot exceed the original infimum.

\subsection{Proof of Proposition~\ref{prop:structured}}
The matrix $Q_T$ is the symmetric idempotent projection orthogonal to $\mathbf1_T$. Therefore $v_i^*=Q_T(D_i-\widetilde m_i)$, and orthogonal decomposition into within-unit and constant components gives \eqref{eq:structured-loss}. Since $\mathbf1_T'v_i^*=0$, multiplying the outcome residual by $Q_T$ leaves its inner product with $v_i^*$ unchanged, proving \eqref{eq:structured-score}.

For the linear comparison, reparameterise the training design as $(X_{it}-\bar X_i,\bar X_i,1)$. Within each training fold, complete units make every within column orthogonal to every between column. Adding the between column $\bar D_i$ thus leaves the estimated within coefficients in both nuisance regressions unchanged. In the treatment regression the between response is exactly $\bar D_i$, so the enlarged between regression reproduces it exactly. On a held-out unit, its treatment prediction is consequently $\bar D_i+(X_{it}-\bar X_i)'\widehat b_m$, also the result of shifting the original prediction. The two outcome predictions have the same within component $(X_{it}-\bar X_i)'\widehat b_\ell$ and may differ only by a unit constant. Equation~\eqref{eq:structured-score} removes that difference. The numerators and denominators coincide.
\subsection{Proof of Theorem~\ref{thm:design}}
By Proposition~\ref{prop:structured}, $v^*_{it}=D_{it}-P_ta_i-\bar D_i+qa_i=(P_t-q)(G_i-a_i)$. For any complete-unit vector $c_i$,
\[
\sum_t(P_t-q)c_{it}=Tq(1-q)
(\bar c_{i,\mathcal T_1}-\bar c_{i,\mathcal T_0}),
\qquad \sum_t(P_t-q)^2=Tq(1-q).
\]
Apply the first equality to $Y_i-\widetilde\ell_i$ and cancel the common factor $Tq(1-q)$ from the numerator and denominator. This proves equation~\eqref{eq:design-collapse} for arbitrary fitted predictions.

For a validation unit, write $G-a=(G-p)+(p-a)$. Conditional on training, the two terms are orthogonal because $\E[G-p\mid Z]=0$. Thus $\E[(G-a)^2]=J_G+\|a-p\|_2^2$, which proves equation~\eqref{eq:design-risk}. The law of total variance gives
\[
J_G=\pi(1-\pi)-\operatorname{Var}(p(Z))\leq\pi(1-\pi).
\]
Dividing $q(1-q)J_G$ by $\operatorname{Var}(D)=\pi q(1-\pi q)$ gives the oracle bound and its equality condition.

Let $\mu_0(Z)=\E[H\mid G=0,Z]$. No anticipation and conditional parallel trends imply $\E[H\mid G=1,Z]=\mu_0(Z)+\tau(Z)$, so $\ell_H(Z)=\mu_0(Z)+p(Z)\tau(Z)$. Conditional on $Z$,
\[
\E[(G-p)(H-b(Z))\mid Z]=p(1-p)\tau(Z).
\]
Taking expectations and dividing by $J_G$ proves the first representation in equation~\eqref{eq:overlap-target}. Bayes' rule gives its representation as a reweighted ATT. Write that numerator as $\E[1-p\mid G=1]\E[\tau\mid G=1]+\operatorname{Cov}(1-p,\tau\mid G=1)$ to obtain equation~\eqref{eq:overlap-att-gap}.

The derivative of the expected score in equation~\eqref{eq:overlap-score} in an outcome direction $h_\ell(Z)$ is $-\E[(G-p)h_\ell]=0$. Its derivative in an assignment direction $h_p(Z)$ is
\[
-\E[h_p(H-\ell_H)]+2\tau_{\mathrm{ov}}\E[(G-p)h_p]=0.
\]
This uses conditional-mean orthogonality and does not require homogeneous effects. For a general random adjustment $b_i$, the numerator with oracle assignment is $\E[p(1-p)\tau]-\E[(G-p)b_i]$, showing why an arbitrary bank-specific adjustment need not preserve the same root.

The robustness to an outcome adjustment $b(Z)$ in part (iii) is weaker than full nuisance orthogonality at that adjustment. At $(\tau_{\mathrm{ov}},b,p)$, the derivative in a propensity direction is $-\E[h_p(Z)\{\ell_H(Z)-b(Z)\}]$, which need not be zero. For fitted functions $a(Z),b(Z)$ held fixed conditional on training and evaluated on independent validation units, conditional-mean orthogonality also gives the exact population root
\begin{equation}
\theta_H(b,a)=
\frac{\tau_{\mathrm{ov}}J_G+\E[(a-p)(b-\ell_H)]}
{J_G+\E[(a-p)^2]}.
\label{eq:overlap-fitted-root}
\end{equation}
Indeed, expand $G-a=(G-p)-(a-p)$ and $H-b=(H-\ell_H)-(b-\ell_H)$; the two linear error terms have zero expectation. With $b=\ell_H$ the product term vanishes, giving attenuation by the factor $J_G/\{J_G+\E[(a-p)^2]\}$. Estimated outcome error can change its sign or magnitude, and a realised sample ratio also contains validation-sample variation. This derivation uses the conditional means of the change-based problem and does not impose the constant-effect PLR model.

For completeness, the overlap score has the pointwise binary-treatment representation
\begin{align*}
\psi_{\mathrm{ov}}={}&G(1-p)(H-\mu_1)-(1-G)p(H-\mu_0)\\
&+\{p(1-p)+(1-2p)(G-p)\}(\mu_1-\mu_0-\theta),
\end{align*}
where $\mu_g(Z)=\E[H\mid G=g,Z]$. It follows from $\ell_H=\mu_0+p(\mu_1-\mu_0)$ and $(G-p)^2=p(1-p)+(1-2p)(G-p)$. The final term accounts for the dependence of the overlap target itself on assignment probabilities.

An extension from exactly factorised fitted predictions to their conditional-mean limits requires consistency of the row-level conditional means on an independent validation unit. Write
\[
\ell_i^0=(\E[Y_{it}\mid Z_i])_{t=1}^T,\qquad
m_i^0=(P_t p(Z_i))_{t=1}^T,\qquad
v_i^0=Q_T(D_i-m_i^0).
\]
With fixed $T$, suppose
\[
\E[\|\widetilde m_i-m_i^0\|^2\mid\mathrm{training}]
+\E[\|\widetilde\ell_i-\ell_i^0\|^2\mid\mathrm{training}]
=o_p(1).
\]
Because $Q_T$ is a contraction, $\widetilde v_i=Q_T(D_i-\widetilde m_i)$ converges to $v_i^0$ in this conditional $L^2$ norm. Cauchy--Schwarz gives convergence of its expected squared norm and of its expected inner product with $Y_i-\widetilde\ell_i$. Their limits are $Tq(1-q)J_G$ and $Tq(1-q)J_G\tau_{\mathrm{ov}}$, respectively, since the difference of the block averages of $\ell_i^0$ is $\ell_H(Z_i)$. The first limit is positive. The validation-population ratio therefore converges to $\tau_{\mathrm{ov}}$. This is a continuity argument for population moments, not an exact fitted-sample identity or a rate assertion for a particular row learner.

\subsection{Proof of Proposition~\ref{prop:fd}}
FWL gives $\widehat\beta_{\mathrm{FD}}=v'By/(v'v)$. Substitute~\eqref{eq:dynamic}; the nuisance component vanishes by the column-space condition, and the assumed conditional error moment removes $v'Bu$. This gives the stated weighted conditional mean. Since $\sum_s Be_s=Bd$ and $v=M_{\mathsf F} Bd$, the weights sum to $v'Bd/(v'v)=1$.

With only saturated period effects, the residual at adoption is $G_i-\widehat\pi$, where $\widehat\pi$ is the realised treated share, and the residual is zero in every other period. The vector $Be_0$ is positive at adoption and negative in the following period; only the first contributes. All later $Be_s$ have support away from adoption. Consequently the adoption weight is one and the others are zero. General control adjustment does not preserve this temporal support, as Example~\ref{ex:fd} demonstrates.

\subsection{Proof of Proposition~\ref{prop:risk}}
Conditional on training, $\widehat m$ is a fixed function on an independent validation unit. With $V=D-m_0(C)$, the cross term $\E[V\delta_m\mid\text{training}]$ is zero. Expanding $(V-\delta_m)^2$ gives~\eqref{eq:risk}. Also $Y-\ell_0(C)=\theta_0V+U$, so
\begin{equation}
\psi(W;\theta_0,\widehat\ell,\widehat m)
=(V-\delta_m)(U-\delta_\ell+\theta_0\delta_m).
\end{equation}
The terms involving $VU$, $V\delta_\ell$, $V\delta_m$, and $\delta_m U$ have zero expectation under the conditional mean restrictions. The remaining terms give~\eqref{eq:remainder}. The derivative of this fitted-nuisance population moment with respect to $\theta$ is $-(J+\E[\delta_m^2\mid\text{training}])$, yielding~\eqref{eq:rootbias}. The identity is algebraic and does not establish finite-sample unbiasedness of a sample ratio.

\subsection{Population double robustness of the DiD score}
Let $\mu_0(Z)=\E[H\mid G=0,Z]$. Under conditional parallel trends and no anticipation, $\E[H-\mu_0(Z)\mid G=1,Z]$ is the conditional average post-period treatment effect. If $\mu_0$ is correct, the control residual has conditional mean zero, so replacing $p$ by any suitably bounded working propensity does not change the score expectation. The treated term then averages the conditional effect to $\bar\tau$.

If $p$ is correct but $b$ is an arbitrary working outcome regression, the difference between the score expectations using $b$ and $\mu_0$ is
\begin{equation}
\E\left[\left\{-\frac{G}{\pi}+
\frac{1-G}{\pi}\frac{p(Z)}{1-p(Z)}\right\}(b(Z)-\mu_0(Z))\right]=0,
\end{equation}
by conditioning on $Z$. The argument requires the relevant support and moments. It cannot identify an untreated mean on treated support where no controls are available without further restrictions.

For the normalised score, fixed clipping can be described explicitly. Let $p_c=\min(1-c,\max(c,p))$, $\rho_c(Z)=(1-p(Z))p_c(Z)/(1-p_c(Z))$, and $\delta(Z)=b(Z)-\mu_0(Z)$ for a working control-outcome regression $b$. Its population root is
\begin{equation}
\tau_c(b)=\bar\tau-
\frac{\E[p\delta]}{\pi}
+\frac{\E[\rho_c\delta]}{\E[\rho_c]}.
\label{eq:clipped-dr-root}
\end{equation}
The treated residual mean is $\bar\tau-\E[p\delta]/\pi$ and the weighted control residual mean is $-\E[\rho_c\delta]/\E[\rho_c]$, which proves the identity. If clipping is inactive, $\rho_c=p$ and the discrepancy cancels. If $b=\mu_0$, it vanishes for any working propensity with well-defined weights. Binding clipping and a misspecified outcome limit generally leave a discrepancy even when the unmodified propensity model is correct.

\subsection{Proofs of Theorem~\ref{thm:collapsed-inference} and Proposition~\ref{prop:dr-inference}}\label{app:collapsed-inference}
We prove the two inference results separately after a common argument for the permitted bank-level folds.

\begin{proof}
For a fitted function, $\mathbb P f$ below integrates it against the law of a fresh independent bank record, holding its training records, fold allocation and learner randomisation fixed. Conditional expectations over a validation fold are distinguished from this population integral. The score moments vanish by Theorem~\ref{thm:design} and the DiD identification argument.

For overlap, $|G-p|\leq1$ and Jensen's inequality gives $\E[\ell_H^2]\leq\E[H^2]<\infty$. Hence the influence function in equation~\eqref{eq:collapsed-if-ov} has finite second moment when $J_G>0$. This uses no uniform upper or lower propensity bound. For DR, the additional upper bound gives $p/(1-p)\leq(1-\varepsilon)/\varepsilon$ and
\[
\E[(1-p)\mu_0^2]\leq\E[(1-G)H^2]<\infty.
\]
Since $1-p\geq\varepsilon$, $\mu_0$ is square integrable on the full covariate distribution. Bounded odds weights and fixed $\pi\in(0,1)$ then give a square-integrable DR influence function. Neither proof uses a moment above order two; the DR proof also needs no lower propensity bound.

We first establish the empirical-process step for both splitting schemes. Any independent learner randomisation is included in the conditioning below. Let $\Delta f_k=f_k-f_0$ be a fitted function difference with $\|\Delta f_k\|_2=o_p(1)$, and let $\mathbb P_{n_k}$ average over held-out fold $k$. For record-independent folds, conditioning on that partition and its complementary training records makes $\Delta f_k$ fixed and the validation observations independent with law $\mathbb P$. Conditional Chebyshev therefore gives
\begin{equation}
(\mathbb P_{n_k}-\mathbb P)\Delta f_k=o_p(N^{-1/2}).
\label{eq:fold-empirical-remainder}
\end{equation}

For treatment-stratified folds, instead condition on the full vector $(G_1,\ldots,G_N)$, the partition and its external randomisation, and the complementary training records. Within the held-out fold, the remaining $(H_i,Z_i)$ are conditionally independent with their original laws given $G_i$. Write $\mathbb P_g f=\E[f(W)\mid G=g]$ and $\widehat\pi_k=n_{1k}/n_k$. Both group probabilities are fixed and positive, so
\[
\mathbb P_g(\Delta f_k)^2
\leq\frac{\|\Delta f_k\|_2^2}{\Pr(G=g)}
=o_p(1),\qquad g=0,1.
\]
The conditional mean of $\mathbb P_{n_k}\Delta f_k$ is
$\widehat\pi_k\mathbb P_1\Delta f_k+(1-\widehat\pi_k)\mathbb P_0\Delta f_k$,
and its conditional variance is $o_p(1)/n_k$. Group balance gives
$\widehat\pi_k=\widehat\pi+O(N^{-1})$, where $\widehat\pi=N_1/N$.
Since $\widehat\pi-\pi=O_p(N^{-1/2})$, the difference between this conditional mean and $\mathbb P\Delta f_k$ is
\[
(\widehat\pi_k-\pi)
  \{\mathbb P_1\Delta f_k-\mathbb P_0\Delta f_k\}
=O_p(N^{-1/2})\,o_p(1).
\]
Conditional Chebyshev proves~\eqref{eq:fold-empirical-remainder} under scheme (b) as well. This argument centres only the fitted-minus-oracle difference. The full oracle sample average is retained, so the final unconditional i.i.d.\ central limit theorem also retains randomness in the group counts.

Apply~\eqref{eq:fold-empirical-remainder} separately within each fold and add the fixed number of contributions; no conditioning on all fitted nuisances is required. The required $L^2$ convergence of the score differences follows from bounded propensity residuals for overlap, bounded odds weights for DR, and the relevant nuisance consistency. To see why no fourth outcome moment is needed, let $A\in L^2$ and $\delta_p=\widehat p-p$. Since $|\delta_p|\leq1$, for any $M>0$,
\[
\mathbb P[\delta_p^2 A^2]
\leq M\mathbb P[\delta_p^2]
   +\mathbb P[A^2\ind\{A^2>M\}].
\]
First use $r_p=o_p(1)$ at fixed $M$, then let $M\to\infty$. Thus this weighted second moment is $o_p(1)$. For overlap take $A=H-\ell_H(Z)$; the other score-difference terms are bounded multiples of $\delta_\ell$ and $\delta_p$. For DR take $A=H-\mu_0(Z)$ and use the Lipschitz bound
$|\widehat p/(1-\widehat p)-p/(1-p)|\leq c^{-2}|\delta_p|$;
the outcome-error terms are bounded multiples of $\widehat\mu_0-\mu_0$. These bounds apply to fixed fold-specific prediction rules. They do not cover outcomes or conditioning variables constructed using shared full-sample quarter statistics.

\emph{Overlap expansion.} Writing $\delta_p=\widehat p-p$ and $\delta_\ell=\widehat\ell_H-\ell_H$, the exact population remainder is
\[
\E[\psi_{\mathrm{ov}}(W;\tau_{\mathrm{ov}},\widehat\ell_H,\widehat p)]
=\E[\delta_p\delta_\ell-\tau_{\mathrm{ov}}\delta_p^2].
\]
Its absolute value is at most $r_p r_\ell+|\tau_{\mathrm{ov}}|r_p^2$. Equation~\eqref{eq:collapsed-rates} and the empirical-process step leave the oracle score average plus $o_p(N^{-1/2})$. Also $\widehat J_G=J_G+o_p(1)$. Solving the affine score and applying the i.i.d.\ central limit theorem proves~\eqref{eq:collapsed-alr} for overlap, without a constant-effect restriction.

\emph{Normalised DR expansion.} Put $R_0=H-\mu_0(Z)$, $w_0=(1-G)p/(1-p)$, $e_i=\widehat\mu_{0,i}-\mu_0(Z_i)$ and $\widehat w_i=(1-G_i)\widehat p_i/(1-\widehat p_i)$. Write $\mathbb P_N a=N^{-1}\sum_i a_i$, $B_1=\mathbb P_NG$ and $\widehat B_0=\mathbb P_N\widehat w$. Conditional on $G=0,Z$, $R_0$ has mean zero. Thus
\[
\mathbb P_N\widehat w R_0
=\mathbb P_N w_0R_0+o_p(N^{-1/2}),\qquad
\widehat B_0-\pi=O_p(r_p+N^{-1/2}),
\]
while $B_1-\pi=O_p(N^{-1/2})$. The nuisance-outcome part of the difference of normalised means is exactly
\[
-\frac{\mathbb P_N(G-\widehat w)e}{B_1}
+\mathbb P_N\widehat w e
 \left(\frac1{\widehat B_0}-\frac1{B_1}\right).
\]
For each training fold,
\[
\E[(G-\widehat w)e]
=\E\left[\frac{p-\widehat p}{1-\widehat p}\,e\right]
=O_p(r_p r_\mu).
\]
The empirical-process step gives the same order plus $o_p(N^{-1/2})$ for its sample counterpart. Further,
$\mathbb P_N\widehat w e=O_p(r_\mu)+o_p(N^{-1/2})$.
The denominator difference is $O_p(r_p+N^{-1/2})$, so the second term is
$O_p(r_p r_\mu+N^{-1/2}r_\mu)+o_p(N^{-1/2})$.
Both terms are therefore $o_p(N^{-1/2})$ by equation~\eqref{eq:dr-rates} and consistency. Expanding the remaining treated ratio and the control ratio around their population denominators $\pi$ gives
\[
\widehat\tau_{\mathrm{DR}}-\bar\tau
=\frac{\mathbb P_N[G(R_0-\bar\tau)]}{\pi}
-\frac{\mathbb P_N[w_0R_0]}{\pi}+o_p(N^{-1/2}).
\]
This proves equation~\eqref{eq:dr-alr} after the i.i.d.\ central limit theorem. The treated denominator contributes the centring term $-G\bar\tau/\pi$. The control numerator has population mean zero, so estimating its denominator contributes no additional first-order term at the true nuisance functions. This accounts for both normalising denominators and yields~\eqref{eq:collapsed-if-dr}.

\emph{Variance and repeated partitions.} For either $j\in\{\mathrm{ov},\mathrm{DR}\}$, write $\theta_{\mathrm{ov}}=\tau_{\mathrm{ov}}$ and $\theta_{\mathrm{DR}}=\bar\tau$. First replace the full-sample coefficient and denominator estimates in the fitted influence values by their probability limits. For DR this also replaces the treated and weighted-control residual centres by $\bar\tau$ and zero, respectively; their individual consistency follows from the preceding ratio argument and nuisance consistency. The resulting change has empirical squared norm $o_p(1)$ by consistency of these scalars, bounded fitted weights and bounded-in-probability empirical second moments of the outcome and nuisance fits. The remaining fitted-minus-oracle difference depends only on each fold's trained nuisance functions. Its conditional $L^2(\mathbb P)$ norm tends to zero by the same score bounds. Under scheme (b), its conditional squared expectation is a mixture of its two group expectations, both tending to zero by the group-norm bound above. Foldwise conditional Markov inequalities under either scheme therefore give
$N^{-1}\sum_i(\widehat\varphi_{j,i}-\varphi_j(W_i))^2=o_p(1)$.
To see the second-moment step explicitly, let $\|f\|_N^2=N^{-1}\sum_i f_i^2$. Cauchy--Schwarz gives
\[
\bigl|\mathbb P_N\widehat\varphi_j^2-\mathbb P_N\varphi_j^2\bigr|
\leq\|\widehat\varphi_j-\varphi_j\|_N
 \{\|\widehat\varphi_j-\varphi_j\|_N+2\|\varphi_j\|_N\}=o_p(1).
\]
The oracle law of large numbers makes $\|\varphi_j\|_N=O_p(1)$ and $\mathbb P_N\varphi_j^2\to_p\E[\varphi_j^2]$. The same empirical-norm bound controls the fitted mean, which converges to zero, proving consistency of the centred sample variance. For fixed $R$, all repetitions have the same oracle term in~\eqref{eq:collapsed-alr} or~\eqref{eq:dr-alr}, as appropriate. The largest of their finitely many remainders is $o_p(N^{-1/2})$, so both their mean and their median have that expansion. Averaging the fitted influence values preserves their empirical $L^2$ convergence and hence variance consistency. For the median convention, $\max_r N(\widehat\theta_{j,r}-\widetilde\theta_j)^2=o_p(1)$ and every $N\widehat s_{j,r}^2\to_p\sigma_j^2$. Their finite median therefore has the same limit, proving~\eqref{eq:median-variance}.
\end{proof}

The upper propensity bound is stronger than the control-support condition needed to identify the ATT. If a propensity fit is consistent at the stated rate, truncating it above at $1-c$, with $c<\varepsilon$, preserves that rate because truncation is a contraction that fixes the true propensity. Adding a fixed lower clip is harmless for this argument only when that threshold does not exceed the true propensity almost surely. No claim is needed that every sample prediction eventually avoids the thresholds. If clipping changes the population limit to $p_c\ne p$, Proposition~\ref{prop:dr-inference} does not apply: equation~\eqref{eq:clipped-dr-root} gives the changed DR root under outcome misspecification, and even a correct outcome limit need not preserve outcome-nuisance orthogonality at $p_c$. The bound is fixed as $N$ increases, so this is not a result for shrinking overlap.

Each result assumes that both relevant nuisances converge to their true conditional means. For example, if the propensity is known but the overlap outcome fit converges in $L^2$ to $b\ne\ell_H$, the target remains $\tau_{\mathrm{ov}}$ but its influence function is
\[
\varphi_{\mathrm{ov},b}
=\frac{(G-p)\{H-b(Z)-\tau_{\mathrm{ov}}(G-p)\}}{J_G}
=\varphi_{\mathrm{ov}}+\frac{(G-p)(\ell_H-b)}{J_G}.
\]
This follows by the same ratio expansion; the extra term has mean zero and generally changes the variance. With an estimated propensity, expansion at $b$ also includes $\E[(\widehat p-p)(b-\ell_H)]$, which is first order in propensity error. Likewise, population double robustness of DR does not by itself justify the displayed intersection-model influence function when one nuisance has a misspecified limit.

\subsection{Proof of Proposition~\ref{prop:finite-basis-inference}}
The projection normal equations give $\E\langle Qv\rangle=0$ and $\E\langle Q(Y^w-Q'\gamma_\ell)\rangle=0$. Hence both nuisance derivatives of the expected score $\E\langle v(Y^w-Q'\gamma_\ell-\theta v)\rangle$ vanish. The fixed-dimensional training Gram matrices converge to $\Gamma$; the fourth moments and either admissible fold scheme give $O_p(N^{-1/2})$ coefficient errors. For stratified folds, group counts change the training mixture by $O_p(N^{-1/2})$, preserving this rate. Expanding the held-out score gives quadratic coefficient-error terms of order $O_p(N^{-1})$ and products of coefficient errors with centred empirical derivatives of the same order. Its leading term is therefore the oracle bank-score average. The denominator converges to $J_B$, and the i.i.d. bank central limit theorem yields equation~\eqref{eq:finite-basis-if}. Fourth moments and consistent fitted coefficients also give empirical $L^2$ convergence of the fitted influence values, proving variance consistency. For a fixed number of repetitions, the leading term is common to all partitions; the median and paired arguments used above apply unchanged.

\subsection{Normalised DR calculation and scaled influence-function dispersion}\label{app:score-dispersion}
For equation~\eqref{eq:normalized}, write $\bar r_1=\sum G_ir_i/\sum G_i$, $\bar r_0^w=\sum w_ir_i/\sum w_i$, $\widehat B_1=N^{-1}\sum G_i$, and $\widehat B_0=N^{-1}\sum w_i$. The quantities $\bar r_1$ and $\bar r_0^w$ are group residual means, distinct from the conditional control regression $\mu_0(Z)$. Holding the fitted nuisances fixed, the fitted influence value obtained by differentiating the two empirical ratios is
\begin{equation}
\widehat\varphi_i=\frac{G_i}{\widehat B_1}(r_i-\bar r_1)
-\frac{w_i}{\widehat B_0}(r_i-\bar r_0^w).\label{eq:normalizedif}
\end{equation}
Its sample mean is zero. The within-partition dispersion is the sample standard deviation of $\widehat\varphi_i$ divided by $\sqrt N$. For the overlap estimator, with $v_i=G_i-\widehat p_i$ and $u_i=H_i-\widehat\ell_i-\widehat\theta v_i$, the fitted influence value is $v_i u_i/(N^{-1}\sum_jv_j^2)$. Table~\ref{tab:covered-uncertainty} applies the median aggregation described in Section~\ref{app:aggregation}; paired comparisons subtract each bank's influence contributions before aggregation. Sampling interpretation requires the joint versions of the conditions in Theorem~\ref{thm:collapsed-inference} or Proposition~\ref{prop:dr-inference}. Numerical score reconstruction alone establishes neither these conditions nor finite-sample coverage.

At the true nuisances with clipping inactive, equation~\eqref{eq:normalizedif} gives the ATT influence function under the identification conditions. Binding population clipping can remove outcome-nuisance orthogonality, and parametric first-stage estimation can contribute additional terms under misspecification \citep{santanna2020}. Observed clipping frequencies do not settle the asymptotic issue. Complete-bank sampling, support, finite score variance and control of nuisance remainders remain necessary. Independent-bank inference also requires justification for any common preprocessing and cross-bank dependence.

For the baseline OLS/logit pair, the median within-partition influence dispersion is 0.0043, compared with a standard deviation of 0.0162 across partition estimates. The corresponding forest figures are 0.0039 and 0.0012. These measure different sources of variation: a control nearly determining its own weighted mean can have a small centred influence residual while the estimate is sensitive to its fitted residual. Table~\ref{tab:mean-score-dispersion} retains the OLS/logit mean--median discrepancy, and the lemma below describes the centring effect for an oracle control component.

\subsection{Weight concentration and a centred variance component}\label{app:weight-centring}
The next elementary calculation describes the control component of a normalised oracle score conditional on the realised covariates and groups. It separates the variance of a weighted mean from the expectation of its centred plug-in estimate.

\begin{lemma}[Centring with concentrated control weights]\label{lem:weight-centring}
Condition on $\mathcal F=\sigma\{(Z_i,G_i):i=1,\ldots,N\}$. Over the controls $\mathcal I_0=\{i:G_i=0\}$, let the nonnegative $\mathcal F$-measurable weights $a_i$ sum to one. Suppose the residuals $\varepsilon_i$ are conditionally independent, with $\E[\varepsilon_i\mid\mathcal F]=0$ and finite conditional variances $\sigma_i^2$. Define
\[
\bar\varepsilon_a=\sum_{i\in\mathcal I_0}a_i\varepsilon_i,
\qquad S_k=\sum_{i\in\mathcal I_0}a_i^k,
\qquad V_0=\sum_{i\in\mathcal I_0}a_i^2\sigma_i^2,
\qquad C_0=\sum_{i\in\mathcal I_0}a_i^2(\varepsilon_i-\bar\varepsilon_a)^2.
\]
Then $\operatorname{Var}(\bar\varepsilon_a\mid\mathcal F)=V_0$ and
\begin{equation}
\E[C_0\mid\mathcal F]
=(1+S_2)V_0-2\sum_{i\in\mathcal I_0}a_i^3\sigma_i^2.
\label{eq:weight-centring-hetero}
\end{equation}
If $\sigma_i^2=\sigma^2>0$ for every control, then
\begin{equation}
V_0=\frac{\sigma^2}{\mathrm{ESS}_0},\qquad
\frac{\E[C_0\mid\mathcal F]}{V_0}
=1+S_2-\frac{2S_3}{S_2}\in[0,1-S_2],
\qquad \mathrm{ESS}_0=S_2^{-1}.
\label{eq:weight-centring-homo}
\end{equation}
Equal weights on $n_0$ controls give the ratio $1-1/n_0$. As one weight tends to one, the ratio tends to zero while $V_0\to\sigma^2$.
\end{lemma}

\begin{proof}
Conditional independence gives $\operatorname{Var}(\bar\varepsilon_a\mid\mathcal F)=V_0$ and $\operatorname{Cov}(\varepsilon_i,\bar\varepsilon_a\mid\mathcal F)=a_i\sigma_i^2$. Expanding each squared centred residual therefore gives
\[
\E[C_0\mid\mathcal F]
=\sum_i a_i^2(\sigma_i^2+V_0-2a_i\sigma_i^2),
\]
which proves~\eqref{eq:weight-centring-hetero}. Under homoskedasticity, substitution yields~\eqref{eq:weight-centring-homo}. Nonnegativity follows from $C_0\geq0$, while Cauchy--Schwarz gives $S_3\geq S_2^2$ and hence the upper bound. Equal weights and a dominant weight give the stated limits.
\end{proof}

For oracle DR, the weights are normalised propensity odds among controls and $\varepsilon_i=H_i-\mu_0(Z_i)$ on controls. If the estimated sampling variance is computed as the sample variance of the fitted influence values divided by $N$, its control component is $N C_0/(N-1)$; the same multiplier must be applied when comparing its conditional expectation with $V_0$. With the oracle residuals available, the uncentred quantity $U_0=\sum_i a_i^2\varepsilon_i^2$ satisfies $\E[U_0\mid\mathcal F]=V_0$ and provides a diagnostic benchmark.

Effective sample size determines $V_0$ under homoskedasticity, but the third weight moment also enters the centring discrepancy. Under heteroskedasticity, the discrepancy in~\eqref{eq:weight-centring-hetero} can have either sign. The lemma concerns a conditional control component with fixed oracle weights; it neither gives the full random-design DR variance nor covers nuisance estimation or establishes a confidence-interval correction. Supplementary Section~\ref{app:collapsed-inference-calibration} evaluates this component using the saved oracle simulations.

\section{Data provenance and replication status}\label{app:status}
The empirical input is the analysis-ready research file supplied by Dimitris K. Chronopoulos and Muhammed H. Yilmaz, coauthors of \citet{chronopoulos2023}. The supplied file, \texttt{given.csv}, contains 1,911 bank-quarter records. Removing 2015Q1 and filling one missing liquidity ratio in 2020Q1 with the analysis-sample median gives the balanced 1,820-observation panel \texttt{panel\_base.csv}. All 24 analysis fields are reconciled to the supplied file, and the additional unit and time means reproduce numerically. The baseline controls are the supplied winsorised deposit-funding, provisions, operating-efficiency, dividend-payout, derivatives and liquidity ratios. They enter contemporaneously in the benchmark and as pre-period averages in the bank-level comparisons.

The supplied outcome equals raw quarterly log-RWA growth clipped at $-0.048185002$ and $0.27956116$. These endpoints equal the nearest-rank first and ninety-ninth percentiles among the 1,820 nonmissing raw changes. Clipping changes 36 observations: eighteen at each tail, twelve in post-enactment quarters and five among treated banks after enactment. A nineteenth observation at each endpoint is already equal to that endpoint. This exactly reproduces the supplied transformation using pooled-sample cutoffs. It does not establish the original preprocessing sequence or recover the executable construction code, extraction vintage, or treatment of missing values upstream. All estimates remain conditional on that supplied research input.

The available files inherit the 91-bank sample and do not identify banks absent because of mergers, reorganisations or the original selection rules. Earlier asset classification within these banks therefore cannot reconstruct a complete population selected on earlier assets. The foreign-activity extension is also conditional on the supplied items: BHCK0397, BHCK1296, BHCK4059 and BHCK4172 are FR Y-9C fields, whereas BHCP2793 is an FR Y-9LP item \citep{fedy9c2016,fedy9lp2015}. The ratios are each divided by total assets and are not separately winsorised in that extension. Missing reports are distinct from zero positions; reporting coverage and any conversion of cumulative income items to quarterly flows remain construction qualifications.

The linear R formulas have been independently estimated in Python, including finite-sample-corrected bank-clustered covariance. Adjacent differences rebuilt on identical bank-quarter keys reproduce the original FD coefficients and standard errors; no differences are taken across gaps. Retained nuisance predictions reconstruct the reported scores, and a fresh fitting run reproduces the retained empirical predictions in the pinned environment. These checks verify the downstream implementation. They do not reconstruct raw regulatory filings or verify conditional parallel trends, independent-bank sampling, or the nuisance convergence rates required for inference.

The replication package contains the detailed field reconciliation, input and script hashes, environment specification, fold assignments, prediction records and execution logs. Requests for access to the original research data should be directed to its providers; permission to redistribute row-level research data has not been confirmed. The manuscript-source package compiles without those data.

\subsection{Consolidated learner settings}\label{app:learner-settings}
Table~\ref{tab:learner-settings} collects the settings of the active comparisons. All outer folds contain complete banks. Forests use bootstrap sampling and all available features at each split; the heterogeneous design has only one feature. A row-level forest bootstraps training rows even when validation holds out complete banks. The seed rules and preprocessing scope below remain part of the specification. LPM denotes a linear probability model.

\begingroup\footnotesize\setlength{\tabcolsep}{4pt}
\begin{longtable}{@{}>{\raggedright\arraybackslash}p{.19\linewidth}>{\raggedright\arraybackslash}p{.36\linewidth}>{\raggedright\arraybackslash}p{.38\linewidth}@{}}
\caption{Learning units, settings and validation rules}\label{tab:learner-settings}\\
\toprule Analysis & Training unit and features & Learners, folds and tuning\\\midrule
\endfirsthead
\toprule Analysis & Training unit and features & Learners, folds and tuning\\\midrule
\endhead
Panel transformations and within-bank adjustment (Table~\ref{tab:new-panel}) & Bank-quarter rows; contemporaneous controls and the means specified by each representation. Demeaning and means use the full panel. The extension with covariate quarter means adds only covariate time means, omitting the treatment time mean. & OLS or 100-tree forests, minimum leaf five rows. Treatment forest is a regressor after demeaning and a classifier on binary responses. Five stratified bank folds, five partitions; no tuning. Seeds 20260961--20260965, nuisance seed equal to partition seed plus $100k$.\\[4pt]
Matched and augmented scores for pre/post changes (Tables~\ref{tab:covered-uncertainty} and~\ref{tab:response-check}) & One bank-level change; six pre-period covariate means, with prior growth added in the augmented fit. The response comparison changes only the outcome response under the same learner settings. Assignment and overlap outcome use all training banks; DR outcome uses training controls. & Training-standardised OLS/$L_2$ logit ($C=1$, tolerance $10^{-10}$), or 200-tree forests, leaf five banks. Five stratified bank folds, ten partitions, seeds 20260905--20260914; nuisance seed equal to partition seed plus $1000k$. No tuning.\\[4pt]
Regularised and boosted bank-level fits (Table~\ref{tab:learner-comparison}) & Same bank responses and six/seven inputs as the matched pre/post comparison. LASSO expands them to 33/42 terms; boosting uses the raw inputs. & Same ten saved outer partitions. Three inner bank folds tune each nuisance by Brier loss or outcome MSE; control outcome uses controls only. Training-only scaling for LASSO. Section~\ref{app:learner-comparison} gives the grids, objectives and tie rules.\\[4pt]
Paired simulation, row forest (Section~\ref{sec:simulation:swaps}) & Rows on $(Z_i,t)$, separate treatment and outcome fits. & 50 trees, leaf five rows. Five shuffled bank folds with seed $r$; forest seed $1000r+k$. No tuning. Draw seeds $[20260906,N,\sigma_a,r]$.\\[4pt]
Paired simulation, bank-CV spline (Section~\ref{sec:simulation:swaps}) & One assignment and separate pre/post mean-outcome responses per training bank; calendar adjustment from training-bank period deviations. & Cubic splines, five knots, linear extrapolation, training scaling. Ridge penalty selected from $\{0.01,0.1,1,10,100,1000\}$ on three inner bank folds by mean squared prediction error for assignment and outcomes. Five outer bank folds.\\[4pt]
Assignment-unit comparison (Section~\ref{app:simulation-calibration}) & Row treatment fits on $(Z_i,t)$; bank assignment fit on $Z_i$, followed by $P_t\widehat p_i$. Row outcome fit held fixed for fitted-treatment rows. & 50 trees; row leaves 5, 20 or 100; bank leaf five. Five bank folds from a shuffled permutation. Forest seed $1000r+k$; fresh draw seeds $[20260917,N,\sigma_a,r]$. Oracle uses both true nuisances.\\[4pt]
Heterogeneous-effects simulation (Table~\ref{tab:heterogeneous-sim}) & One change per bank, feature $Z_i$; all-bank and control-only outcome fits according to the score. & OLS/logit ($C=10^6$, approximately unpenalised) or 200-tree forests, leaf five banks; five bank folds, no tuning. Draw seeds $[20260917,N,r]$; forest seed $1000r+k$. DR clips the propensity to $[0.01,0.99]$.\\[4pt]
Pre/post inference calibration (Table~\ref{tab:collapsed-inference-calibration}) & One change per bank, $N=91$; six independent uniform covariates. Assignment and overlap outcome use all training banks; DR outcome uses training controls. & 200-tree forests, leaf five banks, all six features. Ten treatment-stratified five-fold partitions; no tuning. Draw seeds $[2026092007,91,r]$; partition seed $s$ generated from $[2026092007,r,j,314159]$; forest seed $(s+1000k)\bmod (2^{32}-1)$. DR clips the propensity to $[0.01,0.99]$.\\
\bottomrule
\end{longtable}
\endgroup
Here $k$ is the zero-based fold index, $r$ the zero-based simulation repetition and $j$ the zero-based partition index. Seed lists specify the inputs to NumPy's \texttt{SeedSequence}; the calibration partition seed is its first generated 32-bit integer. The source code and pinned numerical environment specify remaining software settings. Clipping in the pre/post empirical DR comparisons uses $c=0.01$ or $0.05$, with an unclipped forest reference; the overlap score uses unmodified assignment predictions.

\section{Supplementary event-study coefficients}\label{app:legacy}
The complete event-study coefficients below correspond to Figure~\ref{fig:event}; intervals are pointwise with bank-clustered $t_{90}$ critical values under the maintained linear-regression sampling assumptions.

\begin{table}[htbp]
\centering\footnotesize\setlength{\tabcolsep}{4pt}
\caption{Event-study coefficients and pointwise intervals}\label{tab:event-full}
\begin{tabular}{@{}rrrr@{}}
\toprule
Event quarter & Coefficient & Cluster SE & 95\% pointwise interval\\\midrule
-12 & -0.01814 & 0.01172 & [-0.04142, 0.00514]\\
-11 & 0.00180 & 0.01711 & [-0.03219, 0.03579]\\
-10 & -0.01373 & 0.01512 & [-0.04377, 0.01631]\\
-9 & -0.00472 & 0.00930 & [-0.02319, 0.01375]\\
-8 & -0.00861 & 0.00982 & [-0.02812, 0.01090]\\
-7 & 0.01299 & 0.01922 & [-0.02519, 0.05116]\\
-6 & 0.00013 & 0.01048 & [-0.02069, 0.02096]\\
-5 & -0.00506 & 0.01172 & [-0.02836, 0.01823]\\
-4 & -0.02305 & 0.01271 & [-0.04830, 0.00220]\\
-3 & -0.00069 & 0.01017 & [-0.02090, 0.01952]\\
-2 & -0.00851 & 0.01346 & [-0.03525, 0.01824]\\
0 & 0.01152 & 0.00960 & [-0.00755, 0.03060]\\
1 & 0.01012 & 0.00866 & [-0.00709, 0.02733]\\
2 & 0.02516 & 0.01026 & [0.00478, 0.04554]\\
3 & 0.01438 & 0.01268 & [-0.01081, 0.03957]\\
4 & -0.00347 & 0.00933 & [-0.02201, 0.01506]\\
5 & 0.00842 & 0.00839 & [-0.00824, 0.02508]\\
6 & 0.01258 & 0.01619 & [-0.01959, 0.04476]\\
7 & 0.03040 & 0.01309 & [0.00439, 0.05641]\\
\bottomrule\end{tabular}
\par\smallskip\begin{minipage}{.97\linewidth}\footnotesize Notes: 91 banks (20 treated), 20 quarters and 1,820 observations; bank and quarter effects and six contemporaneous controls. CR1 denotes the finite-sample-corrected bank-clustered covariance estimator. Omitted quarter: 2018Q1, $k=-1$. The pre-period joint test is $\chi^2(11)=9.675$, $p=0.560$; the corresponding $F(11,90)=0.880$, $p=0.563$. Intervals use $t_{90}$ and are not simultaneous bands or formal bounds on departures from parallel trends.\end{minipage}\end{table}
\FloatBarrier
\section{Descriptive trajectories and outcome scaling}\label{app:descriptive}
\begin{figure}[htbp]
\centering
\includegraphics[width=\linewidth]{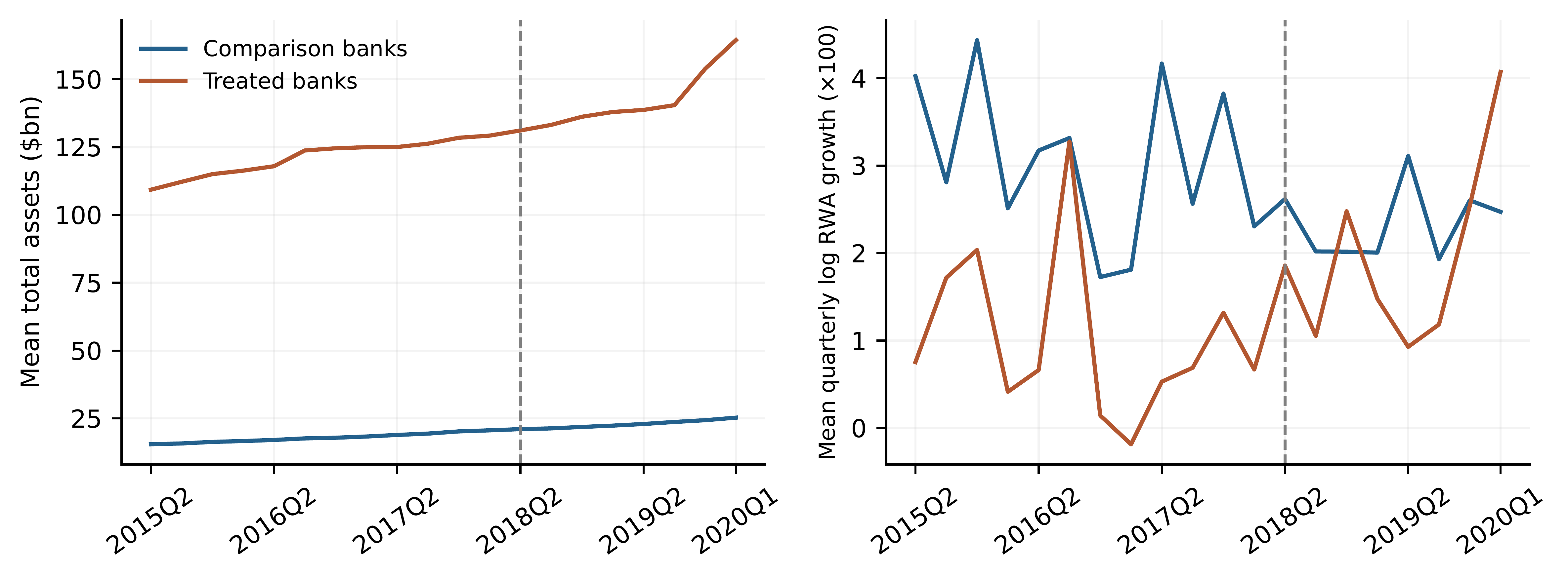}
\caption{Descriptive bank trajectories. Left: unweighted group means of total assets in billions of dollars. Right: group means of the supplied winsorised quarterly log growth in RWA, multiplied by 100. The vertical line marks 2018Q2. The graphs describe the balanced estimation panel and do not adjust for covariates or establish parallel trends.}
\label{fig:descriptive}
\end{figure}

Figure~\ref{fig:descriptive} places the comparison in its economic setting. Treatment groups are defined by different baseline size bands, so differences in asset levels are built into the design. The growth panel displays the outcome actually used in the benchmark regression. Neither similarity in some pre-period movements nor persistent differences in size resolve the counterfactual trend restriction.
\section{Additional simulation diagnostics}\label{app:additional-simulation}
\subsection{Within-bank dependence with known nuisance functions}
There are $N$ independent banks observed for $T=20$ periods, indexed by $t=0,\ldots,19$. Define $P_t=\ind\{t\geq12\}$, so that the final eight periods are after treatment. For each bank, draw a scalar covariate $Z_i\sim N(0,1)$ and a group indicator $G_i$ with conditional probability
\[
 p_i=0.05+0.30\Lambda(Z_i),\qquad
 G_i\mid Z_i\sim\operatorname{Bernoulli}(p_i),
 \qquad \Lambda(z)=\frac{e^z}{1+e^z}.
\]
The treatment indicator is $D_{it}=G_iP_t$. Outcomes follow
\begin{equation}
 Y_{it}=D_{it}+0.3Z_i+0.2Z_i^2+0.1\sin(t)+a_i+\varepsilon_{it},
 \label{eq:simulation-outcome}
\end{equation}
where $a_i\sim N(0,\sigma_a^2)$ is a bank-specific shock and $\varepsilon_{it}\sim N(0,1)$ is an idiosyncratic shock. The shocks are independent of covariates and assignment, and the idiosyncratic shocks are independent across banks and periods. We use $N\in\{91,364\}$ and $\sigma_a\in\{0,1\}$. The coefficient on treatment is one in all four cases.

For the conditioning variables $(Z_i,t)$, the treatment and outcome conditional means are
\[
 m_{0,it}=p_iP_t,
 \qquad
 \ell_{0,it}=m_{0,it}+0.3Z_i+0.2Z_i^2+0.1\sin(t).
\]
These conditioning variables do not determine the realised group indicator. The treatment residual therefore has positive variance in the post-treatment periods. Using the known functions, we calculate the partialling-out estimator
\[
 \widehat\theta
 =\frac{\sum_{i,t}(D_{it}-m_{0,it})(Y_{it}-\ell_{0,it})}
 {\sum_{i,t}(D_{it}-m_{0,it})^2}.
\]
This is an oracle benchmark: no nuisance function is estimated. For each of 1,000 replications per case, we form nominal 95\% intervals using either a bank-clustered variance estimate and a $t_{N-1}$ critical value or a variance estimate treating bank-period observations as independent and a normal critical value.

Writing $V_{it}=D_{it}-m_{0,it}$, $S=\sum_{it}V_{it}^2$ and $\widehat\psi_{it}=V_{it}(Y_{it}-\ell_{0,it}-\widehat\theta V_{it})$, the two variance estimates are
\[
\widehat{\mathrm{se}}_c^{\,2}=\frac{N}{N-1}\frac{\sum_i(\sum_t\widehat\psi_{it})^2}{S^2},\qquad
\widehat{\mathrm{se}}_r^{\,2}=\frac{NT}{NT-1}\frac{\sum_{it}\widehat\psi_{it}^2}{S^2}.
\]

When $\sigma_a=1$, ignoring within-bank dependence gives coverage of 61.5\% at $N=91$ and 65.1\% at $N=364$; the clustered intervals cover in 93.2\% and 94.6\% of replications, respectively. This illustrates the familiar need to account for the sampling unit in panel inference \citep{bertrand2004}. Even with the correct nuisance functions, the clustered interval is not exactly calibrated at $N=91$. The experiment holds the conditional means fixed and varies sampling dependence. Since the bank shock is independent of assignment and covariates, it does not assess the restrictions needed to accommodate correlated bank effects.

\subsection{Joint effects of nuisance-estimation errors}\label{sec:simulation:swaps}
We next estimate both conditional means in the same design, using 400 draws per case and five folds containing complete banks. Every method uses the same draws and folds. A random forest fits each conditional mean to bank-period observations using $(Z_i,t)$, 50 trees and a minimum leaf size of five observations. We call this the row forest to distinguish its training observations from the bank-level responses used by the second procedure.

The second procedure fits a penalised cubic spline to one group indicator per training bank and sets $\widehat m_{it}=P_t\widehat p_i$. It separately fits the pre- and post-treatment mean outcomes and restores a calendar component estimated from training banks. Three inner folds, also formed by bank, select ridge penalties by prediction loss. A third benchmark estimates least-squares coefficients using a basis containing the true functional forms. Its knowledge of those forms makes it an infeasible specification benchmark for most applications. The replication records retain all four cases; Section~\ref{app:learner-settings} gives the estimation rules.

At $N=364$ and $\sigma_a=0$, the row forest has bias $-0.142$ and 18.5\% coverage. The spline has bias $-0.003$ and 94.5\% coverage, compared with 95.5\% coverage for the oracle on the same draws. The spline's structure is well suited to this smooth, low-dimensional design. Its coverage is nevertheless only 90.0\% at $N=91$ without the shared shock, so the comparison does not establish reliable inference for every sample size or a general advantage of one method.

To understand the forest result, replace one fitted conditional mean by its true function while keeping the other fitted predictions unchanged. Define the prediction errors $\delta_{m,it}=\widehat m_{it}-m_{0,it}$ and $\delta_{\ell,it}=\widehat\ell_{it}-\ell_{0,it}$. At the true coefficient of one, the second-order component of the sample estimating equation is
\begin{equation}
 B=\frac{1}{NT}\sum_{i,t}
   \left(\delta_{m,it}\delta_{\ell,it}-\delta_{m,it}^{\,2}\right).
 \label{eq:simulation-remainder}
\end{equation}
The coefficient error contains $B/\widehat J$, where
$\widehat J=(NT)^{-1}\sum_{i,t}(D_{it}-\widehat m_{it})^2$, together with an oracle score and mixed sample terms. Proposition~\ref{prop:risk} gives the corresponding population calculation. This illustrates the established role of joint nuisance-estimation errors in orthogonal estimation \citep{chernozhukov2018}.

Using the true treatment conditional mean reduces the forest's bias from $-0.142$ to $-0.008$. Using only the true outcome conditional mean instead increases its magnitude to $-0.425$. With both functions estimated, the mean error product is 0.0308 and the mean squared treatment error is 0.0458: the former offsets part of the latter. Replacing the outcome fit alone removes that offset. Better estimation of one nuisance function can therefore increase the absolute coefficient error in a finite sample when both were initially estimated poorly. This does not contradict the asymptotic argument for orthogonality, which requires the relevant nuisance errors to become sufficiently small. The exact sample decompositions are verified for every stored replication.

The assignment-unit comparison uses a separate set of 400 draws per case. Within that experiment, the fitted procedures share draws, bank folds and forest outcome predictions, and differ in their treatment fits. Increasing the minimum leaf size from five to one hundred rows, or fitting assignment once per bank, reduces the magnitude of bias. A leaf-size constraint counts training rows; with repeated observations, five rows need not represent five banks. The bank-level forest instead counts banks and imposes $\widehat m_{it}=P_t\widehat p_i$. It therefore changes both the training unit and the treatment specification. At $N=364$ without the shared shock, its bias remains positive at 0.021 (Monte Carlo standard error 0.003), and coverage is 89.2\%. These choices improve some aspects of this design without eliminating finite-sample error.

\subsection{Standard-error calibration}\label{app:simulation-calibration}
The replication file \texttt{simulation\_calibration.csv} places mean bank-clustered standard errors beside Monte Carlo standard deviations across every sample-size and shock cell. In the assignment-fit experiment at $N=364$, $\sigma_a=0$, the mean-SE-to-SD ratios are 0.759 for the baseline row forest, 0.864 for the row forest with leaf size one hundred, and 0.833 for the bank-level forest. The latter two reduce bias but do not fully calibrate the reported uncertainty; the bank-level fit additionally retains positive bias. These diagnostics describe the observed coverage pattern, without decomposing its causes uniquely.

The paired-nuisance and assignment-fit comparisons are separate simulation campaigns. The paired nuisance experiment and its oracle substitutions use seed sequence $[20260906,N,\sigma_a,r]$; its oracle rows match the first 400 draws of the 1,000-draw oracle experiment. The assignment-fit experiment instead uses $[20260917,N,\sigma_a,r]$, with draws and bank folds matched across its own learner variants. Consequently, comparisons between learners within a panel are paired, but comparisons across panels are not.

The supplied forest-leaf records count distinct in-bag banks supporting each held-out post-period tree prediction, not merely rows or bootstrap multiplicities. They cover replication zero only in each cell of the paired nuisance experiment. Summaries weight tree--fold records by the number of held-out post-period rows. For $N=364$, $\sigma_a=0$, the weighted mean support is 7.46 banks, while 10.3\% of these tree predictions use one bank and 26.0\% use at most two. This records a feature of that realised fit; it neither characterises the full distribution over simulation draws nor isolates the effect of bank support from the other changes in the assignment-fit experiment.
\section{Additional empirical comparisons}\label{app:additional-empirical}
\subsection{Original nonlinear panel implementations}
Table~\ref{tab:new-panel} retains the original comparisons. Their time adjustments and nuisance restrictions differ; the matched-basis comparisons in the main text hold those design choices fixed.
\begin{table}[htbp]
\centering\footnotesize\setlength{\tabcolsep}{3pt}
\caption{Panel transformations, time adjustment and nuisance specifications}\label{tab:new-panel}
\begin{tabular}{@{}p{.27\linewidth}p{.15\linewidth}rrrr@{}}
\toprule
Procedure & Time adjustment & OLS & Forest & Partition range & $\widehat R$\\\midrule
Two-way demeaned data & Two-way demeaning & 0.0185 & 0.0181 & [0.0169, 0.0207] & 0.5209\\
Bank means as regressors & None & 0.0178 & 0.0162 & [0.0137, 0.0190] & 0.5425\\
Within-bank adjusted predictions & None & 0.0178 & 0.0130 & [0.0122, 0.0178] & 0.6026\\
Adjusted predictions and covariate quarter means & Covariate quarter means & 0.0188 & 0.0057 & [0.0050, 0.0110] & 0.4562\\
Bank and quarter means as regressors & Treatment and covariate quarter means & 0.0185 & Undefined & -- & 0.0000\\
\bottomrule\end{tabular}
\par\smallskip\begin{minipage}{.98\linewidth}\footnotesize
Notes: Medians and forest ranges use five identical complete-bank partitions, with five folds each; 91 banks and 1,820 observations. Ranges describe partition sensitivity, not sampling uncertainty. Forests use 100 trees and minimum leaf size five bank-quarter observations. Both nuisance regressions use the indicated learner. The demeaned row transforms outcome, treatment and six covariates. The bank-means row adds covariate and treatment bank means to current covariates. The adjusted rows use current covariates and their bank means, followed by the treatment-prediction adjustment in Proposition~\ref{prop:structured}; treatment bank means enter through that adjustment only. The final row supplies current covariates and both bank and quarter means of covariates and treatment. No extra quarter indicators enter the mean-based rows. $\widehat R$ divides the treatment-residual second moment by original binary treatment variance. Means and demeaning use the full sample. Time adjustment and nuisance restrictions vary, so the rows need not have a common causal target.
\end{minipage}\end{table}

\subsection{Detailed benchmark and change-based comparisons}
\begin{table}[htbp]
\centering\small
\caption{Conditional size-group comparison: linear benchmark and sensitivity}\label{tab:linear}
\begin{tabular}{@{}lrrr@{}}
\toprule
Specification & Coefficient & Bank-clustered s.e. & Observations\\\midrule
Baseline & 0.0193 & 0.0045 & 1,820\\
Exclude 2020Q1 & 0.0174 & 0.0044 & 1,729\\
Delete missing liquidity row & 0.0195 & 0.0046 & 1,819\\
Exclude comparison banks: 35--50bn & 0.0214 & 0.0046 & 1,700\\
\bottomrule
\end{tabular}
\par\smallskip\begin{minipage}{.96\linewidth}\footnotesize
Notes: Outcome is winsorised quarterly log growth of risk-weighted assets; all regressions include bank and quarter effects and the six contemporaneous controls. Bank counts are 91 except the last row, which retains 85. The panel covers 20 quarters, or 19 when 2020Q1 is excluded. CR1 denotes the finite-sample-corrected bank-clustered covariance estimator. Standard errors maintain the linear model's sampling assumptions; the coefficient is a conditional size-group comparison. The statutory-timing proxy and the outcome decomposition are in Supplementary Section~\ref{app:timing-outcomes}.
\end{minipage}\end{table}
\begin{table}[htbp]
\centering\footnotesize\setlength{\tabcolsep}{4pt}
\caption{Two outcome responses under the same forest specification and conditioning information $(Z_i,H_i^{\mathrm{pre}})$}\label{tab:response-check}
\begin{tabular}{@{}lrrr@{}}
\toprule
\multicolumn{4}{@{}l}{\emph{(a) Coefficients, forest nuisances}}\\
Score & Learn change & Learn post, subtract pre & Range, post-response fit\\\midrule
Overlap PLR & 0.0015 & 0.0003 & [-0.0016, 0.0012]\\
Normalised DR, $c=0.01$ & 0.0026 & -0.0023 & [-0.0040, -0.0010]\\
Normalised DR, $c=0.05$ & 0.0028 & -0.0022 & [-0.0040, -0.0009]\\
\addlinespace
\multicolumn{4}{@{}l}{\emph{(b) Held-out root mean squared error on the change scale}}\\
Learner & Outcome representation & All-bank regression & Control-only regression\\\midrule
OLS & Either response & 0.0143 & 0.0150\\
Forest & Learn change & 0.0141 & 0.0148\\
Forest & Learn post, subtract pre & 0.0138 & 0.0146\\
\bottomrule\end{tabular}
\par\smallskip\begin{minipage}{.97\linewidth}\footnotesize
Notes: The sample contains 91 banks (20 treated), collapsed from twelve pre- and eight post-enactment quarters. Panel (a) reports medians over ten identical bank partitions with five folds each; both response representations use the same forest specification and the saved propensity predictions, training banks, seeds and settings of the augmented specification and differ only in whether each outcome regression learns $H_i$ directly or learns $H_i^{\mathrm{post}}$ and subtracts the observed $H_i^{\mathrm{pre}}$. For OLS the two predictions coincide to $3.0\times10^{-16}$. Only the DR median changes sign; the overlap median remains positive. Partition ranges describe fitting variation on the same sample, not confidence intervals. The DR label denotes the score whose theoretical target is the ATT under the stated identification conditions. Panel (b) reports median held-out RMSEs against $H_i$, on all 91 held-out banks for the all-bank regression and the 71 held-out controls for the control-only regression. No coefficient is selected from this comparison.
\end{minipage}\end{table}

\subsection{Timing and outcome interpretation}\label{app:timing-outcomes}
The statutory-timing proxy assigns enactment to the five banks below \$100 billion and 2019Q4 to the fifteen larger banks, reflecting the eighteen-month provision. It gives a conditional coefficient of 0.02285 with finite-sample-corrected bank-clustered (CR1) standard error 0.00729. Under that proxy, the fifteen later-switching banks have only 2019Q4 and 2020Q1 as post periods. In a separate single regression splitting enactment exposure by size band, coefficients are 0.0258 (0.0074) for the five banks below \$100 billion and 0.0171 (0.0049) for the other fifteen. Their direct contrast is 0.0086 with CR1 standard error 0.0081, $p=0.29$ on $t_{90}$, and a conventional 95\% interval of $[-0.0075,0.0248]$. These results do not distinguish the bands under that regression's maintained inference assumptions. They neither verify bank-specific implementation dates nor resolve anticipation or direct relief to smaller comparison banks.

Table~\ref{tab:outcomes} applies identical regressors and observations to raw RWA growth, asset growth and risk-density growth. Equation~\eqref{eq:decomp} and regression linearity yield coefficients 0.02403, 0.01678 and 0.00725, whose unrounded asset-growth share is 69.828\%. This decomposes the raw coefficient, not the separately winsorised benchmark of 0.01926. Risk density depends on composition, reporting and regulatory weights and is not a complete measure of economic risk. The decomposition is not a causal mediation analysis.

\begin{table}[htbp]
\centering\small
\caption{Raw RWA growth decomposed into size and risk density}\label{tab:outcomes}
\begin{tabular}{@{}lrrr@{}}
\toprule
Outcome & Coefficient & S.e. & $p$-value\\\midrule
$\Delta\log RWA$ (raw) & 0.02403 & 0.00662 & $<0.001$\\
$\Delta\log A$ (raw) & 0.01678 & 0.00758 & 0.029\\
$\Delta\log(RWA/A)$ (raw) & 0.00725 & 0.00372 & 0.054\\
$\Delta\log(RWA/A)$ (winsorised) & 0.00553 & 0.00333 & 0.100\\
\bottomrule\end{tabular}
\par\smallskip\begin{minipage}{.96\linewidth}\footnotesize
Notes: All rows use 1,820 observations on 91 banks over 20 quarters, identical contemporaneous controls, and bank and quarter effects. Standard errors use the finite-sample-corrected bank-clustered covariance estimator (CR1); $p$-values use $t_{90}$. The first three coefficients obey the exact log identity. Separate winsorisation of the fourth outcome breaks that identity. These are descriptive conditional regressions, and the decomposition is not a causal mediation share.
\end{minipage}\end{table}

The supplied winsorised outcome requires a distinct interpretation from raw growth. A causal transformed-outcome target can be defined by applying the same fixed cutoffs to each potential raw-growth outcome and stating parallel trends for that transformed outcome. The upstream record does not establish the exact cutoff-estimation and preprocessing procedure, so the empirical results remain conditional on the supplied transformed input. Recomputing fold-specific cutoffs would change the analysed outcome rather than merely repair nuisance fitting. The raw-growth comparison above is useful for outcome sensitivity but does not recover the missing preprocessing record.

\subsection{Foreign-activity adjustment and reporting selection}\label{app:foreign}
The five ratios are BHCK0397, BHCK1296, BHCK4059, BHCK4172 and BHCP2793 divided by total assets; Section~\ref{app:status} records their reporting and construction qualifications. All five are observed for 1,486 bank-quarters, or 81.65\% of the sample. There are seven missingness patterns, and reporting status changes for 32 banks. Nonreporting is distinct from a zero foreign position. The full-sample comparison fills missing ratios with zero and adds one all-items-observed indicator, giving twelve inputs with the six baseline controls; it does not use five item-specific missingness indicators. Holding the complete-case sample fixed, adding the foreign ratios moves the coefficient from 0.01927 to 0.02061. The full-sample zero-fill specification gives 0.02026. Explicit bank and quarter effects accommodate the unbalanced complete-case sample.

\subsection{Factorised treatment predictions from full-panel covariates}\label{app:design-full}
A supplementary implementation uses full-period covariate means to predict group membership, imposes the treatment factorisation in Theorem~\ref{thm:design}, and verifies the fitted identity numerically. Its four median coefficients range from 0.00868 to 0.01516. The full-period information prevents interpreting this check as a pre-treatment causal application. For those features $F$, the population risk decomposition is $\E[(G-a(F))^2]=J_F+\E[(a(F)-p_F(F))^2]$, conditional on training, with $p_F(F)=\Pr(G=1\mid F)$ and $J_F=\E[p_F(1-p_F)]$. It involves the same features as the prediction rule; a fitted Brier loss does not bound propensity dispersion conditional on different, pre-treatment features. The linear probability fits also produce predictions outside $[0,1]$. Complete results and loss decompositions are retained in the replication documentation.

\subsection{Implementations of the restricted within and first-difference scores}\label{app:model-linked-fits}
Table~\ref{tab:model-linked-fits} implements two explicit model restrictions. The FD row fits $R_i=Y_{it^*}-Y_{i,t^*-1}$ and $G_i$ on $(X_{it^*},X_{i,t^*-1})$ at $t^*=2018\mathrm{Q}2$, using the five panel partitions. Setting $\widehat m^\Delta_{it}=\mathbf1\{t=t^*\}\widehat a_i$ makes non-adoption treatment residuals zero and implements Corollary~\ref{cor:fd-collapse}. Its median bank-level Brier loss is 0.12324; the all-difference residual second moment is this loss divided by nineteen. Current covariates make this a conditional-model diagnostic, without establishing their treatment invariance or a causal impact effect.

The within rows fit either the six controls or the controls and their squares before bank demeaning, with nineteen centred quarter indicators. Both nuisance projections use this same span, training-only scaling and no intercept. Bank-only demeaning uses each bank's own history; no fitted cross-bank quarter means enter. Adding squares lowers the median from 0.0186 to 0.0154. Training designs have full rank and treatment denominators are positive, but those numerical facts do not establish that either span contains the true transformed outcome nuisance. The matched order-of-transformation comparison is in main-text Table~\ref{tab:matched-basis}.

\begin{table}[htbp]
\centering\small\setlength{\tabcolsep}{3pt}
\caption{Implementations with an explicit score or projection representation}\label{tab:model-linked-fits}
\begin{tabular}{@{}lrrrr@{}}
\toprule
Implementation & Median & Nominal SE & Residual moment & Rank / columns\\\midrule
FD, enactment-quarter fit & 0.0052 & 0.0155 & 0.00649 & --\\
Within, six-term basis & 0.0186 & 0.0046 & 0.04053 & 25 / 25\\
Within, six terms and squares & 0.0154 & 0.0050 & 0.04261 & 31 / 31\\
\bottomrule\end{tabular}
\par\smallskip\begin{minipage}{.98\linewidth}\footnotesize
Notes: Five complete-bank partitions match Table~\ref{tab:new-panel}. The FD row fits group membership and the adoption-quarter growth change on twelve current and lagged controls, using 100-tree forests with minimum leaf size five banks. Treatment predictions are zero outside adoption. The residual moment averages all nineteen differences; its bank-level Brier loss is nineteen times this moment. The score equals the adoption-quarter regression of Corollary~\ref{cor:fd-collapse}. Its nominal standard error is reproduced from Table~\ref{tab:audit-uncertainty}. The within rows evaluate covariates and, where indicated, their squares before bank demeaning. Both include nineteen centred quarter indicators and use OLS without an intercept with training-only scaling. All training designs have full column rank. Their nominal standard errors follow Proposition~\ref{prop:finite-basis-inference} for projection targets; causal interpretation requires additional outcome restrictions. All standard errors omit upstream preprocessing uncertainty. Current covariates may respond to the reform.
\end{minipage}\end{table}

\subsection{Control-outcome regression on the matched features}\label{app:or-att}
The control-outcome regression estimate is $N_1^{-1}\sum_iG_i\{H_i-\widehat\mu_{0,-k(i)}(Z_i)\}$, with $N_1=20$. It uses the DR control regression without propensity weighting. Its OLS and forest medians are 0.0077 and 0.0150. Within a partition, OR minus DR equals the normalised odds-weighted control residual mean. The baseline forest correction has median 0.00288 and range $[0.00169,0.00493]$, indicating underprediction of the outcome change among weighted comparison banks. Its median differs from the difference of the OR and DR medians.

The forest OR median falls to 0.0047 after prior-growth adjustment and to $-0.0017$ when the augmented forest predicts post-period growth before subtracting prior growth. The sensitivity therefore also occurs without propensity odds weights; it does not establish valid control-model extrapolation to treated covariates.

\begin{table}[htbp]
\centering\small\setlength{\tabcolsep}{4pt}
\caption{Group-prediction loss and the outcome-regression-only comparison}\label{tab:propensity-or-diagnostics}
\begin{tabular}{@{}lrrrr@{}}
\toprule
Conditioning and outcome response & Brier loss & OR median & OR range & DR median\\\midrule
\multicolumn{5}{@{}l}{\emph{Forest}}\\
Six means, change & 0.1242 & 0.0150 & [0.0142, 0.0161] & 0.0125\\
Prior growth added, change & 0.1215 & 0.0047 & [0.0037, 0.0058] & 0.0026\\
Prior growth added, translated post & 0.1215 & -0.0017 & [-0.0022, -0.0009] & -0.0023\\
\multicolumn{5}{@{}l}{\emph{OLS/logit}}\\
Six means, change & 0.1245 & 0.0077 & [0.0062, 0.0105] & -0.0048\\
Prior growth added, change & 0.1171 & -0.0029 & [-0.0041, -0.0018] & 0.0065\\
Prior growth added, translated post & 0.1171 & -0.0029 & [-0.0041, -0.0018] & 0.0065\\
\bottomrule\end{tabular}
\par\smallskip\begin{minipage}{.98\linewidth}\footnotesize
Notes: Reconstructed from retained predictions using ten common five-fold bank partitions; no nuisance refitting. Brier loss is $91^{-1}\sum_i(G_i-\widehat p_i)^2$, summarised by its median across partitions. It includes irreducible group variation as well as propensity-estimation error. It is not an estimate of $\|\widehat p-p\|_2^2$ or an attenuation correction. The constant prediction $20/91$ has full-sample loss $0.1715$; this is a descriptive reference, not a cross-fitted competitor. OR averages residuals from the fitted comparison-bank outcome regression over treated banks; DR additionally subtracts the separately normalised control residual mean using odds weights clipped at 0.01 and 0.99. Ranges are over partitions, not confidence intervals. All outcome responses are expressed on the growth-change scale. The last two rows of each learner block share exactly the same propensity predictions.
\end{minipage}\end{table}

\subsection{Pre-enactment growth as an additional conditioning variable}\label{app:pre-growth}
Tables~\ref{tab:covered-uncertainty} and~\ref{tab:mean-score-dispersion} compare adjustment by the six pre-period covariate means $Z_i$ with adjustment by $(Z_i,H_i^{\mathrm{pre}})$, where $H_i^{\mathrm{pre}}$ is average quarterly growth over the twelve pre-enactment quarters. Its group means are 0.01003 for treated banks and 0.03056 for comparison banks. The sole documented liquidity imputation is in 2020Q1 and does not enter these features. Assignment, all-bank outcome and comparison-bank outcome fits are refitted within identical training-bank folds, using matched seeds and hyperparameters. No feature set is selected using the treatment coefficient. The overlap propensity is unmodified; DR clips at 0.01 or 0.05 and separately normalises the two groups.

Let $H_i^{\mathrm{post}}(0)=H_i^{\mathrm{post}}(\mathbf d^0)$ denote average post-enactment growth under the reference exposure path, using the binary-index shorthand of the main text, and assume no anticipation so that observed $H_i^{\mathrm{pre}}$ equals its counterfactual counterpart. Applying this argument to the reform additionally requires the comparison banks to represent that counterfactual, which their direct regulatory relief does not establish. The six-means restriction is
\begin{align*}
\E[H^{\mathrm{post}}(0)-H^{\mathrm{pre}}\mid G=1,Z]
=\E[H^{\mathrm{post}}(0)-H^{\mathrm{pre}}\mid G=0,Z].
\end{align*}
The augmented restriction is
\begin{align*}
&\E[H^{\mathrm{post}}(0)-H^{\mathrm{pre}}\mid G=1,Z,H^{\mathrm{pre}}]\\
&\qquad=\E[H^{\mathrm{post}}(0)-H^{\mathrm{pre}}\mid G=0,Z,H^{\mathrm{pre}}].
\end{align*}
Because the value of $H^{\mathrm{pre}}$ is fixed within an augmented conditioning cell, the second restriction is equivalent to
\[
\E[H^{\mathrm{post}}(0)\mid G=1,Z,H^{\mathrm{pre}}]
=\E[H^{\mathrm{post}}(0)\mid G=0,Z,H^{\mathrm{pre}}].
\]
Neither restriction implies the other without additional outcome-dynamic or distributional assumptions: averaging over different group distributions of prior growth need not preserve equality, and equality averaged over prior growth can conceal conditional differences. Support must be assessed for the information used. Neither conditioning choice establishes comparable size-related growth opportunities or regulatory counterfactuals.

At the population level, each valid ATT identification argument concerns the same treated-bank average effect. The overlap score instead averages conditional effects using the assignment variance associated with its own conditioning set, so its target may change when $H^{\mathrm{pre}}$ is added. For both scores, the fitted changes additionally reflect nuisance estimation, regularisation and finite-sample variation; for ATT, clipping is another distinction from the original population score. The comparison therefore cannot be interpreted as a direct estimate of bias or treatment-effect heterogeneity.

For OLS on a training feature matrix $X_{\mathrm{tr}}$ containing $H_{\mathrm{tr}}^{\mathrm{pre}}$, projection linearity implies
\[
\widehat b_{\Delta}(Z_i,H_i^{\mathrm{pre}})
=\widehat b_{\mathrm{post}}(Z_i,H_i^{\mathrm{pre}})-H_i^{\mathrm{pre}}
\]
at held-out feature rows when the same full-rank design and linear fitting rule are used. The identity applies separately to the all-bank and control-only regressions. It concerns fitted predictions and requires no causal assumption. For arbitrary population conditional means, the analogous identity follows because $H_i^{\mathrm{pre}}$ belongs to the conditioning information. A fitted forest need not satisfy that response-translation identity.

The prediction identity does not select a causal restriction. The connection to DiD and lagged-outcome adjustment in \citet{ding2019bracketing} is useful, but we do not impose their bracketing conditions or read the fitted values as causal bounds. The linear regressions on earlier windows in Section~\ref{app:audit-banking} and the forest comparison in Section~\ref{app:prereform-placebo} examine different procedures and are reported separately.

Prior-growth adjustment gives negative paired changes in every saved forest partition. For DR at clipping 0.01, the median paired change is $-0.0092$, with range $[-0.0111,-0.0070]$, while median control ESS rises from 17.6 to 20.7. The decline therefore does not require worsening weight concentration. Logit remains fragile, with control ESS approximately 1.4 under both feature sets. Partition ranges are fitting-sensitivity summaries; the joint influence calculations in Table~\ref{tab:covered-uncertainty} address a different question and require the stated sampling conditions.

\subsection{Post-period window and forest leaf size}\label{app:prepost-sensitivity}
Table~\ref{tab:prepost-sensitivity} first omits 2020Q1 from the post-period outcome average while retaining all banks, pre-period features and propensity predictions. Refitting the outcome regressions on the same folds makes both augmented medians close to zero. Increasing every forest's minimum leaf size from five to ten banks instead gives larger estimates, but prior-growth adjustment still reduces both scores in every partition. These fixed comparisons change one choice at a time and do not select a preferred causal specification.

\begin{table}[htbp]
\centering\small\setlength{\tabcolsep}{4pt}
\caption{Post-window and minimum-leaf sensitivity of the pre/post forest comparisons}\label{tab:prepost-sensitivity}
\begin{tabular}{@{}lrrrr@{}}
\toprule
Conditioning & \shortstack{Overlap PLR\\median} & \shortstack{Overlap PLR\\range} & DR median & DR range\\\midrule
\multicolumn{5}{@{}l}{\emph{Eight post quarters, leaf 5}}\\
Six means & 0.0107 & [0.0091, 0.0128] & 0.0125 & [0.0096, 0.0131]\\
Add prior growth & 0.0015 & [-0.0001, 0.0026] & 0.0026 & [0.0019, 0.0048]\\
\multicolumn{5}{@{}l}{\emph{Seven post quarters, leaf 5}}\\
Six means & 0.0085 & [0.0077, 0.0109] & 0.0101 & [0.0079, 0.0109]\\
Add prior growth & 0.0000 & [-0.0015, 0.0016] & 0.0003 & [-0.0006, 0.0018]\\
\multicolumn{5}{@{}l}{\emph{Eight post quarters, leaf 10}}\\
Six means & 0.0125 & [0.0107, 0.0145] & 0.0145 & [0.0129, 0.0157]\\
Add prior growth & 0.0020 & [0.0009, 0.0036] & 0.0031 & [0.0016, 0.0047]\\
\bottomrule\end{tabular}
\par\smallskip\begin{minipage}{.98\linewidth}\footnotesize
Notes: Ten common five-fold bank partitions; 91 banks and 200 trees. The first block is the retained baseline. The second excludes 2020Q1 from the post-enactment outcome average, retaining all twelve pre quarters, the same features, and exactly the same assignment predictions; both outcome regressions are refitted on the same training banks. The third retains all eight post quarters and changes the minimum leaf size from five to ten banks for all nuisance forests. Features, partitions, seeds, clipping and other hyperparameters are unchanged. Both scores fit the outcome change directly. DR clips propensities to $[0.01,0.99]$. Changing the post window changes the time average being compared. Each modification was fixed as a sensitivity calculation; the estimates were not used to select learner settings. Ranges describe variation across partitions, not sampling uncertainty.
\end{minipage}\end{table}

\subsection{Matched outcome-forest response comparison}\label{app:response-check}
Table~\ref{tab:response-check} holds the augmented features, assignment predictions, folds and forest settings fixed while changing the outcome response from growth changes to post-period growth followed by subtraction of prior growth. The overlap median changes from 0.0015 to 0.0003, and DR changes from 0.0026 to $-0.0023$ at clipping 0.01. The latter alternative is negative in every partition. OLS satisfies the response-translation identity to numerical precision; separately fitted forests need not.

Both forest procedures represent the same population conditional mean. The translated post-response fit has lower common-change-scale prediction error in nine of ten all-bank partitions and six of ten comparison-bank partitions. Those held-out losses describe fixed procedures. Choosing a procedure would require a prespecified criterion evaluated within the outer training samples. Neither the loss nor the sign establishes the identifying restriction. Section~\ref{app:response-decomposition} gives the exact fitted-score decomposition.

\subsection{Propensity clipping and control-weight concentration}\label{app:clipping}
The saved predictions record crossings of the clipping thresholds. At 0.01, baseline forests have 11--15 comparison-bank crossings and at most one treated crossing; augmented forests have 8--13 comparison-bank crossings and no treated crossing. Neither has predictions above 0.99. Recomputing the normalised DR estimates from the same saved outcome and treatment predictions gives unclipped medians of 0.012479 and 0.002629, compared with clipped medians of 0.012493 and 0.002636. The largest absolute within-partition changes are $2.72\times10^{-5}$ and $2.31\times10^{-5}$, respectively. Clipping at 0.01 therefore has little numerical effect on these estimates and does not explain the prior-growth contraction. These fitted threshold crossings do not establish that the true population propensity lies below 0.01. Conversely, a small observed coefficient change does not establish the nuisance convergence and population clipping conditions required by Proposition~\ref{prop:dr-inference}. Raising the clipping threshold to 0.05 gives forest DR medians 0.0125 before and 0.0028 after prior-growth adjustment, with control ESS 19.2 and 22.4. For OLS/logit the corresponding medians are $-0.0004$ and 0.0025, with ESS approximately 3.6 in both cases. The clipping comparison therefore leaves the qualitative forest contraction intact and does not eliminate logit's weight concentration. Complete threshold-crossing counts and recalculated coefficients are retained with the predictions.

Baseline logit places 77.1--87.9\% of normalised comparison weight on RSSD 1883693. Its median control ESS is 1.40 and remains 1.38 after prior-growth adjustment; forest ESS rises from 17.64 to 20.70. These describe fitted weight concentration. Population support, nuisance convergence and whether clipping binds asymptotically remain separate issues.

\subsection{Covariate ranges and weighted balance}\label{app:covariate-support}
Support must be assessed in the covariates used for adjustment. Table~\ref{tab:covariate-support} reports marginal observed-range checks and standardised mean differences for the six pre-enactment ratios and prior mean growth. Six of the twenty treated banks lie outside at least one baseline marginal range of the 71 comparison banks. The counts by variable are three for deposit funding, three for provisions, one for operating efficiency, zero for dividend payout, one for derivatives and three for liquidity; some banks are counted more than once. All treated values of prior growth lie within its observed comparison range. These statements concern the finite sample and the supplied transformed covariates; inclusion within every observed marginal range neither establishes nor is implied by population overlap in a finite sample.

The balance columns compare the unweighted treated mean with the comparison mean before weighting and with the comparison mean weighted by the normalised control odds. The denominator is the same pooled unweighted within-group standard deviation in every column, so changes are attributable to the weighted mean rather than a changing scale. Forest weighting reduces several baseline differences, but residual differences remain. For prior growth, the unweighted difference is $-1.278$ standard deviations; the median is $-1.336$ under the six-mean forest weights and $-0.880$ when prior growth enters the assignment fit. Logit weights produce a derivatives-ratio difference of approximately $-2.19$ standard deviations under both feature sets, consistent with the concentration documented below. These are descriptive comparisons rather than balance tests or guarantees of counterfactual comparability. Doubly robust estimation does not require exact finite-sample mean balance, and no estimator is selected from this exhibit.

\begin{table}[htbp]
\centering\footnotesize\setlength{\tabcolsep}{4pt}
\caption{Covariate balance and observed comparison-group ranges}\label{tab:covariate-support}
\begin{tabular}{@{}lrrrrrr@{}}
\toprule
& & \multicolumn{2}{c}{Forest weights} & \multicolumn{2}{c}{$L_2$ logit weights} & Outside\\
Covariate & Unweighted & Six means & Augmented & Six means & Augmented & range\\\midrule
Deposit funding & -0.821 & -0.593 & -0.659 & -1.126 & -1.154 & 3/20\\
Provisions & 0.578 & 0.496 & 0.366 & 0.595 & 0.555 & 3/20\\
Operating efficiency & 0.184 & -0.145 & -0.083 & -0.248 & -0.246 & 1/20\\
Dividend payout & -0.516 & 0.013 & -0.053 & -0.465 & -0.450 & 0/20\\
Derivatives & 0.719 & 0.262 & 0.321 & -2.188 & -2.190 & 1/20\\
Liquidity & 0.669 & 0.443 & 0.360 & -0.135 & -0.144 & 3/20\\
Prior mean growth & -1.278 & -1.336 & -0.880 & -0.358 & -0.232 & 0/20\\
\bottomrule
\end{tabular}
\begin{minipage}{.99\linewidth}\scriptsize
Notes: Entries in the five balance columns are treated-minus-comparison mean differences divided by the pooled unweighted within-group sample standard deviation, $\{(s_1^2+s_0^2)/2\}^{1/2}$. The treated mean remains unweighted. Weighted columns use normalised propensity-odds weights for controls, with propensities clipped to $[0.01,0.99]$, and report the median signed difference over the ten complete-bank partitions. Six means uses the six pre-enactment covariate means; augmented adds prior mean growth. Prior growth is displayed for both fits although it is absent from the six-mean feature set. Outside range counts treated banks outside the full comparison sample's marginal minimum--maximum range on that variable. Six distinct treated banks are outside at least one of the six baseline ranges; the counts across rows overlap. All diagnostics are computed on the supplied transformed variables. They describe finite-sample covariate geometry and fitted balance, not a test of population overlap or parallel trends.
\end{minipage}
\end{table}

\subsection{Exact decomposition of the outcome-response comparison}\label{app:response-decomposition}
The comparison in Table~\ref{tab:response-check} can be decomposed without introducing a new target or an uncertainty approximation. Let $a$ denote the change-response fit and $b$ the post-response fit minus observed prior growth. Within one bank partition, write $\Delta\widehat\mu_i=\widehat\mu_{0i}^{b}-\widehat\mu_{0i}^{a}$, $\Delta\widehat\ell_i=\widehat\ell_i^{b}-\widehat\ell_i^{a}$, and $v_i=G_i-\widehat p_i$. With the assignment predictions held fixed, define the normalised propensity-odds weights for controls
\[
 w_i^c=\frac{(1-G_i)\widehat p_i^c/(1-\widehat p_i^c)}
 {\sum_j(1-G_j)\widehat p_j^c/(1-\widehat p_j^c)},
 \qquad \widehat p_i^c=\min\{1-c,\max(c,\widehat p_i)\}.
\]
Then the fitted-score changes satisfy the exact identities
\begin{align}
 \widehat\tau_{\mathrm{DR}}^b-\widehat\tau_{\mathrm{DR}}^a
 &=-\frac{\sum_iG_i\Delta\widehat\mu_i}{\sum_iG_i}
   +\sum_iw_i^c\Delta\widehat\mu_i,\label{eq:response-decomposition-dr}\\
 \widehat\tau_{\mathrm{ov}}^b-\widehat\tau_{\mathrm{ov}}^a
 &=-\frac{\sum_iG_iv_i\Delta\widehat\ell_i}{\sum_i v_i^2}
   -\frac{\sum_i(1-G_i)v_i\Delta\widehat\ell_i}{\sum_i v_i^2}.
 \label{eq:response-decomposition-overlap}
\end{align}
These are algebraic comparisons of fitted scores. They require neither conditional parallel trends nor consistent nuisance estimation.

Table~\ref{tab:response-decomposition} reports the resulting contributions. At $c=0.01$, the control-outcome prediction for treated banks increases by an average of 0.00633 across partitions under the translated post-response forest. Its negative contribution to the DR change is partly offset by the weighted-control contribution of 0.00099, giving a mean score change of $-0.00534$. The paired change is negative in every partition, with median $-0.00564$ and range $[-0.00646,-0.00364]$. Thus the reversal is mainly attributable to changed control-model predictions at treated-bank feature values. This is a statement about the fitted decomposition, not evidence that either prediction is closer to the unobserved reference-path outcome. The overlap change has negative treated and control contributions on average. The decomposition is performed before summarising over partitions: differences of medians need not equal median paired differences, and median components need not add.

\begin{table}[t]
\centering
\caption{Exact decomposition of the outcome-response comparison}\label{tab:response-decomposition}
\small
\begin{tabular}{lrrrr}
\toprule
 & \multicolumn{3}{c}{Mean contribution across partitions} & Paired change\\
Forest score & Treated & Controls & Total & Median [range]\\
\midrule
Overlap PLR & -0.00103 & -0.00042 & -0.00145 & -0.00138 [-0.00217, -0.00071]\\
DR, $c=0.01$ & -0.00633 & 0.00099 & -0.00534 & -0.00564 [-0.00646, -0.00364]\\
DR, $c=0.05$ & -0.00633 & 0.00090 & -0.00543 & -0.00568 [-0.00655, -0.00381]\\
\bottomrule
\end{tabular}
\begin{minipage}{0.98\linewidth}\footnotesize
Notes: DR denotes the normalised doubly robust estimator. Each change is the estimate using post-response predictions minus observed prior growth, less the estimate using change-response predictions. Features, assignment predictions, bank folds and forest seeds are held fixed. Treated and control contributions sum exactly within every partition. Their means sum to the mean total; medians need not be additive. Ranges describe the ten fixed partitions and are not confidence intervals.
\end{minipage}
\end{table}

The replication record retains the predictions, bank contributions and partition summaries needed to reproduce the fitted-score identities and Table~\ref{tab:response-decomposition}. No prediction-based selection rule is applied.

\subsection{Propensity distributions, weight concentration and deletion sensitivity}\label{app:propensity-weight-diagnostics}
\begin{figure}[t]
\centering
\includegraphics[width=\linewidth]{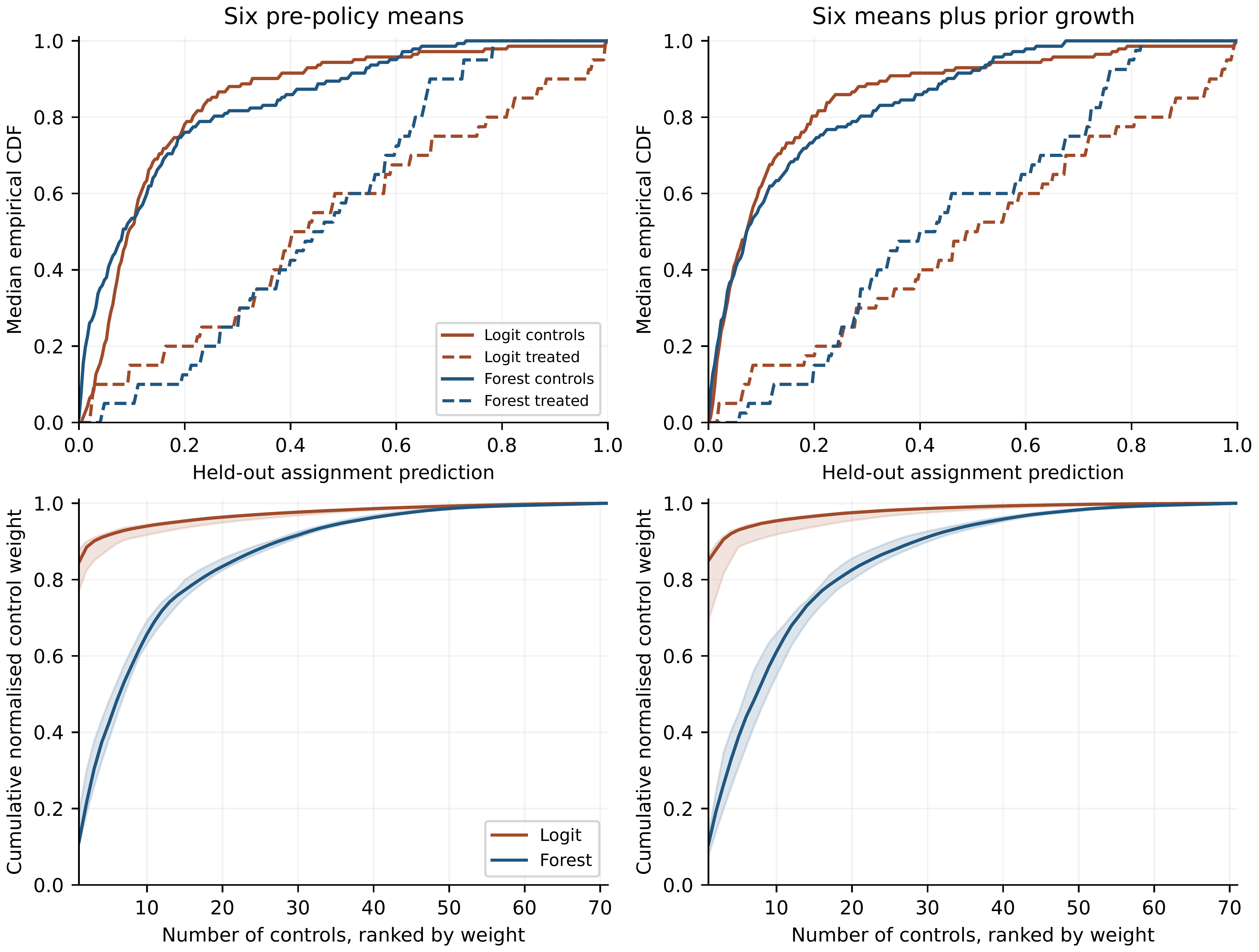}
\caption{Held-out assignment predictions and concentration of control weights. Upper panels: pointwise median empirical distribution functions over ten bank partitions, with treatment status distinguished by line style. Lower panels: median cumulative normalised control odds-weight shares after ranking controls within each partition, with clipping at 0.01 and 0.99. Shaded regions are the full partition ranges, not confidence bands.}\label{fig:propensity-weight-diagnostics}
\end{figure}

Figure~\ref{fig:propensity-weight-diagnostics} displays held-out assignment predictions separately by treatment status, together with the concentration of normalised propensity-odds weights for controls. Each empirical distribution is first computed within a partition; the plotted curve is its pointwise median over the ten partitions. For the lower panels, controls are ranked separately within each partition by their fitted weight before calculating cumulative shares. This preserves the distinction between group separation in predictions and concentration in the weights used by the score. Neither fitted distribution establishes population overlap.

\begin{table}[t]
\centering
\caption{Normalised control-weight concentration and paired regression correction}\label{tab:weight-diagnostics}
\small
\begin{tabular}{llrrrr}
\toprule
Features & Learner & ESS & Largest weight & Top-five share & OR $-$ DR\\
\midrule
Six means & Logit / OLS & 1.4 & 84.5\% & 91.7\% & 0.01395\\
Six means & Forest & 17.6 & 11.4\% & 42.6\% & 0.00288\\
Plus prior growth & Logit / OLS & 1.4 & 85.0\% & 93.0\% & -0.00941\\
Plus prior growth & Forest & 20.7 & 10.7\% & 38.9\% & 0.00184\\
\bottomrule
\end{tabular}
\begin{minipage}{0.98\linewidth}\footnotesize
Notes: Entries are medians across ten bank partitions, with assignment predictions clipped at 0.01 and 0.99. ESS is the inverse sum of squared normalised propensity-odds weights for controls and measures weight concentration; all comparison banks remain in the calculation. OR denotes the control-outcome regression estimator and DR the normalised doubly robust estimator. OR $-$ DR is calculated within each partition from the same control-outcome regression; it equals the weighted control mean residual before taking the median. These are fitted diagnostics, not population-overlap tests or calibrated uncertainty measures.
\end{minipage}
\end{table}

At $c=0.01$, the baseline logit fit places a median 84.5\% of control weight on one bank and 91.7\% on the five most heavily weighted controls; the corresponding forest shares are 11.4\% and 42.6\%. With prior growth added, the shares are 85.0\% and 93.0\% for logit, against 10.7\% and 38.9\% for the forest. RSSD~1883693 is the largest-weight control in every logit partition under both feature sets. Table~\ref{tab:weight-diagnostics} also reports the matched OR-minus-DR correction, calculated within each partition from the common control-outcome fit.

A complementary deletion diagnostic holds all nuisance predictions fixed, removes the largest-weight control from the control residual average, and renormalises the remaining control weights. If $r_i=H_i-\widehat\mu_{0i}$ and $\bar r_w=\sum_iw_i^cr_i$, the resulting score change is $w_j^c(r_j-\bar r_w)/(1-w_j^c)$. For the baseline logit specification its partition range is $[-0.02954,0.01863]$, compared with $[-0.00041,0.00162]$ for the forest. This measures sensitivity of the fitted weighted average. It is not a deletion-and-refitting exercise, an alternative reported estimate, or a correction for lack of overlap.

\subsection{Event-study diagnostics}
The event study retains the benchmark covariates and omits 2018Q1. Pre-period non-rejection is not a test establishing parallel trends. The simultaneous bands account for multiplicity across the nineteen displayed coefficients under the stated bank-clustered approximation.

\subsection{Matched comparisons of basis transformations and first-difference treatment predictions}
\label{sec:matched-specification-diagnostics}

Two comparisons hold the fitting choices fixed while changing the specific
restriction being examined. They isolate numerical consequences of the
implementation within this sample; they do not establish which specification
has a causal interpretation.

\paragraph{Order of basis evaluation and bank demeaning.}
For any panel variable $V_{it}$, define
$(MV)_{it}=V_{it}-T^{-1}\sum_{s=1}^{T}V_{is}$, where $T=20$.
Let $X_{it}$ contain the six benchmark controls and let
$b(x)=(x_1,\ldots,x_6,x_1^2,\ldots,x_6^2)'$.
The first design evaluates the common nonlinear basis in levels and then
demeans each resulting column, giving $Mb(X)$.
The second evaluates the same basis on demeaned covariates, giving $b(MX)$.
The third additionally demeans the resulting basis columns, giving $Mb(MX)$.
The final centring step matters because $(MX_{it,j})^2$ generally has a
nonzero bank mean. Thus the first-to-third comparison preserves bank
centring of the design as well as its dimension while changing the order
in which the nonlinear basis is evaluated.

Every design includes the same nineteen bank-centred quarter indicators.
Outcome and treatment are bank-demeaned in every case. Both nuisance
functions are fitted by OLS without an intercept, using the same complete-bank
folds and partition seeds as the principal panel comparison. Each column is
scaled using its training-sample root mean square; this changes neither its
linear span nor the information in the held-out observations. All training
designs have rank 31. No tuning or choice among the three designs is based
on the resulting treatment coefficient.

Main-text Table~\ref{tab:matched-basis} reports a median of $0.01535$ when the basis is
evaluated in levels and then demeaned, compared with $0.01897$ when it is
evaluated on demeaned covariates and then re-centred. The uncentred
$b(MX)$ version gives $0.01890$. The difference therefore persists after
removing bank means from all design columns. Because the design enters both
nuisance regressions, this comparison does not assign the coefficient
change exclusively to the outcome regression. It shows that the
noncommutation of basis evaluation and demeaning has observable
consequences even with the basis dimension, calendar adjustment, folds,
and fitting method held fixed. These finite-basis fits do not implement
the principal forest procedures.

\paragraph{Predictions outside the adoption quarter.}
Write $R_{it}=\Delta Y_{it}$, $S_{it}=\Delta D_{it}$, and let
$\widehat\ell_{it}$ and $\widehat m_{it}$ denote the saved held-out
outcome and treatment predictions in the first-difference comparison.
For adoption quarter $t^*$, define
\[
 \begin{aligned}
 A_*&=\sum_i(S_{it^*}-\widehat m_{it^*})
                    (R_{it^*}-\widehat\ell_{it^*}),&
 B_*&=\sum_i(S_{it^*}-\widehat m_{it^*})^2,\\
 A_{\mathrm{off}}&=\sum_i\sum_{t\ne t^*}
       (S_{it}-\widehat m_{it})(R_{it}-\widehat\ell_{it}),&
 B_{\mathrm{off}}&=\sum_i\sum_{t\ne t^*}
       (S_{it}-\widehat m_{it})^2.
 \end{aligned}
\]
All time sums here run over the nineteen observed adjacent differences.
Writing $A=A_*+A_{\mathrm{off}}$ and $B=B_*+B_{\mathrm{off}}$,
the original coefficient is $A/B$.
Replacing only the treatment prediction outside adoption by zero,
\[
 \widehat m^{\mathrm{restricted}}_{it}
 =\mathbf 1\{t=t^*\}\widehat m_{it},
\]
gives coefficient $A_*/B_*$, because $S_{it}=0$ outside adoption.
This operation preserves every outcome prediction, every adoption-quarter
treatment prediction, the training sample, folds, and original fitting
settings. It requires no refitting.

Table~\ref{tab:fd-support} shows that the operation leaves both forest
medians unchanged at six decimal places. For the differenced-controls
forest, treatment predictions outside adoption are already exactly zero
in every partition. For the current-and-lagged-controls forest, the maximum
outside-adoption denominator contribution is $0.01125\%$ and the largest
absolute coefficient change across partitions is $0.000074$.
The corresponding OLS medians change from $0.01252$ to $0.01346$ and from
$0.01427$ to $0.01246$. The signed numerator contributions and nonnegative
denominator contributions are retained separately for every partition.

Thus prediction error outside the adoption quarter does not materially
explain the difference between the two principal forest estimates in
these saved fits. In particular, the separate adoption-quarter forest
estimate of $0.0052$ also changes the training sample and the unit to which
the minimum leaf size applies. Its difference from the principal forest
estimates cannot be attributed to zeroing predictions outside adoption
alone. Neither this diagnostic nor the matched basis comparison resolves
the causal identifying assumptions.

\begin{table}[htbp]
\centering\small\setlength{\tabcolsep}{4pt}
\caption{Removing treatment predictions outside the adoption quarter}\label{tab:fd-support}
\begin{tabular}{@{}lrrrr@{}}
\toprule
 & Original & Restricted & \multicolumn{2}{c}{Maximum outside-adoption share (\%)}\\
Covariates, learner & median & median & $|A_{\mathrm{off}}/A|$ & $B_{\mathrm{off}}/B$\\
\midrule
Differences, Forest & 0.018590 & 0.018590 & 0.0000 & 0.0000 \\
Differences, OLS & 0.012524 & 0.013459 & 10.3897 & 0.2136 \\
Current/lagged, Forest & 0.009165 & 0.009165 & 0.6079 & 0.0112 \\
Current/lagged, OLS & 0.014268 & 0.012459 & 23.0111 & 4.3398 \\
\bottomrule
\end{tabular}
\par\smallskip\begin{minipage}{.97\linewidth}\footnotesize
Notes: The original column reproduces the saved first-difference fits. The restricted column retains all outcome predictions and all adoption-quarter treatment predictions and replaces only treatment predictions outside adoption by zero. No model is refitted. $A$ and $B$ are the original score numerator and denominator sums, and the subscript ``off'' denotes contributions from outside adoption. The last two columns report the maximum across the five partitions; the numerator percentage is the absolute value of a signed contribution, not a share of absolute observation-level contributions. Signed numerator and nonnegative denominator contributions, on both sum and average scales, are retained separately for every specification and partition in the replication records. The restricted coefficient need not have a causal interpretation.
\end{minipage}
\end{table}

\paragraph{Substitution of the saved outcome and treatment fits.}
Table~\ref{tab:fd-nuisance-swaps} substitutes the saved outcome and treatment predictions separately. Both substitutions move the coefficient, so the difference between the two principal forest estimates cannot be assigned exclusively to either nuisance fit. This comparison holds the data, quarter indicators, folds and all fitted predictions fixed; it requires no new estimation and does not identify which fit has less causal bias.
\begin{table}[htbp]
\centering\small\setlength{\tabcolsep}{5pt}
\caption{First-difference coefficients after substituting saved forest predictions}\label{tab:fd-nuisance-swaps}
\begin{tabular}{@{}p{.42\linewidth}rr@{}}
\toprule
& \multicolumn{2}{c}{Covariates used for treatment prediction}\\
\cmidrule(l){2-3}
Covariates used for outcome prediction & First differences & Current and lagged\\\midrule
Six first differences & 0.01859 & 0.01560\\
Six current and six lagged values & 0.01240 & 0.00917\\
\bottomrule
\end{tabular}
\par\smallskip\begin{minipage}{.97\linewidth}\footnotesize
Notes: Each cell recomputes the through-origin residual regression with the indicated saved outcome and treatment predictions from Table~\ref{tab:fd-nuisance-comparison}. All predictions are held out by bank, and both feature specifications include the same quarter indicators. The sample, five bank partitions, folds, random seeds and fitted predictions remain fixed; no nuisance model is refitted. Entries are medians of the five resulting coefficients. Diagonal cells reproduce the original forest estimates. The substitutions diagnose the fitted procedures and do not decompose causal effects or establish which specification has less causal bias.
\end{minipage}\end{table}
\subsection{A pre-enactment comparison using the same conditioning choices}\label{app:prereform-placebo}
The twelve pre-enactment quarters permit a descriptive comparison of the same size groups before the coded reform exposure changes. Let $\mathcal T_E$ contain 2015Q2--2016Q3 and $\mathcal T_L$ contain 2016Q4--2018Q1, with six quarters in each set. Define earlier average growth, later average growth, and their difference by
\[
B_i^E=\frac{1}{6}\sum_{t\in\mathcal T_E}Y_{it},\qquad
B_i^L=\frac{1}{6}\sum_{t\in\mathcal T_L}Y_{it},\qquad
K_i=B_i^L-B_i^E.
\]
Here $Y_{it}$ remains quarterly log-RWA growth. The vector $Z_i^E$ contains the six covariate averages over $\mathcal T_E$ only. The group label $G_i$ retains the paper's original 2018Q1 asset classification. It is a retrospective label for this exercise, not an assignment observed at the start of $\mathcal T_E$. The supplied winsorised variables are inherited without reconstructing their original cutoffs; no post-enactment observation enters the block averages or subsequent fitting.

Table~\ref{tab:prereform-placebo} applies the same partialling-out and normalised DR formulas to $K_i$. The first row uses $Z_i^E$ and fits the change directly. The second adds $B_i^E$ to the features. The third retains those augmented features and exactly the same fitted propensities, but fits $B_i^L$ and subtracts each held-out bank's observed $B_i^E$ from its predicted value. This construction is applied separately to the all-bank and comparison-bank outcome regressions. All three rows use the same ten stratified five-fold bank partitions, fold-specific seeds and forest settings as the main pre/post analysis. No specification is selected using these estimates.

The two augmented response constructions represent the same population conditional mean because $B_i^E$ is included in the conditioning information:
\[
E[K_i\mid Z_i^E,B_i^E]
=E[B_i^L\mid Z_i^E,B_i^E]-B_i^E.
\]
The identity also holds conditional on $G_i=0$ for the comparison-bank regression. It need not hold after fitting separate forests, and the corresponding population identity would require $E[B_i^E\mid Z_i^E]$ in place of the observed $B_i^E$ if earlier growth were omitted from the conditioning set.

The unadjusted difference between groups in $K_i$ is $-0.0030$. With the six early covariate means, the partialling-out median is $-0.0057$ and the normalised DR median is $-0.0076$. Adding early growth moves these medians to $-0.0063$ and $-0.0105$. The paired change is negative in six of ten partitions for partialling out and nine of ten for DR; the median paired changes are $-0.0006$ and $-0.0022$, respectively. Median comparison-bank DR ESS rises from 17.3 to 19.1. The corresponding nominal paired standard errors are 0.0041 and 0.0057, with 95\% intervals $[-0.0087,0.0074]$ and $[-0.0133,0.0090]$. The adjustments are small relative to their uncertainty and do not establish a pre-enactment counterpart to the post-enactment contraction.

Fitting later growth under the same augmented information set lowers the partialling-out median further to $-0.0098$ and the DR median to $-0.0148$. Both changes are negative in every paired partition. The comparison-bank outcome-regression median, which does not use propensity odds weights, similarly moves from $-0.0064$ to $-0.0107$ after early-growth adjustment and to $-0.0185$ under the later-growth response. The median paired changes from fitting later growth instead of the change are $-0.0036$ (nominal standard error 0.0025) and $-0.0048$ (0.0030). Both nominal paired intervals include zero. Consistency of signs across partitions is a fitting diagnostic, not independent replication or a sampling test.

These comparisons retain evidence about earlier growth and fitted-response sensitivity, but do not establish that prior-growth adjustment had the same effect before enactment. They neither select a growth-dynamics model nor supply a causal bound. The retrospective group labels and inherited preprocessing distinguish the exercise from a prospective placebo. The table reports nominal sampling uncertainty separately from partition ranges, under the same maintained inference conditions as the post-enactment calculations.

\begin{table}[htbp]
\centering\small\setlength{\tabcolsep}{3.5pt}
\caption{Pre-enactment comparison of earlier and later average growth}\label{tab:prereform-placebo}
\begin{tabular}{@{}lrrrr@{}}
\toprule
Specification & \shortstack{Overlap PLR\\median} & \shortstack{Overlap PLR\\range} & DR median & DR range\\\midrule
Six early covariate means & -0.0057 & [-0.0071, -0.0041] & -0.0076 & [-0.0124, -0.0038]\\
 & (0.0072) & & (0.0077) & \\
Add early growth; fit change & -0.0063 & [-0.0071, -0.0053] & -0.0105 & [-0.0128, -0.0087]\\
 & (0.0055) & & (0.0061) & \\
Add early growth; fit late growth & -0.0098 & [-0.0112, -0.0088] & -0.0148 & [-0.0167, -0.0133]\\
 & (0.0052) & & (0.0052) & \\
\midrule
\multicolumn{5}{l}{Median within-partition changes (nominal standard errors)}\\
Add early growth & -0.0006 & & -0.0022 & \\
 & (0.0041) & & (0.0057) & \\
Fit late growth instead of change & -0.0036 & & -0.0048 & \\
 & (0.0025) & & (0.0030) & \\
\bottomrule\end{tabular}
\par\smallskip\begin{minipage}{.98\linewidth}\footnotesize
Notes: The earlier block is 2015Q2--2016Q3 and the later block is 2016Q4--2018Q1. Both contain six quarters for the same 91 banks; $G_i$ retains the original 2018Q1 size classification. The unadjusted difference between groups in the late-minus-early growth change is -0.0030. The six covariate means and the added growth average use only the earlier block. All rows use ten common five-fold bank partitions, 200 trees, minimum leaf size five banks, and identical fold-specific seeds. The second and third level specifications use identical propensity predictions and features. The third fits later growth and subtracts observed earlier growth from each held-out outcome prediction. DR clips propensities to $[0.01,0.99]$. Supplied winsorised variables are inherited; their original preprocessing is not reconstructed. Ranges describe partition sensitivity. Parentheses give nominal influence-based standard errors using the median aggregation convention in Section~\ref{app:inference-statement}. Paired rows aggregate within-partition differences, not differences of displayed medians. Their interpretation requires the stated joint nuisance and independent-bank conditions and excludes upstream preprocessing uncertainty. These are descriptive comparisons before enactment, not estimates of an earlier reform effect.
\end{minipage}\end{table}
\section{Sample construction and prior-growth diagnostics}\label{app:audit-banking}
\subsection{Asset classification and observation exclusions}
The endpoint classification in Table~\ref{tab:sample-sensitivity} uses assets in 2015Q1 and 2018Q1. Thirty banks change asset band: 27 enter the \$10--50 billion comparison band from below \$10 billion, and three enter the \$50--250 billion band from below \$50 billion. Allowing any crossing during the intervening quarters adds one bank that temporarily exceeds \$50 billion before returning to the comparison band. Excluding all 31 gives a two-way fixed-effects coefficient of 0.0124 (CR1 standard error 0.0046). Excluding only the 27 comparison-bank crossers gives 0.0133 (0.0042); excluding only the three larger-bank crossers gives 0.0176 (0.0049). Thus the influence of classification is concentrated in the composition of the comparison group. These exercises condition on the retained 91 banks; missing entrants, exits and banks outside the original asset range cannot be recovered from this file.

For observation exclusions, bank and quarter effects and all slope coefficients are re-estimated on the retained observations. FD sensitivity calculations pair observations only when both are present and their dates are consecutive. Deleting BB\&T's 2019Q4 observation therefore removes both the 2019Q3--2019Q4 and 2019Q4--2020Q1 differences, leaving 1,727 differences. Its quarter-adjusted FD coefficient is 0.0134 (0.0094), compared with 0.0135 (0.0095) in the full sample. Excluding quarters with $|\Delta\log A|>0.15$ leaves 1,762 level observations and 1,620 consecutive differences; the FD coefficient is 0.0002 (0.0059). These adjacent-change estimates have a different horizon from the level coefficients and cannot be interpreted as alternative estimates of the same average post-enactment effect.

Table~\ref{tab:audit-prepost-samples} extends the prior-growth comparison to selected sample restrictions. Original learner settings and seeds are retained; stratified folds are regenerated after bank exclusions. The direction of the prior-growth adjustment persists. This documents sensitivity within each retained population; comparisons across rows also change that population or the outcome horizon.

\begin{table}[htbp]
\centering\small\setlength{\tabcolsep}{4pt}
\caption{Prior-growth sensitivity across banking samples}\label{tab:audit-prepost-samples}
\begin{tabular}{@{}p{.40\linewidth}rrrr@{}}
\toprule
 & \multicolumn{2}{c}{Overlap PLR} & \multicolumn{2}{c}{DR}\\
Sample & Six means & Add prior growth & Six means & Add prior growth\\\midrule
Original sample & 0.0107 & 0.0015 & 0.0125 & 0.0026 \\
Exclude banks in different endpoint bands & 0.0074 & 0.0000 & 0.0088 & 0.0016 \\
Exclude BB\&T & 0.0090 & -0.0001 & 0.0114 & 0.0014 \\
Exclude banks with post-period $|\Delta\log A|>0.25$ & 0.0065 & -0.0009 & 0.0091 & 0.0012 \\
End post-period average in 2019Q4 & 0.0085 & 0.0000 & 0.0101 & 0.0003 \\
\bottomrule\end{tabular}
\par\smallskip\begin{minipage}{.97\linewidth}\footnotesize
Notes: Medians over ten five-fold bank partitions, using 200-tree forests and minimum leaf size five banks. DR clips propensities at $[0.01,0.99]$. Excluding endpoint-band crossers leaves 61 banks, including seventeen larger banks; excluding BB\&T leaves ninety, including nineteen larger banks; excluding banks with a post-period absolute log asset change above 0.25 leaves 85, including nineteen larger banks. Seeds and learner settings are held fixed, but deleting banks changes the stratified fold assignments. The shortened post period keeps all banks and the original propensity predictions; its outcome regressions are refitted. These specifications change the sample or outcome horizon and do not identify a preferred causal effect.
\end{minipage}\end{table}

\subsection{Earlier windows and the linear coefficient decomposition}
For any window, let $E_i$ and $L_i$ denote its early and late average growth, let $Z_i^E$ contain the six covariate averages formed only over its early block, and let $H_i^E=L_i-E_i$. Table~\ref{tab:prior-growth-linear} compares the coefficient $b_0$ on $G_i$ from regressing $H_i^E$ on $(1,G_i,Z_i^E)$ with the coefficient $b_1$ after adding $E_i$. If $c_E$ is the coefficient on $E_i$ in the augmented regression and $d_G$ is the coefficient on $G_i$ in the regression of $E_i$ on $(1,G_i,Z_i^E)$, the omitted-variable formula gives
\begin{equation}
 b_1-b_0=-c_Ed_G.
 \label{eq:prior-growth-fwl}
\end{equation}
For the post-enactment window, $c_E=-0.7343$ and $d_G=-0.0166$, yielding a coefficient change of $-0.0122$. The same linear decomposition holds in the two earlier windows. These full-sample OLS calculations establish that the qualitative post-enactment contraction is not specific to machine learning. HC3 standard errors in Table~\ref{tab:prior-growth-linear} describe uncertainty for the linear projection coefficients under independent-bank sampling. They are distinct from cross-fitted forest estimates and do not attribute either adjustment to a single mechanism.

\begin{table}[htbp]
\centering\small\setlength{\tabcolsep}{4pt}
\caption{Linear prior-growth adjustment across time windows}\label{tab:prior-growth-linear}
\begin{tabular}{@{}lrrrr@{}}
\toprule
Window & Six means & Add early growth & Change & Early-growth slope\\\midrule
Post-enactment, 12/8 quarters & 0.0110 & -0.0012 & -0.0122 & -0.7343 \\
 & (0.0047) & (0.0040) & (0.0039) & \\
Pre-enactment, 6/6 quarters & -0.0073 & -0.0161 & -0.0088 & -0.7349 \\
 & (0.0080) & (0.0053) & (0.0053) & \\
Before 2017, 4/3 quarters & 0.0110 & -0.0025 & -0.0135 & -0.7707 \\
 & (0.0112) & (0.0101) & (0.0051) & \\
\bottomrule\end{tabular}
\par\smallskip\begin{minipage}{.97\linewidth}\footnotesize
Notes: Coefficients on the fixed size-group indicator from OLS regressions of late minus early average growth, using an intercept and six early-period covariate means. The augmented specification adds early growth. The last column is its coefficient on early growth, conditional on the group indicator and the six means. These are full-sample linear diagnostics, distinct from cross-fitted forest estimates. Parentheses report HC3 standard errors under an independent analysis-ready-bank working model. The change uses a stacked covariance calculation with equation-specific HC3 leverage corrections, retaining covariance between the two group coefficients; equation~\eqref{eq:linear-paired-hc3} gives the convention. Inherited preprocessing uncertainty is excluded. All rows retain 91 banks, including twenty in the larger group. The respective blocks are 2015Q2--2018Q1 versus 2018Q2--2020Q1; 2015Q2--2016Q3 versus 2016Q4--2018Q1; and 2015Q2--2016Q1 versus 2016Q2--2016Q4. These pre-enactment windows can contain regulatory developments or anticipation.
\end{minipage}\end{table}

For the paired linear uncertainty calculation, let $\mathsf X_r$ be the full-rank design matrix for specification $r\in\{0,1\}$, with bank row $x_{ri}'$, and let $\widehat u_{ri}$ and $h_{ri}=x_{ri}'(\mathsf X_r'\mathsf X_r)^{-1}x_{ri}$ be its OLS residual and leverage. Let $e_G$ select the group coefficient. The reported paired standard error uses the stacked estimating equations with an HC3 correction in each equation:
\begin{equation}
c_{ri}=e_G'(\mathsf X_r'\mathsf X_r)^{-1}x_{ri}
\frac{\widehat u_{ri}}{1-h_{ri}},\qquad
\widehat s_{\mathrm{pair}}^2=\sum_i(c_{1i}-c_{0i})^2.
\label{eq:linear-paired-hc3}
\end{equation}
This retains the covariance between estimates on the same banks. Nominal normal intervals for the changes are $[-0.0198,-0.0046]$ after enactment, $[-0.0193,0.0017]$ in the six/six comparison and $[-0.0235,-0.0035]$ in the four/three comparison. The calculation concerns linear projection coefficients under independent-bank sampling; it neither establishes causal identification nor tests a difference between changes across windows.

With identical additional regressors, the coefficient on $E_i$ in a regression of $L_i-E_i$ is one less than its coefficient in a regression of $L_i$. The negative early-growth slopes therefore partly reflect the subtraction in the outcome. Their economic interpretation depends on persistence, transitory variation, measurement error, acquisitions and covariate adjustment. A stationary independent-noise calculation does not account for arbitrary serial dependence in bank growth. The data are consistent with mean reversion but do not uniquely identify it as the explanation for the contraction. We do not impose the restrictions needed for the causal bracketing interpretation in \citet{ding2019bracketing}.

The shorter pre-2017 forest comparison uses 2015Q2--2016Q1 covariate averages and compares those four growth quarters with the three quarters 2016Q2--2016Q4. Its overlap medians are 0.0013 and $-0.0076$ before and after prior-growth conditioning; DR medians are $-0.0009$ and $-0.0103$. The twelve-quarter pre-enactment comparison in Table~\ref{tab:prereform-placebo} instead gives smaller changes under the direct-change forest response. The shorter-window linear coefficient before growth adjustment is 0.0110 with HC3 standard error 0.0112, whereas its overlap forest median is 0.0013. The linear estimate is imprecise; no formal difference between these procedures is established. Window length, fitted functions and initial growth distributions all differ. Neither window is demonstrably free of policy developments or anticipation: the shorter one contains the September 2016 CCAR proposal, while the longer one also includes the January 2017 final rule \citep{fed2016ccar,fed2017ccar}.

\subsection{Two interpretations of prior-growth adjustment}\label{app:prior-growth-interpretation}
Consider an illustrative comparison with early growth $E_i$, untreated later growth $L_i(0)$, and observed later growth $L_i=L_i(0)+\delta G_i$, where $\delta$ is a constant exposure effect. Let $\eta_g=\E[E_i\mid G_i=g]$, let $0\leq\rho<1$ describe persistence between block averages, and suppose the disturbance $\epsilon_i$ has conditional mean zero given $(E_i,G_i)$. Additional covariates are suppressed, and the indicated regression designs are assumed to have full rank. Two untreated-outcome models give different interpretations:
\begin{align}
L_i(0)&=\eta_{G_i}+\rho(E_i-\eta_{G_i})+\epsilon_i,
\label{eq:prior-growth-persistent}\\
L_i(0)&=a+\rho E_i+\epsilon_i,
\label{eq:prior-growth-transitory}
\end{align}
where $a$ is a common intercept. Under~\eqref{eq:prior-growth-persistent}, untreated mean changes are zero in both groups. The unadjusted group difference in $L_i-E_i$ equals $\delta$, whereas adding $E_i$ to a linear regression gives group coefficient $\delta+(1-\rho)(\eta_1-\eta_0)$. With $\eta_1<\eta_0$, the adjustment lowers the coefficient even when unadjusted parallel trends holds. Under~\eqref{eq:prior-growth-transitory}, the adjusted group coefficient equals $\delta$, while the unadjusted difference equals $\delta+(\rho-1)(\eta_1-\eta_0)$ and is upward displaced when $\eta_1<\eta_0$.

These models explain why an observed downward adjustment alone cannot select an identifying restriction. They are illustrative, not fitted structural models of quarterly bank growth. In particular, $1+c_E$ in Table~\ref{tab:prior-growth-linear} is a coefficient for block averages and is not an estimate of quarterly autoregressive persistence. The six/six forest placebo has a DR difference of medians of about $-0.0029$ and a median paired difference of $-0.0022$, rather than the approximately one-percentage-point linear shift. The latter paired difference has nominal standard error 0.0057, so it is not evidence of a common pre- and post-enactment adjustment mechanism. Alternative windows and intervening regulatory developments further limit that interpretation.

\subsection{Raw growth and alternative pre/post windows}\label{app:outcome-timing}
Table~\ref{tab:final-outcome-timing} holds the bank sample, classification, partitions and forest settings fixed. The two supplied-outcome rows with the original pre-period reuse retained predictions. Four additional scenarios freshly fit the outcome and group regressions: raw growth under both post-period endpoints, and the five-quarter pre-period under both endpoints. The shorter pre-period starts in 2017Q1. The supplied winsorised variables are retained without recalculating cutoffs on each window. For raw-outcome comparisons, both early and late growth averages and the prior-growth feature use untrimmed growth; the six control variables remain as supplied. Baseline treatment predictions are unchanged by replacing the outcome because their features are unchanged. Augmented treatment predictions can change because prior growth is itself a feature.

\begin{table}[p]
\centering
\caption{Outcome construction and pre/post windows}
\label{tab:final-outcome-timing}
\small
\begin{tabular}{lrrrr}
\toprule
& \multicolumn{2}{c}{Six pre-period means} & \multicolumn{2}{c}{Plus pre-period growth}\\
Comparison & Overlap PLR & DR & Overlap PLR & DR\\
\midrule
Supplied outcome, full window & 0.0107 & 0.0125 & 0.0015 & 0.0026\\
 & (0.0042) & (0.0040) & (0.0037) & (0.0035)\\
Raw outcome, full window & 0.0145 & 0.0167 & 0.0035 & 0.0050\\
 & (0.0066) & (0.0057) & (0.0063) & (0.0052)\\
Supplied outcome, end 2019Q4 & 0.0085 & 0.0101 & 0.0000 & 0.0003\\
 & (0.0042) & (0.0043) & (0.0038) & (0.0036)\\
Raw outcome, end 2019Q4 & 0.0135 & 0.0156 & 0.0037 & 0.0036\\
 & (0.0069) & (0.0060) & (0.0068) & (0.0054)\\
2017Q1 pre-window, end 2019Q4 & 0.0130 & 0.0189 & -0.0021 & -0.0031\\
 & (0.0070) & (0.0080) & (0.0043) & (0.0043)\\
2017Q1 pre-window, end 2020Q1 & 0.0161 & 0.0213 & -0.0003 & 0.0003\\
 & (0.0066) & (0.0080) & (0.0042) & (0.0041)\\
\midrule
\multicolumn{5}{l}{Paired change after adding pre-period growth}\\
Comparison & \multicolumn{2}{c}{Overlap PLR} & \multicolumn{2}{c}{DR}\\
Supplied outcome, full window & \multicolumn{2}{c}{-0.0103 (0.0040)} & \multicolumn{2}{c}{-0.0092 (0.0040)}\\
Raw outcome, full window & \multicolumn{2}{c}{-0.0120 (0.0043)} & \multicolumn{2}{c}{-0.0112 (0.0043)}\\
Supplied outcome, end 2019Q4 & \multicolumn{2}{c}{-0.0088 (0.0042)} & \multicolumn{2}{c}{-0.0099 (0.0040)}\\
Raw outcome, end 2019Q4 & \multicolumn{2}{c}{-0.0096 (0.0045)} & \multicolumn{2}{c}{-0.0119 (0.0046)}\\
2017Q1 pre-window, end 2019Q4 & \multicolumn{2}{c}{-0.0148 (0.0064)} & \multicolumn{2}{c}{-0.0214 (0.0078)}\\
2017Q1 pre-window, end 2020Q1 & \multicolumn{2}{c}{-0.0159 (0.0064)} & \multicolumn{2}{c}{-0.0210 (0.0078)}\\
\bottomrule
\end{tabular}
\par\vspace{3pt}
\begin{minipage}{0.98\textwidth}\footnotesize
Notes: All rows use the same 91 banks, treatment classification, ten five-fold bank partitions, and 200-tree forests with minimum leaf size five. Entries are medians over partitions; parentheses report the nominal influence-based median standard error defined in equation~\eqref{eq:median-variance}. These calculations omit uncertainty from inherited preprocessing and require the stated sampling, nuisance-rate and overlap conditions; they are not finite-sample coverage guarantees. The doubly robust score clips fitted group probabilities to $[0.01,0.99]$. Full-window rows use 2015Q2--2018Q1 before enactment and 2018Q2--2020Q1 afterwards. End-2019Q4 rows remove 2020Q1. The last two rows use only 2017Q1--2018Q1 before enactment; their six covariate means and prior-growth feature are recomputed from those five quarters. They change conditioning and the comparison window and do not isolate the January 2017 regulatory change. Raw-outcome rows replace supplied winsorised quarterly growth with the supplied untrimmed growth, including in the prior-growth feature; the six supplied control variables and their inherited preprocessing remain unchanged. The lower panel reports medians of paired differences, which need not equal differences of the upper-panel medians.
\end{minipage}
\end{table}

The table reports nominal standard errors and paired changes using the same fixed-repetition aggregation as the main estimates. A change in the pre-period also changes the proposed conditional-trends restriction. Removing 2020Q1 changes the post-period horizon. Consequently these rows answer distinct sensitivity questions and are not alternative estimates of an invariant causal parameter.

\subsection{Paired uncertainty and aggregation}
Table~\ref{tab:audit-uncertainty} retains both nuisance families and both aggregation conventions. For each bank and partition, the fitted influence difference for a prior-growth comparison is the augmented influence value minus its baseline counterpart. Its sample dispersion divided by $\sqrt N$ gives the within-partition paired standard-error calculation. Both estimators must satisfy their own asymptotic conditions jointly for this calculation to have the proposed sampling interpretation.

For overlap, the population contrast is the augmented-conditioning target minus the baseline-conditioning target; these targets can differ under heterogeneous effects. For DR, both specifications identify the same ATT only if both conditional parallel-trends restrictions and the respective nuisance requirements hold. The paired interval therefore does not validate either identifying restriction. Binding clipping can also invalidate the stated orthogonal influence calculation, as discussed in Proposition~\ref{prop:dr-inference}.

\begin{table}[htbp]
\centering\small\setlength{\tabcolsep}{4pt}
\caption{Mean and median aggregation of bank-level uncertainty calculations}\label{tab:audit-uncertainty}
\begin{tabular}{@{}lrrrr@{}}
\toprule
Score and specification & Median & $s_{\mathrm{med}}$ & Mean & $s_{\mathrm{mean}}$\\\midrule
\multicolumn{5}{l}{\textit{Forests}} \\
Overlap PLR, six means & 0.0107 & 0.0042 & 0.0111 & 0.0038 \\
Overlap PLR, plus prior growth & 0.0015 & 0.0037 & 0.0015 & 0.0035 \\
Overlap PLR, paired change & -0.0103 & 0.0040 & -0.0096 & 0.0035 \\
DR, six means & 0.0125 & 0.0040 & 0.0121 & 0.0037 \\
DR, plus prior growth & 0.0026 & 0.0035 & 0.0029 & 0.0032 \\
DR, paired change & -0.0092 & 0.0040 & -0.0092 & 0.0033 \\
\midrule
\multicolumn{5}{l}{\textit{OLS/logit}} \\
Overlap PLR, six means & 0.0084 & 0.0046 & 0.0085 & 0.0044 \\
Overlap PLR, plus prior growth & -0.0003 & 0.0037 & -0.0003 & 0.0035 \\
Overlap PLR, paired change & -0.0087 & 0.0034 & -0.0087 & 0.0031 \\
DR, six means & -0.0048 & 0.0069 & 0.0013 & 0.0038 \\
DR, plus prior growth & 0.0065 & 0.0095 & 0.0105 & 0.0041 \\
DR, paired change & 0.0087 & 0.0067 & 0.0092 & 0.0045 \\
\midrule
Adoption-quarter FD forest & 0.0052 & 0.0155 & 0.0043 & 0.0152 \\
\bottomrule\end{tabular}
\par\smallskip\begin{minipage}{.97\linewidth}\footnotesize
Notes: All quantities use retained held-out predictions for the same 91 banks. Pre/post estimators use ten partitions; the adoption-quarter FD estimator uses five and 100-tree forests. Median aggregation follows Table~\ref{tab:covered-uncertainty}. For mean aggregation, $s_{\mathrm{mean}}$ is the sample standard deviation across banks of the mean fitted influence value across partitions, divided by $\sqrt{91}$. It retains dependence across partitions. Paired changes are augmented minus baseline estimates using the same banks and folds. OLS/logit DR has median control effective sample size approximately 1.4 in both specifications; these rows retain the resulting instability. No aggregation rule corrects weak overlap, nuisance bias or failure of the joint inference conditions.
\end{minipage}\end{table}

The OLS/logit DR mean--median differences persist under these calculations. Control weights are concentrated throughout the partitions, with median effective sample size approximately 1.4. The summaries do not distinguish variation in those weights from variation in the fitted residuals, and partition aggregation does not remove this fragility. All underlying partition estimates, sample definitions and numerical records are retained with the replication code.
\section{Scores for pre/post changes under heterogeneous effects}\label{app:heterogeneous-sim}
To complement Example~\ref{ex:overlap-att}, let $Z\sim N(0,1)$, $p(Z)=\Lambda(-1+1.2Z)$, $G\mid Z\sim\mathrm{Bernoulli}\{p(Z)\}$, $\mu_0(Z)=0.3Z+0.2Z^2$ and $\tau(Z)=1+Z$. Generate $H=\mu_0(Z)+G\tau(Z)+\varepsilon$ with independent standard normal $\varepsilon$. The two scores share assignment fits and five bank folds but use all-bank and comparison-bank outcome regressions, respectively. The nuisance specifications are oracle functions, logit and OLS linear in $Z$, or 200-tree forests with minimum leaf size five. DR clips the propensity to $[0.01,0.99]$.
\begin{table}[htbp]
\centering\footnotesize\setlength{\tabcolsep}{4pt}
\caption{Collapsed-score estimators under heterogeneous effects: bias against each reference target}\label{tab:heterogeneous-sim}
\begin{tabular}{@{}rllrrrr@{}}
\toprule $N$ & Nuisances & Score & Mean (MCSE) & Bias vs $\tau_{\mathrm{ov}}$ & Bias vs $\bar\tau$ & RMSE\\\midrule
91 & Oracle & Overlap PLR & 1.341 (0.013) & 0.032 & -0.312 & 0.290\\
91 & Oracle & DR & 1.680 (0.014) & 0.371 & 0.027 & 0.303\\
91 & Logit/OLS & Overlap PLR & 1.339 (0.012) & 0.031 & -0.313 & 0.277\\
91 & Logit/OLS & DR & 1.632 (0.018) & 0.323 & -0.021 & 0.409\\
91 & Forest & Overlap PLR & 1.365 (0.014) & 0.056 & -0.288 & 0.322\\
91 & Forest & DR & 1.658 (0.019) & 0.350 & 0.006 & 0.430\\
\addlinespace
364 & Oracle & Overlap PLR & 1.317 (0.007) & 0.008 & -0.336 & 0.155\\
364 & Oracle & DR & 1.663 (0.007) & 0.354 & 0.010 & 0.166\\
364 & Logit/OLS & Overlap PLR & 1.316 (0.007) & 0.008 & -0.336 & 0.148\\
364 & Logit/OLS & DR & 1.658 (0.008) & 0.349 & 0.006 & 0.186\\
364 & Forest & Overlap PLR & 1.317 (0.008) & 0.009 & -0.335 & 0.170\\
364 & Forest & DR & 1.628 (0.015) & 0.319 & -0.025 & 0.333\\
\bottomrule\end{tabular}
\par\smallskip\begin{minipage}{.97\linewidth}\footnotesize Notes: 500 draws per sample size, one pre/post change per bank. $Z\sim N(0,1)$, $p(Z)=\Lambda(-1+1.2Z)$, $\mu_0(Z)=0.3Z+0.2Z^2$, $\tau(Z)=1+Z$, $H=\mu_0(Z)+G\tau(Z)+\varepsilon$ with standard normal $\varepsilon$; Assumptions~\ref{ass:noanticipation} to \ref{ass:support} hold exactly. Population targets, evaluated by deterministic quadrature: $\bar\tau=\E[\tau(Z)\mid G=1]\approx1.653$ and $\tau_{\mathrm{ov}}=\E[p(1-p)\tau]/\E[p(1-p)]\approx1.309$. Bias and RMSE use the quadrature values before rounding. The RMSE reference is $\tau_{\mathrm{ov}}$ for Overlap PLR and $\bar\tau$ for DR. Nuisances are cross-fitted on five bank folds: logit and OLS linear in $Z$ (the outcome regressions are then misspecified), or forests with 200 trees and minimum leaf size five. The DR score clips propensities to $[0.01,0.99]$. MCSE denotes Monte Carlo standard error.\end{minipage}
\end{table}

Over 500 replications per sample size, Table~\ref{tab:heterogeneous-sim} evaluates both estimators against both targets. Deterministic quadrature gives $\bar\tau=1.6525764974$ and $\tau_{\mathrm{ov}}=1.3085977274$. At $N=91$, the oracle overlap mean exceeds its target by 0.0319, or 2.48 Monte Carlo standard errors; the oracle DR discrepancy is 0.0273, or 2.02 Monte Carlo standard errors. Known nuisance functions therefore do not eliminate finite-sample discrepancies for these ratio estimators. The discrepancies diminish at $N=364$: the overlap means are within 0.01 of their target and the DR means within 0.03 of the ATT, while cross-target discrepancies are approximately 0.34. Forest DR has the largest RMSE: 0.43 at $N=91$ and 0.33 at $N=364$.

The linear outcome regressions are misspecified. At the true propensity, overlap still targets $\tau_{\mathrm{ov}}$, but propensity-direction orthogonality generally fails at a misspecified outcome limit. Unmodified correctly specified logit would identify the DR ATT despite that misspecification. Fixed clipping changes this exact statement because the generating propensity has positive probability outside $[0.01,0.99]$; equation~\eqref{eq:clipped-dr-root} gives the discrepancy. A correct control-outcome regression still gives the ATT under clipping. The experiment illustrates target differences and finite-sample behaviour, without establishing consistency of the clipped misspecified case or validating confidence intervals.
\section{Finite-sample assessment of pre/post-score intervals}\label{app:collapsed-inference-calibration}
Each of two designs uses $N=91$ independent banks and 400 replications. The six covariates are independent uniforms on $[-\sqrt3,\sqrt3]$, and
\begin{align*}
\mu_0(Z)&=0.4Z_1+0.2(Z_1^2-1)+0.25Z_2+0.2Z_3Z_4+0.15\sin(Z_5)+0.1Z_6,\\
\tau(Z)&=1+0.6Z_1+0.15Z_2,\qquad H=\mu_0(Z)+G\tau(Z)+\varepsilon,
\end{align*}
where $\varepsilon$ is independently standard normal and $G\mid Z\sim\mathrm{Bernoulli}\{p(Z_1)\}$. Let $p(z)=L+(U-L)\operatorname{expit}(a+bz)$, with $a$ chosen to give $\E[p]=20/91$. The ordinary-overlap design uses $(L,U,b)=(0.05,0.70,1)$; weaker overlap uses $(0.011,0.989,3)$. The true propensity ranges are approximately $[0.081,0.450]$ and $[0.0112,0.883]$, both strictly inside the clipping interval $[0.01,0.99]$. Thus neither design violates population support or imposes binding population clipping. Quadrature gives ATT and overlap targets $(1.2974,1.2140)$ and $(1.6673,1.4784)$, respectively.

Assignment, all-bank outcome and comparison-bank outcome fits use 200-tree forests, minimum leaf size five and all six candidate features. The two scores share assignment fits and ten treatment-stratified five-fold bank partitions. The overlap propensity is unmodified; DR clips at 0.01 and 0.99. For mean aggregation, the variance estimate is the sample variance of each bank's averaged influence contribution divided by $N$. For median aggregation it is
\[
\widehat V_{\mathrm{med}}=\operatorname{median}_{s}\left\{\widehat{\mathrm{se}}_s^{\,2}+(\widehat\theta_s-\operatorname{median}_{r}\widehat\theta_r)^2\right\},
\]
where $\widehat{\mathrm{se}}_s$ is the within-partition influence standard error. The median of an even number of values averages the two central order statistics. All intervals use the standard normal critical value. Oracle scores use the true nuisance functions on the same draws. All 800 draws completed without failed fits or replacement draws.

\begin{table}[tbp]
\centering
\small
\setlength{\tabcolsep}{4pt}
\caption{Finite-sample assessment of pre/post-score intervals}
\label{tab:collapsed-inference-calibration}
\begin{tabular}{llrrrrrr}
\toprule
Estimator & Target & Bias & SD & Mean SE & SE/SD & Coverage (MCSE), \% & Length\\
\midrule
\multicolumn{8}{l}{\textit{Ordinary overlap; $N=91$}}\\
\multicolumn{8}{l}{\footnotesize True propensity: 10th, 50th and 90th percentiles 0.093, 0.192, 0.394}\\
\multicolumn{8}{l}{\footnotesize Mean control ESS: oracle 48.46; forest 35.76}\\
\addlinespace[2pt]
Oracle & $\tau_{\mathrm{ov}}$ & 0.025 & 0.296 & 0.282 & 0.95 & 94.00 (1.19) & 1.104\\
Oracle & ATT & 0.018 & 0.308 & 0.291 & 0.94 & 94.25 (1.16) & 1.140\\
Forest mean & $\tau_{\mathrm{ov}}$ & 0.007 & 0.331 & 0.288 & 0.87 & 90.25 (1.48) & 1.127\\
Forest mean & ATT & 0.005 & 0.340 & 0.337 & 0.99 & 94.25 (1.16) & 1.319\\
Forest median & $\tau_{\mathrm{ov}}$ & 0.007 & 0.330 & 0.303 & 0.92 & 92.25 (1.34) & 1.187\\
Forest median & ATT & 0.005 & 0.341 & 0.355 & 1.04 & 95.50 (1.04) & 1.393\\
\addlinespace
\multicolumn{8}{l}{\textit{Weaker overlap; $N=91$}}\\
\multicolumn{8}{l}{\footnotesize True propensity: 10th, 50th and 90th percentiles 0.012, 0.053, 0.735}\\
\multicolumn{8}{l}{\footnotesize Mean control ESS: oracle 9.42; forest 8.50}\\
\addlinespace[2pt]
Oracle & $\tau_{\mathrm{ov}}$ & 0.020 & 0.373 & 0.356 & 0.95 & 93.50 (1.23) & 1.394\\
Oracle & ATT & 0.001 & 0.430 & 0.365 & 0.85 & 88.75 (1.58) & 1.432\\
Forest mean & $\tau_{\mathrm{ov}}$ & 0.033 & 0.405 & 0.360 & 0.89 & 90.75 (1.45) & 1.411\\
Forest mean & ATT & -0.011 & 0.541 & 0.422 & 0.78 & 87.00 (1.68) & 1.655\\
Forest median & $\tau_{\mathrm{ov}}$ & 0.033 & 0.406 & 0.385 & 0.95 & 92.75 (1.30) & 1.508\\
Forest median & ATT & -0.002 & 0.539 & 0.454 & 0.84 & 89.50 (1.53) & 1.779\\
\addlinespace
\bottomrule
\end{tabular}
\par\vspace{3pt}
\begin{minipage}{0.98\linewidth}\footnotesize
Notes: Each design has 400 Monte Carlo replications. Propensity quantiles pool the 36,400 true bank-level propensities per design. Control ESS is the inverse sum of squared normalised propensity-odds weights for controls, averaged over 400 oracle draws or 4,000 forest draw--partition combinations; these are the weights of the DR score. SD is the empirical standard deviation across replications; Mean SE is the mean of the candidate standard errors. Coverage is reported as a percentage for nominal 95\% normal intervals for the indicated population target; its Monte Carlo standard error, in percentage points, is in parentheses. Length is mean interval length. Forest mean uses bankwise averaged influence contributions over ten complete partitions. Forest median uses the median of the within-partition variance plus the squared point-estimate deviation from the median, as defined in the text. Forest settings and treatment-stratified folds match the application; this experiment does not establish coverage for the observed banking data.
All 800 data draws completed successfully.
\end{minipage}
\end{table}

Table~\ref{tab:collapsed-inference-calibration} evaluates each score against its own target. The forest mean's standard-error-to-sampling-SD ratios are 0.87 and 0.99 for overlap and DR under ordinary overlap, and 0.89 and 0.78 under weaker overlap. For the forest estimates, absolute biases range from 0.002 to 0.033, each within 1.64 Monte Carlo standard errors of zero. The oracle overlap estimator under ordinary overlap has a bias of 0.025, or 1.70 Monte Carlo standard errors. For the forest estimates, the coverage shortfall accompanies low estimated dispersion and limited evidence of bias at the precision of this experiment. Median aggregation changes both the point estimate and variance summary, lengthens intervals and raises coverage. Yet weaker-overlap DR coverage remains 89.50\%, compared with oracle coverage of 88.75\%. These fixed-$N$ experiments do not establish convergence or coverage for the banking application.

\subsection{Weight concentration and the control variance component}\label{app:calibration-weight-diagnostics}
Mean control ESS falls from 48.46 to 9.42 for the oracle and from 35.76 to 8.50 for the forest. The respective largest normalised control weights rise from 0.040 to 0.235 and from 0.075 to 0.275. Although population clipping is inactive, fitted predictions requiring clipping rise from 1.63\% to 26.08\%.
\begin{table}[htbp]
\centering
\caption{Control-weight concentration in the pre/post-score experiment}
\label{tab:calibration-weights}
\small\setlength{\tabcolsep}{4pt}
\begin{tabular}{llrrrrr}
\toprule
& & \multicolumn{2}{c}{Control ESS} & \multicolumn{2}{c}{Largest weight} & Clipped\\
Design & Propensity & Mean & 10th pct. & Mean & 90th pct. & (\%)\\
\midrule
Ordinary & Oracle & 48.46 & 44.17 & 0.040 & 0.044 & 0.00\\
Ordinary & Forest & 35.76 & 22.20 & 0.075 & 0.114 & 1.63\\
Weaker & Oracle & 9.42 & 6.23 & 0.235 & 0.323 & 0.00\\
Weaker & Forest & 8.50 & 3.49 & 0.275 & 0.461 & 26.08\\
\bottomrule
\end{tabular}
\par\smallskip
\begin{minipage}{.97\linewidth}\footnotesize
Notes: The control weights are proportional to $(1-G_i)\widehat p_i/(1-\widehat p_i)$ after clipping the propensity to $[0.01,0.99]$, and are normalised to sum to one. ESS is the inverse sum of their squares. Oracle rows summarise 400 independent data draws; forest rows summarise the 4,000 draw--partition combinations in each design. Forest quantiles describe that pooled distribution, whose partitions within a draw are dependent. Clipped is the mean percentage of all bank-level propensity predictions outside $[0.01,0.99]$, before clipping. No true propensity is clipped. The entries are calculated from the same predictions used in Table~\ref{tab:collapsed-inference-calibration}.
\end{minipage}
\end{table}

Conditional on covariates and groups, oracle control residuals are independent with mean zero and variance one. In Lemma~\ref{lem:weight-centring}, their weighted-mean variance is therefore $V_0=S_2$. The implemented centred variance component is $c_N C_0$, with $c_N=91/90$, and its conditional expectation relative to $V_0$ is $c_N(1-2S_3/S_2+S_2)$. The uncentred oracle quantity $U_0=\sum_i a_i^2\varepsilon_i^2$ is a conditionally unbiased benchmark.
\begin{table}[htbp]
\centering
\caption{Oracle control-variance estimates relative to the weighted mean's conditional variance}
\label{tab:oracle-control-variance}
\small\setlength{\tabcolsep}{5pt}
\begin{tabular}{lrrr}
\toprule
Design & \shortstack{Conditional expected\\centred ratio} & \shortstack{Realised\\centred ratio} & \shortstack{Realised\\uncentred ratio}\\
\midrule
Ordinary & 0.979 (0.0002) & 0.984 (0.014) & 1.008 (0.015)\\
Weaker & 0.757 (0.005) & 0.767 (0.030) & 1.010 (0.040)\\
\bottomrule
\end{tabular}
\par\smallskip
\begin{minipage}{.97\linewidth}\footnotesize
Notes: Means over the 400 saved oracle draws in each design, with Monte Carlo standard errors of those means in parentheses. Conditional on all covariates and assignments, control residuals are independent with variance one. With $S_k=\sum_i a_i^k$, $V_0=S_2$ is the conditional variance of their weighted mean. The first column averages $c_N(1-2S_3/S_2+S_2)$, where $c_N=N/(N-1)=91/90$ is the multiplier in the implemented influence-variance estimate. The second averages $c_N C_0/V_0$, and the third averages $U_0/V_0$, using the notation of Lemma~\ref{lem:weight-centring}. The uncentred oracle diagnostic $U_0$ has no finite-sample multiplier. These are ratios for the control component, not the full ATT variance or interval coverage.
\end{minipage}
\end{table}

The expected ratio averages 0.979 under ordinary overlap and 0.757 under weaker overlap; realised ratios average 0.984 and 0.767, while the uncentred ratios are close to one. This identifies one source of downward bias in the control component for the homoskedastic oracle design. It does not decompose the full estimator's variance, which also involves treated outcomes, random covariates and assignments, and nuisance estimation. Under heteroskedasticity the component bias can have either sign.

Reporting ESS, propensity distributions, control-weight concentration, clipping and partition sensitivity helps describe this problem; none is a validated inference threshold. A different critical value or bootstrap based on the same centred influence contributions is not automatically a correction. Trimming changes the target, while clipping can change the population root under outcome misspecification. Valid small-sample intervals with concentrated ATT weights remain unresolved here.
\FloatBarrier
\section{Aggregation across bank partitions}\label{app:aggregation}
The main pre/post-score tables report medians across ten treatment-stratified five-fold bank partitions. Table~\ref{tab:mean-score-dispersion} also gives their means and the scaled influence-function dispersion calculated after averaging each bank's fitted influence value across partitions. All summaries use the same held-out predictions. The OLS/logit doubly robust estimates have a substantial mean--median difference. In the baseline fits, however, the same comparison bank receives 77.1--87.9\% of normalised control weight across the ten partitions, as documented in Section~\ref{app:clipping}. Weight concentration is therefore persistent across partitions. Both fitted weights and outcome residuals vary with the training sample, and these summaries do not isolate their respective contributions to the mean--median difference.

\begin{table}[htbp]
\centering\footnotesize\setlength{\tabcolsep}{4pt}
\caption{Means across bank partitions and scaled influence-function dispersion}\label{tab:mean-score-dispersion}
\begin{tabular}{@{}llrrrr@{}}
\toprule
Features and nuisance & Score & Mean & Median & \shortstack{Dispersion of\\averaged IF} & Partition range\\
\midrule
Six means, forest & Overlap PLR & 0.0111 & 0.0107 & 0.0038 & [0.0091, 0.0128]\\
Six means, forest & DR, $c=0.01$ & 0.0121 & 0.0125 & 0.0037 & [0.0096, 0.0131]\\
Six means, OLS/logit & Overlap PLR & 0.0085 & 0.0084 & 0.0044 & [0.0068, 0.0113]\\
Six means, OLS/logit & DR, $c=0.01$ & 0.0013 & -0.0048 & 0.0038 & [-0.0123, 0.0373]\\
\addlinespace
Plus prior growth, forest & Overlap PLR & 0.0015 & 0.0015 & 0.0035 & [-0.0001, 0.0026]\\
Plus prior growth, forest & DR, $c=0.01$ & 0.0029 & 0.0026 & 0.0032 & [0.0019, 0.0048]\\
Plus prior growth, OLS/logit & Overlap PLR & -0.0003 & -0.0003 & 0.0035 & [-0.0014, 0.0018]\\
Plus prior growth, OLS/logit & DR, $c=0.01$ & 0.0105 & 0.0065 & 0.0041 & [-0.0082, 0.0449]\\
\bottomrule\end{tabular}
\par\smallskip\begin{minipage}{.98\linewidth}\footnotesize
Notes: Means and medians use the same ten complete-bank partitions, with five folds stratified by treatment status in each. For each bank, the fitted influence-function (IF) value is averaged over partitions; the sample standard deviation across banks is then divided by $\sqrt{91}$. This dispersion of averaged IF values differs from the median of within-partition dispersions and from the median aggregation in Table~\ref{tab:audit-uncertainty}. It describes the fitted influence values. It is not a validated sampling standard error for the empirical procedures: independent-bank sampling, nuisance consistency and rates, score moments, and the relevant clipping conditions have not been established. The columns do not test differences between specifications. Partition ranges describe variation in fitting on the same sample and are not confidence intervals. DR denotes the normalised doubly robust estimator. Main-text point summaries use medians throughout.
\end{minipage}\end{table}

For partition estimates $\widehat\theta_r$, median $\widetilde\theta$ and within-partition scaled influence-function dispersions $s_r$, the median adjustment considered in \citet{chernozhukov2018} is
\[
s_{\mathrm{med}}^2=\operatorname{med}_r\{s_r^2+(\widehat\theta_r-\widetilde\theta)^2\}.
\]
The arithmetic bound
\[
s_{\mathrm{med}}^2\leq\operatorname{med}_r(s_r^2)+\max_r(\widehat\theta_r-\widetilde\theta)^2
\]
is computed from full-precision partition records in the replication materials. Each median of ten values is the arithmetic mean of the two central order statistics. Consequently $\operatorname{med}(s_r^2)$ need not equal $\{\operatorname{med}(s_r)\}^2$, and replacing one by the other need not preserve the bound. The exact median adjustment is computed before rounding. Neither the adjustment nor its upper bound controls nuisance bias or establishes the sampling assumptions for the banking data. Table~\ref{tab:covered-uncertainty} uses the exact aggregation and labels its sampling interpretation as conditional on the stated inference assumptions. Theorem~\ref{thm:collapsed-inference} and Proposition~\ref{prop:dr-inference} state sufficient conditions for an asymptotic interpretation under fixed $R$. Their permitted allocation schemes include the balanced treatment-stratified partitions used here; the required sampling, support and nuisance-rate conditions remain separate restrictions.
\section{Known-target validation of panel transformations}\label{app:transformed-panel-validation}

This experiment implements Example~\ref{ex:quadratic-trend} with repeated samples and separately examines the effect horizons in Corollary~\ref{cor:dynamic-panel}. We generate 1,000 independent balanced panels at each of $N=91$ and $N=364$, with $T=20$, $\mathcal T_0=\{1,\ldots,12\}$ and $\mathcal T_1=\{13,\ldots,20\}$. Banks are independent. Their group indicator is $G_i\sim\operatorname{Bernoulli}(0.25)$, and their covariates satisfy
\begin{align}
\nu_i&=0.8G_i+\eta_i,&\eta_i&\sim N(0,0.7^2),\nonumber\\
X_{it}&=\nu_i+0.025t+\varepsilon_{it},&\varepsilon_{it}&\sim N(0,0.6^2).
\label{eq:transform-sim-covariates}
\end{align}
The disturbances $\eta_i$ and $\varepsilon_{it}$ are mutually independent and independent of $G_i$. This group-conditional construction specifies a joint observational distribution with different covariate levels across groups. The potential covariate history is identical under either exposure path; the construction does not posit a causal effect of treatment on $X_{it}$.

Let $P_t=\ind\{t\geq13\}$ and $D_{it}=G_iP_t$. Untreated outcomes are
\begin{equation}
Y_{it}(0)=\alpha_i+\lambda_t+X_{it}^2+U_{it},\qquad
\alpha_i=\nu_i+0.5G_i+\zeta_i,\qquad
\lambda_t=0.03\sin(2\pi t/20),
\label{eq:transform-sim-outcome}
\end{equation}
where $\zeta_i\sim N(0,0.3^2)$. The error follows a stationary Gaussian AR(1): $U_{i1}\sim N(0,0.5^2)$ and $U_{it}=0.4U_{i,t-1}+\xi_{it}$ for $t\geq2$, with $\xi_{it}\sim N(0,0.5^2(1-0.4^2))$. The $\zeta_i$, initial errors and innovations are independent of the covariate and group histories and of one another. The bank effect may therefore be correlated with group membership and covariates, while the required strict conditional mean restriction on $U_{it}$ holds. The conditional group probabilities are strictly between zero and one, and both transformed scores have positive residual treatment variance.

\subsection{Holding the fitted basis and sample division fixed}
The constant-effect experiment observes $Y_{it}=Y_{it}(0)+0.2D_{it}$. Define $M_iR_{it}=R_{it}-T^{-1}\sum_sR_{is}$, applied componentwise, and $b(x)=(x,x^2)'$. We compare two nuisance-regression designs:
\begin{equation}
B^{\mathrm{level}}_{it}=M_i b(X_{it}),\qquad
B^{\mathrm{input}}_{it}=M_i b(M_iX_{it}).
\label{eq:transform-sim-basis}
\end{equation}
The additional demeaning in $B^{\mathrm{input}}$ gives both designs zero bank means. Thus differences do not arise merely because squaring a demeaned covariate reintroduces a bank-level component. Each design contains two covariate columns and the same 19 bank-demeaned quarter indicators, with no separate intercept. The dimension, time adjustment and fitting rule are identical.

Within each draw, a random permutation divides banks into five approximately equal folds, without stratification on group membership. The same folds are used for both designs. Each training regression fits $M_iY_{it}$ and $M_iD_{it}$ by OLS on its 21 columns, retaining all quarters of a bank together. Held-out predictions are combined in the partialling-out ratio. The correctly transformed basis contains $M_iX_{it}^2$, so its restricted outcome class is correctly specified and its population coefficient is 0.2. This argument uses the restricted projection result; it does not identify the fitted treatment projection with the unrestricted full-history propensity.

For the second design, $M_iX_{it}$ removes $\nu_i$, while $M_iX_{it}^2$ contains the omitted group-dependent trend. The population coefficient is
\[
0.2+2(0.025)(16.5-6.5)(0.8)=0.6.
\]
The terms 16.5 and 6.5 are the mean post- and pre-treatment dates. The corresponding bias relative to the structural coefficient is 0.4. In Table~\ref{tab:transformed-panel-validation}, the fitted means are 0.203 and 0.198 for $B^{\mathrm{level}}$, and 0.602 and 0.597 for $B^{\mathrm{input}}$, at $N=91$ and $N=364$, respectively. The paired differences average 0.3991 and 0.3989, with Monte Carlo standard errors 0.0047 and 0.0022. Increasing the number of banks reduces dispersion without removing the misspecification bias. All training matrices have rank 21.

\subsection{Oracle first-difference and within effect horizons}
On the same underlying draws, replace the constant effect by
\[
Y_{it}=Y_{it}(0)+G_i h_t,\qquad
h_t=P_t\{0.2+0.04(t-13)\}.
\]
Effects are constant across banks at a given date. Their adoption-quarter value is 0.2 and their average over the eight post-treatment quarters is 0.34. This part isolates the difference in effect horizons. It does not evaluate the effect of heterogeneous bank-level responses on the two weighting distributions.

The conditional means are available exactly. For any $k$ observed dates, let $s_i$ be the mean of the corresponding detrended covariates $X_{it}-0.025t$. Gaussian conditioning gives
\begin{equation}
p_k(s_i)=\Lambda\left\{\log\frac{0.25}{0.75}
 +\frac{0.8(s_i-0.4)}{0.7^2+0.6^2/k}\right\},
\label{eq:transform-sim-propensity}
\end{equation}
where $\Lambda(v)=(1+e^{-v})^{-1}$. For FD, use $k=2$ and the pair at each date, denoting the resulting probability by $p_{2,it}$. For within demeaning, use the full history with $k=20$, denoting its probability by $p_{20,i}$. With $q=8/20$, the oracle nuisance functions are
\begin{align}
m^\Delta_{it}&=\ind\{t=13\}p_{2,it},&
\ell^\Delta_{it}&=\Delta\lambda_t+\Delta(X_{it}^2)+p_{2,it}\Delta h_t,
\label{eq:transform-sim-fd-oracle}\\
m^W_{it}&=(P_t-q)p_{20,i},&
\ell^W_{it}&=\lambda_t-\bar\lambda+M_i(X_{it}^2)
 +p_{20,i}(h_t-\bar h).
\label{eq:transform-sim-within-oracle}
\end{align}
Here $\Delta(X_{it}^2)=X_{it}^2-X_{i,t-1}^2$, $\bar\lambda=T^{-1}\sum_t\lambda_t$ and $\bar h=T^{-1}\sum_th_t$. In particular, the within outcome mean uses the entire dynamic path, and the FD outcome mean includes its subsequent increments even though the treatment residual is zero outside adoption. These are the true conditional means for the respective information sets, rather than finite-basis approximations to them.

The oracle FD means are 0.203 and 0.200, and the oracle within means are 0.341 and 0.338, at the two sample sizes. All four mean estimates are within two Monte Carlo standard errors of their respective population targets. The paired within-minus-FD differences average 0.1387 and 0.1383, against the population difference of 0.14. Panel-score estimates equal their corresponding adoption-quarter or pre/post collapsed ratios to within $4.5\times10^{-16}$ in every draw. No confidence-interval coverage claim is made in this experiment; it assesses specification bias and known effect horizons.

\subsection{Reproduction records}
The program \texttt{code/simulations/transformed\_panel\_validation.py} generates the experiment. It uses NumPy seed sequences $[2026100913,N,r]$, where $r=0,\ldots,999$ indexes the draw. The replication records contain all 8,000 coefficient estimates, numerical summaries, paired differences, design metadata and verification results. The table is generated directly from these records. The program's \texttt{--verify} option rebuilds the summaries, checks the exact score identities, checks the training ranks and reproduces the first draw at each sample size. Its \texttt{--smoke} option writes a separate ten-draw experiment without replacing the retained full results.
\section{Regularised and boosted nuisance fits}\label{app:learner-comparison}
Table~\ref{tab:learner-comparison} adds polynomial LASSO and XGBoost to the existing bank-level comparison. The 91 banks, size-group assignment, growth-change outcome, six pre-period covariate means and prior-growth augmentation are unchanged. The OLS/$L_2$-logit and random-forest predictions are reused with their fixed settings; only LASSO and XGBoost undergo the nested hyperparameter search described below. The comparison therefore concerns specified fitting procedures, including their tuning rules, rather than isolating learner families under equal tuning. These cross-fitted OLS/logit entries differ from the full-sample linear regressions discussed in the main text. No estimate is selected by its significance or proximity to the replicated panel coefficient.

The polynomial dictionary contains the $d$ levels, all $d(d+1)/2$ quadratic products and the $d$ pure cubes, giving 33 terms for $d=6$ and 42 for $d=7$. It contains no mixed cubic terms. Each column is standardised using only the relevant training banks, with an unpenalised intercept. Gaussian LASSO minimises mean squared loss divided by two plus $\lambda\|\beta\|_1$; binomial LASSO minimises mean log loss plus the same penalty. The grid contains 50 logarithmically spaced values from $10^{-5}$ to 1. XGBoost uses binary logistic loss for assignment and squared-error loss for each outcome nuisance. Its eight settings cross depth $\{1,2\}$, tree count $\{100,300\}$ and learning rate $\{0.03,0.1\}$, with all banks and features available at each boosting step, minimum child weight 1, zero split penalty and $L_1$ penalty, and $L_2$ penalty 1. The replication files pin XGBoost 3.0.5 and record all settings and fitted models.

Both learners reuse the ten saved, treatment-stratified five-fold bank partitions. The partition seeds range from 20260905 to 20260914. Within each outer training set, three inner bank folds select each nuisance setting by mean held-out Brier loss for assignment or mean squared error for the outcome. Assignment and the all-bank outcome fit use stratified inner folds; the control-outcome fit uses only training comparison banks. The inner and learner seed is the outer partition seed plus $1000k$, where $k$ is the zero-based outer fold index. Ties within $10^{-12}$ favour the larger penalty or, for boosting, smaller depth, fewer trees and then the smaller learning rate. Scaling is recomputed within every inner training fold. The inherited preprocessing of the supplied panel is preserved; all additional scaling and tuning are training-only.

The outer training samples contain 72--73 banks, including 16 larger banks, and 56--57 comparison banks for the control regression. Inner control regressions have 37--38 banks, fewer than the 42 augmented dictionary terms. Regularisation is therefore relevant to these fitted approximations even though both dictionaries are smaller than the full bank count. The exercise does not isolate the effect of regularisation from dictionary choice or establish superior performance under high-dimensional asymptotics.

\begin{table}[!htbp]
\centering\footnotesize
\caption{Matched bank-level comparison of four nuisance-learner families}
\label{tab:learner-comparison}
\setlength{\tabcolsep}{4pt}
\begin{tabular}{@{}llrrrrrr@{}}
\toprule
Learner & Features & \shortstack{Overlap\\PLR} & \shortstack{Normalised\\DR-DiD} & Brier & \shortstack{Control\\RMSE} & \shortstack{Control\\ESS} & \shortstack{Largest\\weight} \\
\midrule
OLS/$L_2$ logit & Six means & \shortstack{$0.0084$\\$(0.0046)$} & \shortstack{$-0.0048^{\dagger}$\\$(0.0069)$} & 0.1245 & 0.0208 & 1.4 & 0.845 \\[3pt]
 & $+$ prior growth & \shortstack{$-0.0003$\\$(0.0037)$} & \shortstack{$0.0065^{\dagger}$\\$(0.0095)$} & 0.1171 & 0.0150 & 1.4 & 0.850 \\[3pt]
\addlinespace[4pt]
Random forest & Six means & \shortstack{$0.0107$\\$(0.0042)$} & \shortstack{$0.0125$\\$(0.0040)$} & 0.1242 & 0.0200 & 17.6 & 0.114 \\[3pt]
 & $+$ prior growth & \shortstack{$0.0015$\\$(0.0037)$} & \shortstack{$0.0026$\\$(0.0035)$} & 0.1215 & 0.0148 & 20.7 & 0.107 \\[3pt]
\addlinespace[4pt]
Polynomial LASSO & Six means & \shortstack{$0.0095$\\$(0.0045)$} & \shortstack{$-0.0033^{\dagger}$\\$(0.0040)$} & 0.1247 & 0.0201 & 1.3 & 0.873 \\[3pt]
 & $+$ prior growth & \shortstack{$0.0015$\\$(0.0033)$} & \shortstack{$-0.0053^{\dagger}$\\$(0.0036)$} & 0.1242 & 0.0145 & 1.3 & 0.866 \\[3pt]
\addlinespace[4pt]
XGBoost & Six means & \shortstack{$0.0106$\\$(0.0045)$} & \shortstack{$0.0117$\\$(0.0044)$} & 0.1438 & 0.0200 & 19.8 & 0.129 \\[3pt]
 & $+$ prior growth & \shortstack{$0.0024$\\$(0.0036)$} & \shortstack{$0.0046$\\$(0.0035)$} & 0.1278 & 0.0155 & 16.1 & 0.189 \\[3pt]
\bottomrule
\end{tabular}
\vspace{3pt}
\begin{minipage}{\linewidth}\scriptsize
\emph{Notes:} The same 91 banks (20 larger and 71 comparison banks), outcome change, five bank folds and ten saved partitions are used throughout. Entries are partition medians; parentheses contain nominal standard errors combining influence-function and partition variation as in the existing aggregation rule. OLS/$L_2$ logit and forest predictions are reused without refitting. LASSO and XGBoost use nested three-fold bank validation. Brier loss concerns group assignment; control RMSE concerns held-out comparison-bank outcome changes. ESS and largest weight refer only to normalised DR control weights, with propensities clipped to $[0.01,0.99]$. Overlap PLR uses unmodified predictions and targets an overlap-weighted effect under its identifying restrictions. Normalised DR-DiD targets the ATT under its separate conditional-trends, nuisance and support conditions; active clipping can alter its probability limit if the control regression is misspecified. The nominal errors do not validate these conditions or finite-sample coverage. $\dagger$ marks inference-fragile DR diagnostics with very low control ESS. The point estimates and nominal standard errors are retained, but their nominal intervals are not empirically calibrated for such concentrated weights.
\end{minipage}
\end{table}

\paragraph{Estimates and predictive performance.}
The overlap medians are 0.0084--0.0107 with the six means and $-0.0003$--0.0024 after adding prior growth. All ten paired changes are negative for each of the four learner families. Their median changes (nominal standard errors) are $-0.0087$ (0.0034) for OLS/logit, $-0.0103$ (0.0040) for forests, $-0.0078$ (0.0034) for LASSO and $-0.0072$ (0.0038) for XGBoost. The XGBoost paired interval includes zero. Thus the direction of the overlap adjustment extends across regularised polynomial and tree approximations, without uniform inferential precision. Conditioning on prior growth can change both the identifying restriction and the overlap-weighted target.

All four families improve assignment Brier loss over the training-fold constant benchmark of 0.1715. Baseline outcome prediction is less informative: control RMSE is about 0.0200--0.0208, close to the constant benchmark of 0.0201. Adding prior growth lowers control RMSE to 0.0145--0.0155. Predictive losses alone do not establish the nuisance convergence rates or support conditions needed for causal inference.

\paragraph{DR estimates and concentrated control weights.}
XGBoost gives DR medians 0.0117 and 0.0046, compared with forest medians 0.0125 and 0.0026. The loss-minimising polynomial LASSO instead gives $-0.0033$ and $-0.0053$. Its median control effective sample sizes are only 1.31 and 1.33 despite clipping at 0.01; the largest normalised weights are 0.873 and 0.866. A comparison bank with identifier 1883693 has median fitted assignment probabilities 0.9988 and 0.9930 under LASSO, versus 0.5736 and 0.5684 under forests. The reused $L_2$-logit fits also concentrate control weights and yield unstable DR estimates. These rows demonstrate sensitivity of odds weighting to the fitted assignment function. Table~\ref{tab:learner-comparison} marks them as inference-fragile diagnostics: their nominal intervals have not been empirically calibrated for such concentrated weights.

Saved-prediction substitutions help diagnose the discrepancy without selecting a preferred estimator. Averaging the two orders of substitution attributes 0.0122 of the 0.0156 mean baseline LASSO--forest DR gap in magnitude to the propensity fit. This is an arithmetic decomposition, not evidence that either estimate is less biased. Raising the clipping threshold from 0.01 to 0.05 changes the LASSO baseline median from $-0.0033$ to 0.0022, while its augmented median remains negative at $-0.0037$. These are clipping diagnostics, not alternative estimates selected for agreement.

\paragraph{Penalty sensitivity and practical implications.}
The largest LASSO penalty within one estimated standard error of the minimum inner-fold loss can be evaluated from the same fitted paths. This rule gives overlap medians 0.0138 and 0.0040 and DR medians 0.0084 and 0.0027. Median DR control effective sample sizes increase to 9.2 and 15.6, while Brier loss rises to 0.1421 and 0.1442. Twenty-two of the 300 loss-minimising LASSO selections lie at the upper grid endpoint, all in baseline outcome regressions and all selecting an intercept-only fit; none lies at the lower endpoint. No effect-driven grid expansion is undertaken.

The comparison supports broader learner robustness of the overlap contraction and agreement between the two tree families for DR. It does not support learner-invariant ATT estimates. A useful implementation must examine predictive performance together with fitted odds, weight concentration and partition variation. The saved outputs retain all ten partitions, fitted nuisance paths, alternative penalty choices, clipping diagnostics, influence values and paired comparisons. Nominal uncertainty uses the existing bank-level influence formulas and repeated-partition aggregation. The existing forest simulation does not validate coverage for the new learners, and no additional simulation is used here.

\end{document}